\documentclass[10pt,letterpaper]{article}

\usepackage[utf8]{inputenc}
\usepackage{amsmath}
\usepackage{amsfonts}
\usepackage{amssymb}
\usepackage{graphicx}
\usepackage{hyperref}
\usepackage{authblk}
\usepackage[numbers,sort&compress,square]{natbib}

\usepackage{subfig}
\usepackage{morefloats}

\usepackage[lined,boxed,linesnumbered,ruled,vlined]{algorithm2e}

\SetCommentSty{mycommfont}

\usepackage{multirow}
\usepackage{hhline}
\usepackage{url}
\usepackage[usenames,dvipsnames]{color}

\title{Quantifying randomness in real networks}
\date{}

\author[1,2]{Chiara~Orsini}
\author[3,4]{Marija~Mitrovi\'{c} Dankulov}
\author[5]{Pol~Colomer-de-Sim\'{o}n}
\author[6]{Almerima~Jamakovic}
\author[7]{Priya~Mahadevan}
\author[8]{Amin~Vahdat}
\author[9,10]{Kevin~E.~Bassler}
\author[11]{Zolt\'{a}n~Toroczkai}
\author[5]{Mari\'{a}n~Bogu\~{n}\'{a}}
\author[12]{Guido~Caldarelli}
\author[13]{Santo~Fortunato}
\author[1,14]{Dmitri~Krioukov}

\affil[1]{\normalsize CAIDA, University of California San Diego, San Diego, USA}
\affil[2]{Information Engineering Department, University of Pisa, Pisa, Italy}
\affil[3]{Scientific Computing Laboratory, Institute of Physics Belgrade, University of Belgrade, Belgrade, Serbia}
\affil[4]{Department of Biomedical Engineering and Computational Science, Aalto University School of Science, Helsinki, Finland}
\affil[5]{Departament de Física Fonamental, Universitat de Barcelona, Barcelona, Spain}
\affil[6]{Communication and Distributed Systems group, Institute of Computer Science and Applied Mathematics, University of Bern, Bern, Switzerland}
\affil[7]{Palo Alto Research Center, Palo Alto, CA, USA}
\affil[8]{Google, Mountain View, CA, USA}
\affil[9]{Department of Physics and Texas Center for Superconductivity, University of Houston, Houston, TX, USA}
\affil[10]{Max Planck Institut f{\"u}r Physik komplexer Systeme, Dresden, Germany}
\affil[11]{Department of Physics and Interdisciplinary Center for Network Science and Applications, University of Notre Dame, Notre Dame, IN, USA}
\affil[12]{IMT Alti Studi, Lucca, Italy}
\affil[13]{Department of Computer Science, Aalto University School of Science, Helsinki, Finland}
\affil[14]{Department of Physics, Department of Mathematics, Department of Electrical\&Computer Engineering, Northeastern University, Boston, MA, USA}

\begin{document}

\maketitle

\begin{abstract}

Represented as graphs, real networks are intricate combinations of order and disorder.
Fixing some of the structural properties of network models to their values observed in real networks, many other properties
appear as statistical consequences of these fixed observables, plus randomness in other respects.
Here we employ the $dk$-series, a complete set of basic characteristics of the network structure, to study the statistical dependencies between different network properties. We consider six real networks---the Internet, US airport network, human protein interactions, technosocial web of trust, English word network, and an fMRI map of the human brain---and find that many important local and global structural properties of these networks are closely reproduced by $dk$-random graphs whose degree distributions, degree correlations, and clustering are as in the corresponding real network. We discuss important conceptual, methodological, and practical implications of this evaluation of network randomness, and release software to generate $dk$-random graphs.

\end{abstract}

\section*{Introduction}

Network science studies complex systems by representing them as networks~\cite{Newman10-book}. This approach has proven quite fruitful because in many cases the network representation achieves a practically useful balance between simplicity and realism: while always grand simplifications of real systems, networks often encode some crucial information about the system. Represented as a network, the system structure is fully specified by the network adjacency matrix, or the list of connections, perhaps enriched with some additional attributes. This (possibly weighted) matrix is then a starting point of research in network science.

One significant line of this research studies various (statistical) properties of adjacency matrices of real networks. The focus is often on properties that convey useful information about the global network structure that affects the dynamical processes in the system that this network represents~\cite{BarratVespignani-book08}. A common belief is that a self-organizing system should evolve to a network structure that makes these dynamical processes, or network functions, efficient~\cite{newman03c-review,MotterOptimization2007,BurdaMotifs2011}. If this is the case, then given a real network, we may ``reverse engineer'' it by showing that its structure optimizes its function. In that respect the problem of interdependency between different properties becomes particularly important~\cite{VazquezTopologicalRelationship2004,GuimeraClasses2007,TakemotoStructure2007,FosterClustering2011,ColomerClustering2013}.

Indeed, suppose that the structure of some real network has property~$X$---some statistically over- or under-represented subgraph, or motif~\cite{MiSh02-motifs}, for example---that we believe is related to a particular network function. Suppose also that the same network has in addition property~$Y$---some specific degree distribution or clustering, for example---and that all networks that have property~$Y$ necessarily have property~$X$ as a consequence. Property~$Y$ thus enforces or ``explains'' property~$X$, and attempts to ``explain''~$X$ by itself, ignoring~$Y$, are misguided.
For example, if a network has high density (property~$Y$), such as the interarial cortical network in the primate brain where 66\% of edges that could exist do exist~\cite{Markov2013}, then it will necessarily have short path lengths and high clustering, meaning it is a small-world network (properties~$X$). However, unlike social networks where the small-world property is an independent feature of the network, in the brain this property is a simple consequence of high density.

The problem of interdependencies among network properties has been long understood~\cite{AmaralLies2006,CoFlSeVe06}. The standard way to address it, is to generate many graphs that have property~$Y$ and that are random in all other respects---let us call them {\it $Y$-random graphs}---and then to check if property~$X$ is a typical property of these $Y$-random graphs. In other words, this procedure checks if graphs that are sampled uniformly at random from the set of all graphs that have property~$Y$, also have property~$X$ with high probability. For example, if graphs are sampled from the set of graphs with high enough edge density, then all sampled graphs will be small worlds. If this is the case, then $X$ is not an interesting property of the considered network, because the fact that the network has property~$X$ is a statistical consequence of that it also has property~$Y$. In this case we should attempt to explain~$Y$ rather than~$X$. In case $X$ is not a typical property of $Y$-random graphs, one cannot really conclude that property $X$ is interesting or important (for some network functions). The only conclusion one can make is that $Y$ cannot explain~$X$, which does not mean however that there is no other property~$Z$ from which $X$ follows.

In view of this inherent and unavoidable relativism with respect to a null model, the problem of structure-function relationship requires an answer to the following question in the first place: what is the right base property or properties~$Y$ in the null model ($Y$-random graphs) that we should choose to study the (statistical) significance of a given property~$X$ in a given network~\cite{Zscore2015}? For most properties~$X$ including motifs~\cite{MiSh02-motifs}, the choice of~$Y$ is often just the degree distribution. That is, one usually checks if $X$ is present in random graphs with the same degree distribution as in the real network. Given that scale-free degree distributions are indeed the striking and important features of many real networks~\cite{Newman10-book}, this null model choice seems natural, but there are no rigorous and successful attempts to justify it. The reason is simple: the choice cannot be rigorously justified because there is nothing special about the degree distribution---it is one of infinitely many ways to specify a null model.

Since there exists no unique preferred null model, we have to consider a series of null models satisfying certain requirements. Here we consider a particular realization of such series---the $dk$-series~\cite{MaKrFaVa06-phys}, which provides a complete systematic basis for network structure analysis, bearing some conceptual similarities with a Fourier or Taylor series in mathematical analysis. The $dk$-series is a converging series of basic interdependent degree- and subgraph-based properties that characterize the local network structure at an increasing level of detail, and define a corresponding series of null models or random graph ensembles. These random graphs have the same distribution of differently sized subgraphs as in a given real network. Importantly, the nodes in these subgraphs are labeled by node degrees in the real network. Therefore this random graph series is a natural generalization of random graphs with fixed average degree, degree distribution, degree correlations, clustering, and so on. Using $dk$-series we analyze six real networks, and find that they are essentially random as soon as we constrain their degree distributions, correlations, and clustering to the values observed in the real network ($Y$=degrees+correlations+clustering). In other words, these basic local structural characteristics almost fully define not only local but also global organization of the considered networks. These findings have important implications on research dealing with network structure-function interplay in different disciplines where networks are used to represent complex natural or designed systems. We also find that some properties of some networks cannot be explained by just degrees, correlations, and clustering. The $dk$-series methodology thus allows one to detect which particular property in which particular network is non-trivial, cannot be reduced to basic local degree- or subgraph-based characteristics, and may thus be potentially related to some network function.

\section*{Results}

The introductory remarks above instruct one to look not for a single base property~$Y$, which cannot be unique or universal, but for a systematic series of base properties $Y_0, Y_1,\ldots$. By ``systematic'' we mean the following conditions: 1)~\textit{inclusiveness}, that is, the properties in the series should provide strictly more detailed information about the network structure, which is equivalent to requiring that networks that have property~$Y_d$ ($Y_d$-random graphs), $d>0$, should also have properties~$Y_{d'}$ for all $d'=0,1,\ldots,d-1$; and
2)~\textit{convergence}, that is, there should exist property $Y_D$ in the series that fully characterizes the adjacency matrix of any given network, which is equivalent to requiring that $Y_D$-random graphs is only one graph---the given network itself.
If these $Y$-series satisfy the conditions above, then whatever property~$X$ is deemed important now or later in whatever real network, we can always standardize the problem of explanation of~$X$ by reformulating it as the following question: what is the minimal value of~$d$ in the above $Y$-series such that property~$Y_d$ explains~$X$? By convergence, such~$d$ should exist; and by inclusiveness, networks that have property $Y_{d'}$ with any $d'=d,d+1,\ldots,D$, also have property~$X$. Assuming that properties~$Y_d$ are once explained, the described procedure provides an explanation of any other property of interest~$X$.

The general philosophy outlined above is applicable to undirected and directed networks, and it is shared by different approaches, including motifs~\cite{MiSh02-motifs}, graphlets~\cite{PrzuljHiddenLanguage2014}, and similar constructions~\cite{NewmanDseries2010}, albeit they violate the inclusiveness condition as we show below. Yet one can still define many different $Y$-series satisfying both conditions above. Some further criteria are needed to focus on a particular one. One approach is to use degree-based tailored random graphs as null models for both undirected~\cite{Coolen2011,CoMa09,AnCo09} and directed~\cite{Roberts2011a,Roberts2012} networks. The criteria that we use to select a particular $Y$-series in this study are simplicity and the importance of subgraph- and degree-based statistics in networks. Indeed, in the network representation of a system, subgraphs, their frequency and convergence are the most natural and basic building blocks of the system, among other things forming the basis of the rigorous theory of graph family limits known as graphons~\cite{LovaszBook2012}, while the degree is the most natural and basic property of individual nodes in the network. Combining the subgraph- and degree-based characteristics leads to \emph{$dk$-series}~\cite{MaKrFaVa06-phys}.

\subsection*{$dk$-series}

\begin{figure*}[h]
\centerline{\includegraphics[width=.75\textwidth]{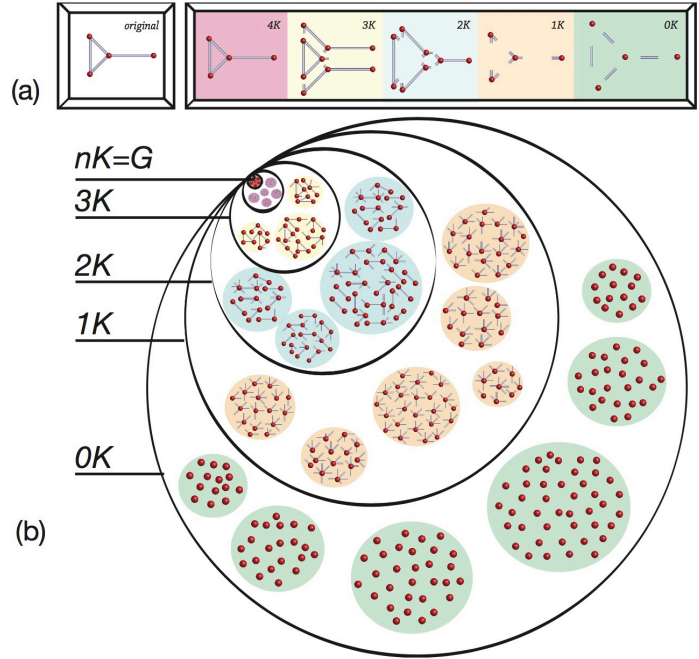}}
\caption{
\textbf{The $dk$-series illustrated. (a)} shows the $dk$-distributions for a graph of size $4$. The $4k$-distribution is the graph itself. The $3k$-distribution consists of its three subgraphs of size $3$: one triangle connecting nodes of degrees $2$, $2$, and $3$, and two wedges connecting nodes of degrees $2$, $3$, and $1$. The $2k$-distribution is the joint degree distribution in the graph. It specifies the number of links (subgraphs of size $2$) connecting nodes of different degrees: one link connects nodes of degrees $2$ and $2$, two links connect nodes of degrees $2$ and $3$, and one link connects nodes of degree $3$ and $1$. The $1k$-distribution is the degree distribution in the graph. It lists the number of nodes (subgraphs of size $1$) of different degree: one node of degree $1$, two nodes of degree $2$, and one node of degree $3$. The $0k$-distribution is just the average degree in the graph, which is $2$. \textbf{(b)} illustrates the inclusiveness and convergence of $dk$-series by showing the hierarchy of $dk$-graphs, which are graphs that have the same $dk$-distribution as a given graph $G$ of size $n$. The black circles schematically shows the sets of $dk$-graphs. The set of $0k$-graphs, that is, graphs that have the same average degree as $G$, is largest. Graphs in this set may have a structure drastically different from $G$'s. The set of $1k$-graphs is a subset of $0k$-graphs, because each graph with the same degree distribution as in $G$ has also the same average degree as $G$, but not {\it vice versa}. As a consequence, typical $1k$-graphs, that is, $1k$-random graphs, are more similar to $G$ than $0k$-graphs. The set of $2k$-graphs is a subset of $1k$-graphs, also containing $G$. As $d$ increases, the circles become smaller because the number of different $dk$-graphs decreases. Since all the $dk$-graph sets contain $G$, the circles ``zoom-in'' on it, and while their number decreases, $dk$-graphs become increasingly more similar to $G$. In the $d=n$ limit, the set of $nk$-graphs consists of only one element, $G$ itself.
\label{fig:dk-series-appendix}
}
\end{figure*}

In $dk$-series, properties $Y_d$ are {\it $dk$-distributions}. For any given network~$G$ of size~$N$, its $dk$-distribution is defined as a collection of distributions of $G$'s subgraphs of size~$d=0,1,\ldots,N$ in which nodes are labeled by their degrees in~$G$. That is, two isomorphic subgraphs of~$G$ involving nodes of different degrees---for instance, edges ($d=2$) connecting nodes of degrees $1,2$ and $2,2$---are counted separately. The $0k$-``distribution'' is defined as the average degree of~$G$. Figure~\ref{fig:dk-series-appendix} illustrates the $dk$-distributions of a graph of size~$4$.

Thus defined the $dk$-series subsumes all the basic degree-based characteristics of networks of increasing detail. The zeroth element in the series, the $0k$-``distribution,'' is the coarsest characteristic, the average degree.  The next element, the $1k$-distribution, is the standard degree distribution, which is the number of nodes---subgraphs of size $1$---of degree $k$ in the network. The second element, the $2k$-distribution, is the joint degree distribution, the number of subgraphs of size $2$---edges---between nodes of degrees $k_1$ and $k_2$. The $2k$-distribution thus defines $2$-node degree correlations and network's assortativity. For $d=3$, the two non-isomorphic subgraphs are triangles and wedges, composed of nodes of degrees $k_1$, $k_2$, and $k_3$, which defines clustering, and so on. For arbitrary~$d$ the $dk$-distribution characterizes the `$d$'egree `$k$'orrelations in $d$-sized subgraphs, thus including, on the one hand, the correlations of degrees of nodes located at hop distances below~$d$, and, on the other hand, the statistics of $d$-cliques in~$G$. We will also consider $dk$-distributions with fractional $d\in(2,3)$ which in addition to specifying $2$-node degree correlations ($d=2$), fix some $d=3$ substatistics related to clustering.

The $dk$-series is inclusive because the $(d+1)k$-distribution contains the same information about the network as the $dk$-distribution, plus some additional information. In the simplest $d=0$ case for example, the degree distribution $P(k)$ ($1k$-distribution) defines the average degree $\bar{k}$ ($0k$-distribution) via $\bar{k}=\sum_kkP(k)$. The analogous expression for $d=1,2$ are derived in Supplementary Note 1.

It is important to note that if we omit the degree information, and just count the number of $d$-sized subgraphs in a given network regardless their node degrees, as in motifs~\cite{MiSh02-motifs}, graphlets~\cite{PrzuljHiddenLanguage2014}, or similar constructions~\cite{NewmanDseries2010}, then such degree-$k$-agnostic $d$-series (versus $dk$-series) would not be inclusive (Supplementary Discussion). Therefore preserving the node degree (`$k$') information is necessary to make a subgraph-based (`$d$') series inclusive. The $dk$-series is clearly convergent because at $d=N$ where $N$ is the network size, the $Nk$-distribution fully specifies the network adjacency matrix.

A sequence of $dk$-distributions then defines a sequence of random graph ensembles (null models). The {\it $dk$-graphs} are a set of all graphs with a given $dk$-distribution, for example, with the $dk$-distribution in a given real network. The {\it $dk$-random graphs} are a maximum-entropy ensemble of these graphs~\cite{MaKrFaVa06-phys}. This ensemble consists of all $dk$-graphs, and the probability measure is uniform (unbiased): each graph $G$ in the ensemble is assigned the same probability $P(G)=1/{\mathcal N}_d$, where ${\mathcal N}_d$ is the number of $dk$-graphs. For $d=0,1,2$ these are well studied classical random graphs~${\mathcal G}_{N,M}$~\cite{ErRe59}, configuration model~\cite{BeCa78,NewStrWat01,ChatterjeeDegreeSequences2011}, and random graphs with a given joint degree distribution~\cite{StantonJDD2012}, respectively. Since a sequence of $dk$-distributions is increasingly more informative and thus constraining, the corresponding sequence of the sizes of $dk$-random graph ensembles is non-increasing and shrinking to~$1$, ${\mathcal N}_0 \geq {\mathcal N}_1 \geq \ldots \geq {\mathcal N}_N = 1$, Fig.~\ref{fig:dk-series-appendix}. At low $d=0,1,2$ these numbers ${\mathcal N}$ can be calculated either exactly or approximately~\cite{BianconiEntropy2008,BarvinokNumberOfGraphs2013}.

We emphasize that in $dk$-graphs the $dk$-distribution constraints are sharp, that is, they hold exactly---all $dk$-graphs have exactly the same $dk$-distribution. An alternative description uses soft maximum-entropy ensembles belonging to the general class of exponential random graph models~\cite{HollandExponentialFamily1981,PaNe04,ChatterjeeERGM2013,Horvat2014Degeneracy} in which these constraints hold only on average over the ensemble---the expected $dk$-distribution in the ensemble (not in any individual graph) is fixed to a given distribution. This ensemble consists of all possible graphs~$G$ of size~$N$, and the probability measure $P(G)$ is the one maximizing the ensemble entropy $S=-\sum_GP(G) \ln P(G)$ under the $dk$-distribution constraints.
Using analogy with statistical mechanics, sharp and soft ensemble are often called microcanonical and canonical, respectively.

As a consequence of the convergence and inclusiveness properties of $dk$-series, any network property~$X$ of any given network~$G$ is guaranteed to be reproduced with any desired accuracy by high enough~$d$. At $d=N$ all possible properties are reproduced exactly, but the $Nk$-graph ensemble trivially consists of only one graph, $G$self, and has zero entropy. In the sense that the entropy of $dk$-ensembles $S_d = \ln {\mathcal N}_d$ is a non-increasing function of~$d$, the smaller the~$d$, the more random the $dk$-random graphs, which also agrees with the intuition that $dk$-random graphs are ``the less random and the more structured,'' the higher the~$d$. Therefore the general problem of explaining a given property~$X$ reduces to the general problem of how random a graph ensemble must be so that $X$ is statistically significant. In the $dk$-series context, this question becomes: how much local degree information, that is, information about concentrations of degree-labeled subgraphs of what minimal size~$d$, is needed to reproduce a possibly global property~$X$ with a desired accuracy?

Below we answer this question for a set of popular and commonly used structural properties of some paradigmatic real networks. But to answer this question for any property in any network, we have to be able to sample graphs uniformly at random from the sets of $dk$-graphs---the problem that we discuss next.

\clearpage
\newpage

\clearpage
\newpage

\subsection*{$dk$-random graph sampling}

Soft $dk$-ensembles tend to be more amenable for analytic treatment, compared to sharp ensembles, but even in soft ensembles the exact analytic expressions for expected values are known only for simplest network properties in simplest ensembles~\cite{GarlaschelliLikelihood2011,Squartini2015Sampling}. Therefore we retreat to numeric experiments here. Given a real network $G$, there exist two ways to sample $dk$-random graphs in such experiments: $dk$-randomize $G$ generalizing the randomization algorithms in~\cite{MaSne03,MaSneZa04}, or construct random graphs with $G$'s $dk$-sequence from scratch~\cite{MaKrFaVa06-phys,Gjoka2.5K2013}, also called direct construction~\cite{Kim2009,ZoltanSampling2010,Kim2012,Bassler2015}.

\begin{figure*}
\centerline{\includegraphics[trim=0cm 2cm 0cm 7cm,width=0.75\textwidth]{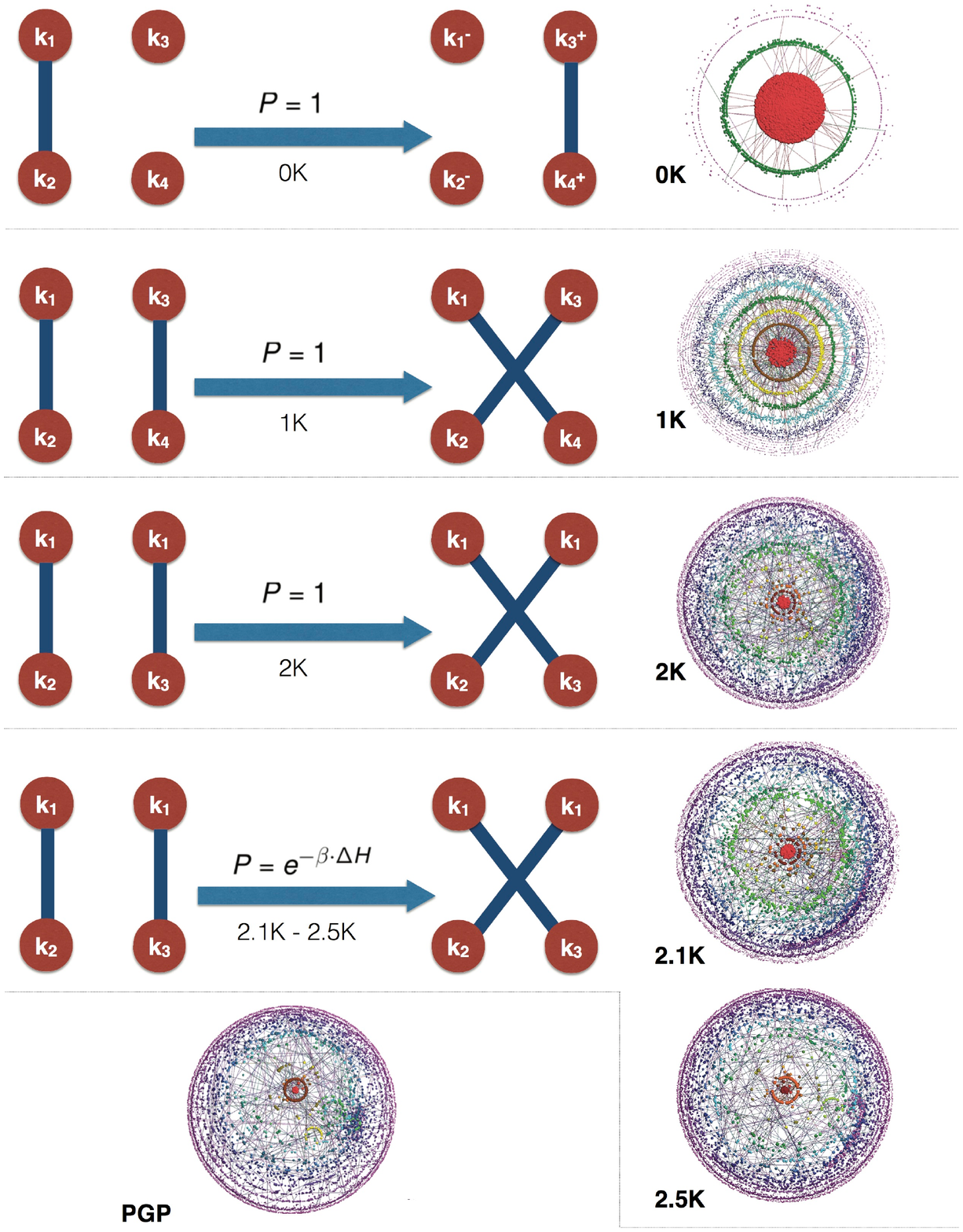}}
\caption{\textbf{The $dk$-sampling and convergence of $dk$-series illustrated.} The left column shows the elementary swaps of $dk$-randomizing (for $d=0,1,2$) and $dk$-targeting (for $d=2.1,2.5$) rewiring. The nodes are labeled by their degrees, and the arrows are labeled by the rewiring acceptance probability. In $dk$-randomizing rewiring, random (pairs of) edges are rewired preserving the graph's $dk$-distribution (and consequently its $d'K$-distributions for all $d'<d$). In $2.1k$- and $2.5k$-targeting rewiring, the moves preserve the $2k$-distribution, but each move is accepted with probability $p$ designed to drives the graph closer to a target value of average clustering~$\bar{c}$ ($2.1k$) or degree-dependent clustering~$c(k)$ ($2.5k$): $p=\min(1,e^{-\beta \Delta H})$, where $\beta$ the inverse temperature of this simulated annealing process, $\Delta H = H_{a} - H_{b}$, and $H_{a,b}$ are the distances, after and before the move, between the current and target values of clustering: $H_{2.1k}=\vert\overline{c}_{\texttt{current}}-\overline{c}_{\texttt{target}}\vert$ and $H_{2.5k}=\sum_i\vert\overline{c}_{\texttt{current}}[k_i]-\overline{c}_{\texttt{target}}[k_i]\vert$. The right column shows LaNet-vi~\cite{AlvarezLaNet-vi2008} visualizations of the results of these $dk$-rewiring processes (Supplementary Methods), applied to the PGP network, visualized at the bottom of the left column. The node sizes are proportional to the logarithm of their degrees, while the color reflects node coreness~\cite{AlvarezLaNet-vi2008}. As $d$ grows, the shown $dk$-random graphs quickly become more similar to the real PGP network.
\label{fig:methods}
}
\end{figure*}

The first option, $dk$-randomization, is easier. It accounts for swapping random (pairs of) edges, starting from $G$, such that the $dk$-distribution is preserved at each swap, Fig.~\ref{fig:methods}. There are many concerns with this prescription~\cite{Zlatic2009RichClub}, two of which are particularly important. The first concern is if this process ``ergodic,'' meaning that if any two $dk$-graphs are connected by a chain of $dk$-swaps. For $d=1$ the 2-edge swap is ergodic~\cite{MaSne03,MaSneZa04}, while for $d=2$ it is not ergodic. However the so-called restricted 2-edge swap, when at least one node attached to each edge has the same degree, Fig.~\ref{fig:methods}, was proven to be ergodic~\cite{ErdosJDD2015}. It is now commonly believed that there is no edge-swapping operation, of this or other type, that is ergodic for the $3k$-distribution, although a definite proof is lacking at the moment. If there exists no ergodic $3k$-swapping, then we cannot really rely on it in sampling $dk$-random graphs because our real network $G$ can be trapped on a small island of atypical $dk$-graphs, which is not connected by any $dk$-swap chain to the main land of many typical $dk$-graphs. Yet we note that in an unpublished work~\cite{JaMa09} we showed that five out of six considered real networks were virtually indistinguishable from their $3k$-randomizations across all the considered network properties, although one network (power grid) was very different from its $3k$-random counterparts.

The second concern with $dk$-randomization is about how close to uniform sampling the $dk$-swap Markov chain is after its mixing time is reached---its mixing time is yet another concern that we do not discuss here, but according to many numerical experiments and some analytic estimates, it is $O(M)$~\cite{MaSne03,MaSneZa04,MaKrFaVa06-phys,Gjoka2.5K2013,StantonJDD2012,ErdosJDD2015}. Even for $d=1$ the swap chain does not sample $1k$-graphs uniformly at random, yet if the edge-swap process is done correctly, then the sampling is uniform~\cite{CoMa09,AnCo09}.

A simple algorithm for the second $dk$-sampling option, constructing $dk$-graphs from scratch, is widely known for $d=1$: given $G$'s degree sequence $\{k_i\}$, build a $1k$-random graph by attaching $k_i$ half-edges (``stubs'') to node $i$, and then connect random pairs of stubs to form edges~\cite{NewStrWat01}. If during this process a self-loop (both stubs are connected to the same node) or double-edge (two edges between the same pair of nodes) is formed, one has to restart the process from scratch since otherwise the graph sampling is not uniform~\cite{MiloUniform2003}. If the degree sequence is power-law distributed with exponent close to $-2$ as in many real networks, then the probability that the process must be restarted approaches $1$ for large graphs~\cite{BasslerSparse2011}, so that this construction process never succeeds. An alternative greedy algorithm is described in~\cite{ZoltanSampling2010}, which always quickly succeeds and gives an efficient way of testing if a given sequence of integers is graphical, that is, if it can be realized as a degree sequence of a graph. The base sampling procedure does not sample graphs uniformly, but then an importance sampling procedure is used to account for the bias, which results in uniform sampling. Yet again, if the degree distribution is a power law, one can show that even without importance sampling, the base sampling procedure is uniform, since the distribution of sampling weights that one can compute for this greedy algorithm approaches a delta function. Extensions of the naive $1k$-construction above to $2k$ are less known, but they exist~\cite{MaKrFaVa06-phys,StantonJDD2012,Gjoka2015JDD,Bassler2015}. Most of these $2k$-constructions do not sample $2k$-graphs exactly uniformly either~\cite{ErdosJDD2015}, but importance sampling~\cite{CoMa09,Bassler2015} can correct for the sampling biases.

Unfortunately, to the best of our knowledge, there currently exists no $3k$-construction algorithm that can be successfully used in practice to generate large $3k$-graphs with $3k$-distributions of real networks. The $3k$-distribution is quite constraining and non-local, so that the $dk$-construction methods described above for $d=1,2$ cannot be readily extended to $d=3$~\cite{MaKrFaVa06-phys}. There is yet another option---$3k$-targeting rewiring, Fig.~\ref{fig:methods}. It is $2k$-preserving rewiring in which each $2k$-swap is accepted not with probability $1$, but with probability equal to $\min(1,\exp(-\beta\Delta H))$, where $\beta$ is the inverse temperature of this simulated-annealing-like process, and $\Delta H$ is the change in the $L^1$ distance between the $3k$-distribution in the current graph and the target $3k$-distribution before and after the swap. This probability favors and, respectively, suppresses $2k$-swaps that move the graph closer or farther from the target $3k$-distribution. Unfortunately we report that in agreement with~\cite{Gjoka2.5K2013} this $2k$-preserving $3k$-targeting process never converged for any considered real network---regardless of how long we let the rewiring code run,
after the initial rapid decrease, the $3k$-distance, while continuing to slowly decrease, remained substantially large.
The reason why this process never converges is that the $3k$-distribution is extremely constraining, so that the number of $3k$-graphs ${\mathcal N}_3$ is infinitesimally small compared to the number of $2k$-graphs ${\mathcal N}_2$, ${\mathcal N}_3/{\mathcal N}_2\ll1$~\cite{MaKrFaVa06-phys,BianconiEntropy2008}. Therefore it is extremely difficult for the $3k$-targeting Markov chain to find a rare path to the target $3k$-distribution, and the process gets hopelessly trapped in abundant local minima in distance $H$.

Therefore, on the one hand, even though $3k$-randomized versions of many real networks are indistinguishable from the original networks across many metrics~\cite{JaMa09}, we cannot use this fact to claim that at $d=3$ these metrics are not statistically significant in those networks, because the $3k$-randomization Markov chain may be non-ergodic. On the other hand, we cannot generate the corresponding $3k$-random graphs from scratch in a feasible amount of compute time. The $3k$-random graph ensemble is not analytically tractable either. Given that $d=2$ is not enough to guarantee the statistical insignificance of some important properties of some real networks, see~\cite{JaMa09} and below, we, as in~\cite{Gjoka2.5K2013}, retreat to numeric investigations of $2k$-random graphs in which in addition to the $2k$-distribution, some substatistics of the $3k$-distribution is fixed. Since strong clustering is a ubiquitous feature of many real networks~\cite{Newman10-book}, one of the most interesting such substatistics is clustering.

Specifically we study $2.1k$-random graphs, defined as $2k$-random graphs with a given value of average clustering $\bar{c}$, and $2.5k$-random graphs, defined as $2k$-random graphs with given values of average clustering $\bar{c}(k)$ of nodes of degree $k$~\cite{Gjoka2.5K2013}. The $3k$-distribution fully defines both $2.1k$- and $2.5k$-statistics, while $2.5k$ defines $2.1k$. Therefore $2k$-graphs are a superset of $2.1k$-graphs, which are a superset of $2.5k$-graphs, which in turn contain all the $3k$-graphs, ${\mathcal N}_2 > {\mathcal N}_{2.1} > {\mathcal N}_{2.5} >{\mathcal N}_{3}$. Therefore if a particular property is not statistically significant in $2.5k$-random graphs, for example, then it is not statistically significant in $3k$-random graphs either, while the converse is not generally true.

We thus generate 20 $dk$-random graphs with $d=0,1,2,2.1,2.5$ for each considered real network. For $d=0,1,2$ we use the standard $dk$-randomizing swapping, Fig.~\ref{fig:methods}. We do not use its modifications to guarantee exactly uniform sampling~\cite{CoMa09,AnCo09}, because: (1)~even without these modifications the swapping is close to uniform in power-law graphs, (2)~these modifications are non-trivial to efficiently implement, and (3)~we could not extend these modifications to the $2.1k$ and $2.5k$ cases. As a consequence, our sampling is not exactly uniform, but we believe it is close to uniform for the reasons discussed above. To generate $dk$-random graphs with $d=2.1,2.5$, we start with a $2k$-random graph, and apply to it the standard $2k$-preserving $2.xk$-targeting ($x=1,5$) rewiring process, Fig.~\ref{fig:methods}. The algorithms that do that, as described in~\cite{Gjoka2.5K2013}, did not converge on some networks, so that we modified the algorithm in~\cite{ColomerClustering2013} to ensure the convergence in all cases. The details of these modifications are in Supplementary Methods (the parameters used are listed in Supplementary Table 4), along with the details of the software package implementing these algorithms that we release to public~\cite{polcode}.

\clearpage
\newpage

\clearpage
\newpage

\subsection*{Real versus  $dk$-random networks}

\begin{figure*}[t]
\centering
\includegraphics[width=\textwidth]{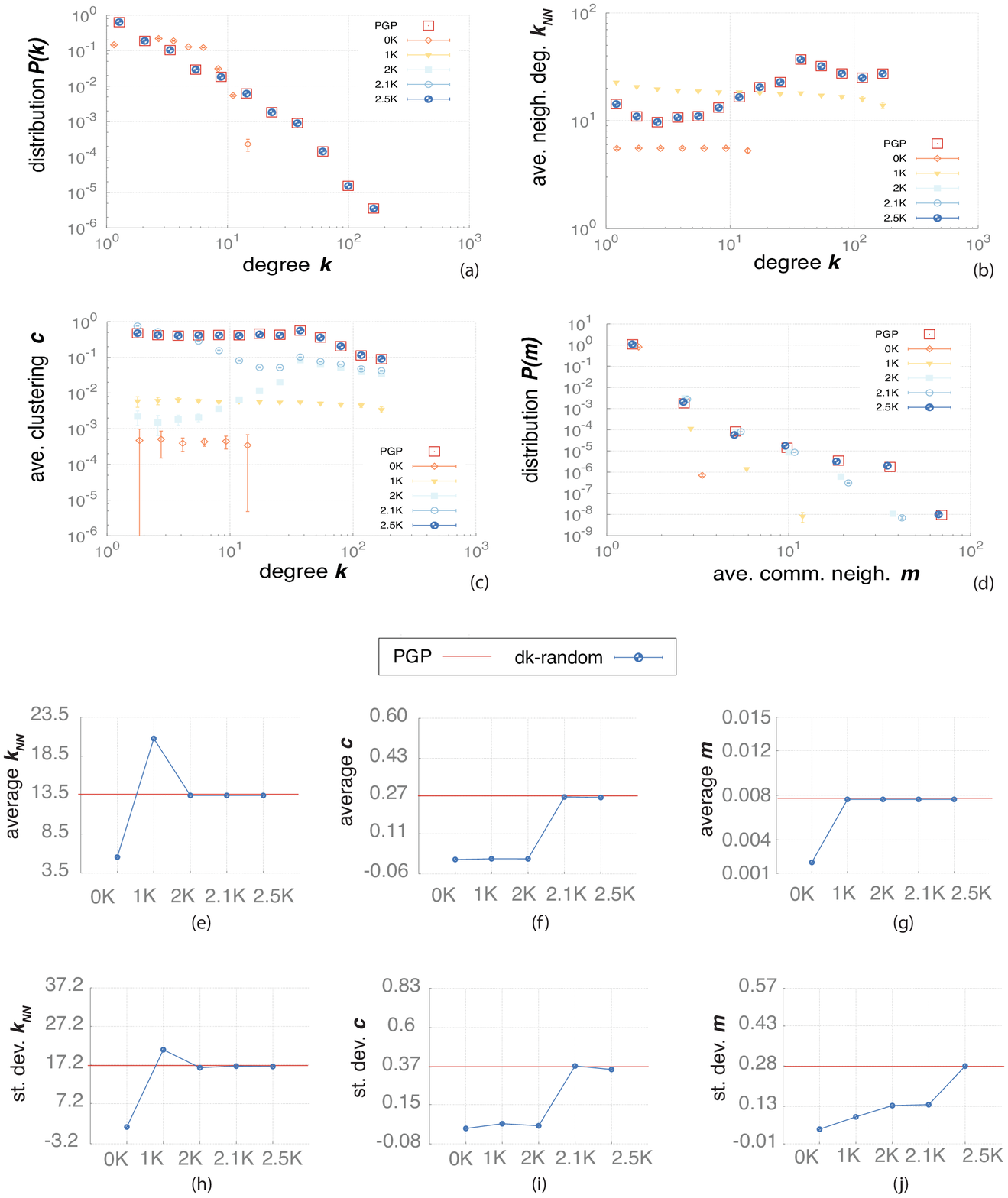}
\caption{\textbf{Microscopic properties of the PGP network and its $dk$-random graphs.} The figure shows \textbf{(a)} the degree distribution $P(k)$, \textbf{(b)} average degree $\bar{k}_{nn}(k)$ of nearest neighbors of nodes of degree $k$, \textbf{(c)} average clustering $\bar{c}(k)$ of nodes of degree $k$, \textbf{(d)} the distribution  $P(m)$ of the number $m$ of common neighbors between all connected pairs of nodes, and \textbf{(e,f,g)} the means and \textbf{(h,i,j)} standard deviations of the corresponding distributions. The error bars indicate the standard deviations of the metrics across different graph realizations.}
\label{fig:local}
\end{figure*}

\begin{figure*}[t]
\centering
\includegraphics[trim=6cm 0cm 0cm 0cm,width=\textwidth]{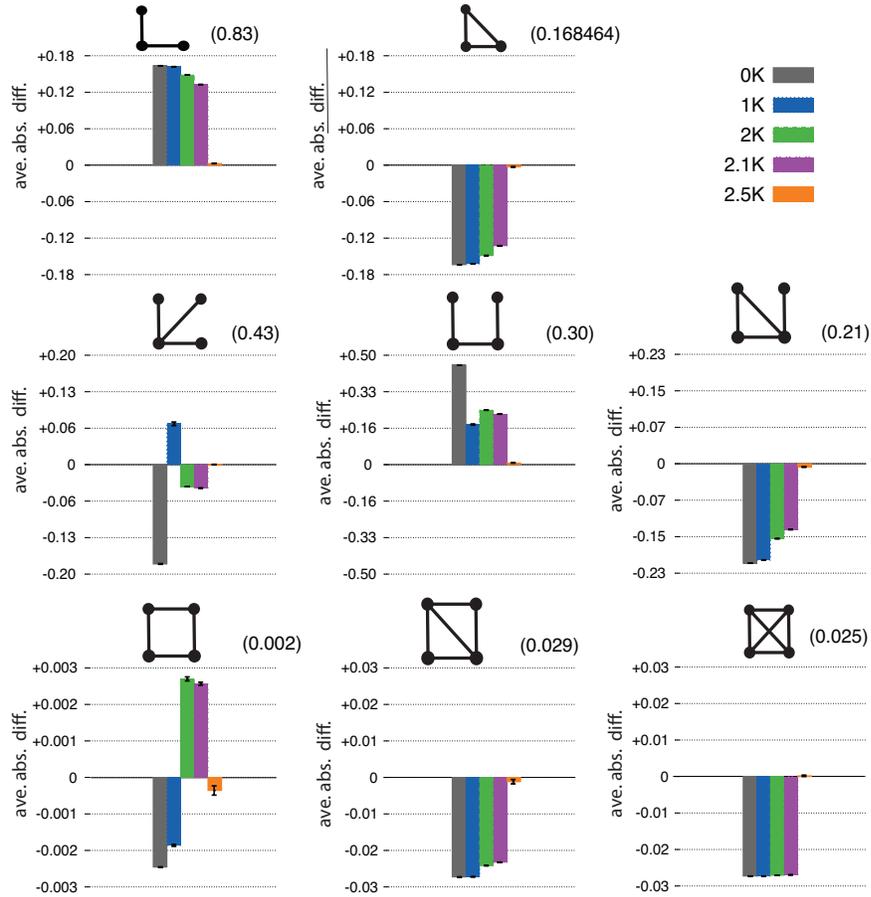}
\caption{\textbf{The densities of subgraphs of size $3$ and $4$ in the PGP network and its $dk$-random graphs.} The two different graphs of size $3$ and six different graphs of size $4$ are shown on each panel. The numbers on top of panels are the concentrations of the corresponding subgraph in the PGP network, while the histogram heights indicate the average absolute difference between the subgraph concentration in the $dk$-random graphs and its concentration in the PGP network. The subgraph concentration is the number of given subgraphs divided by the total number of subgraphs of the same size. The error bars are the standard deviations across different graph realizations.}
\label{fig:motifs}
\end{figure*}

\begin{figure*}
\centering
\includegraphics[width=\textwidth]{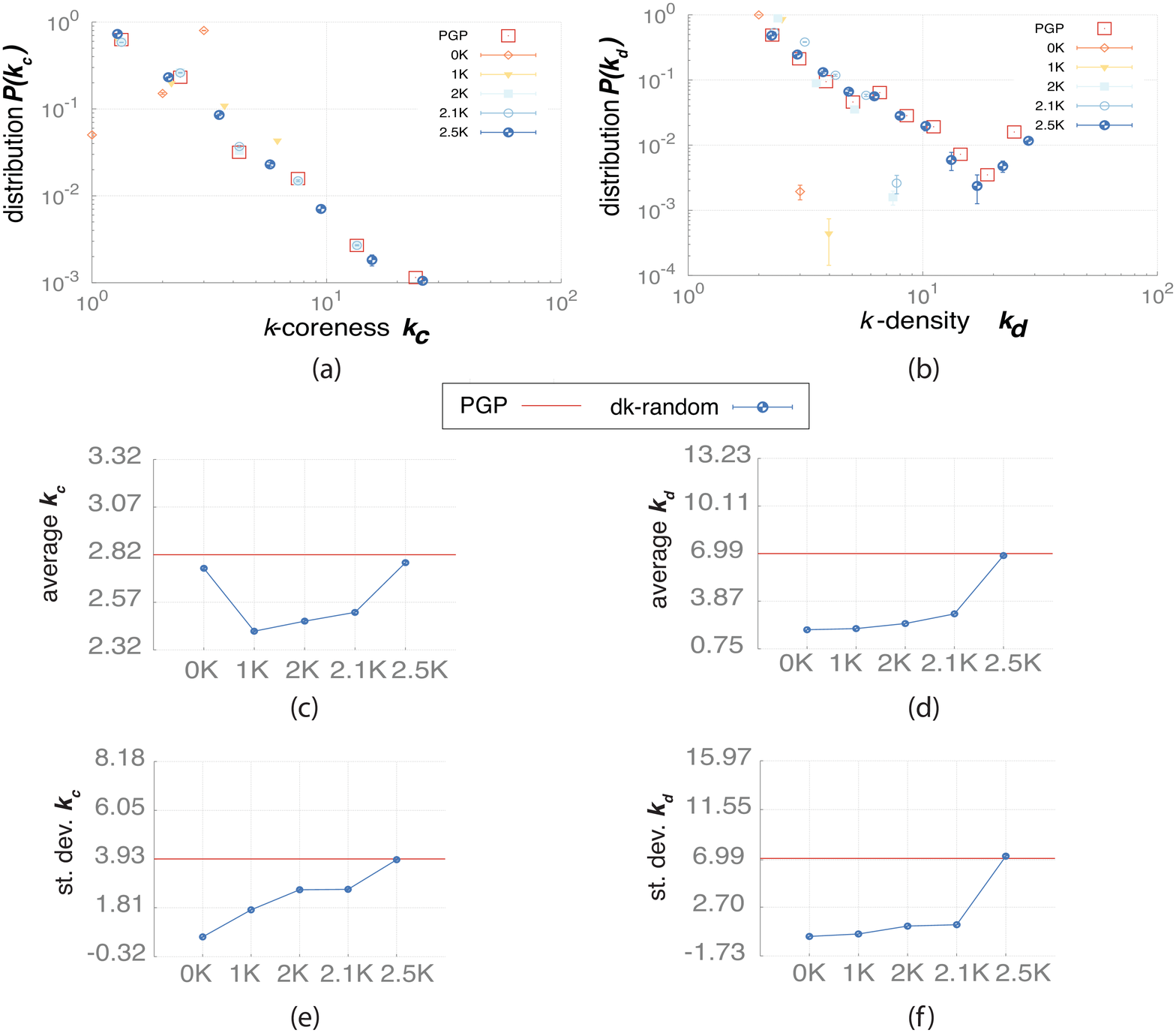}
\caption{\textbf{Mesoscopic properties, the $k$-coreness and $k$-density distributions, in the PGP network and its $dk$-random graphs.} The figure shows the distributions $P(k_{c,d})$ of \textbf{(a)}  node $k$-coreness~$k_c$ and  \textbf{(b)} edge $k$-density~$k_d$, and their \textbf{(c,d)} means and \textbf{(e,f)} standard deviations. The $k_c$-core of a graph is its maximal subgraph in which all nodes have degree at least $k_c$. The $k_d$-core of a graph is its maximal subgraph in which all edges have multiplicity at least $k_d-2$; the multiplicity of an edge is the number of common neighbors between the nodes that this edge connects, or equivalently the number of triangles that this edge belongs to. A node has $k$-coreness $k_c$ if it belongs to the $k_c$-core but not to the $k_c+1$-core. An edge has $k$-density $k_d$ if it belongs to the $k_d$-core but not to the $k_d+1$-core. The error bars indicate the standard deviations of the metrics across different graph realizations.}
\label{fig:decomposition}
\end{figure*}

\begin{figure*}[t]
\centering
\includegraphics[width=\textwidth]{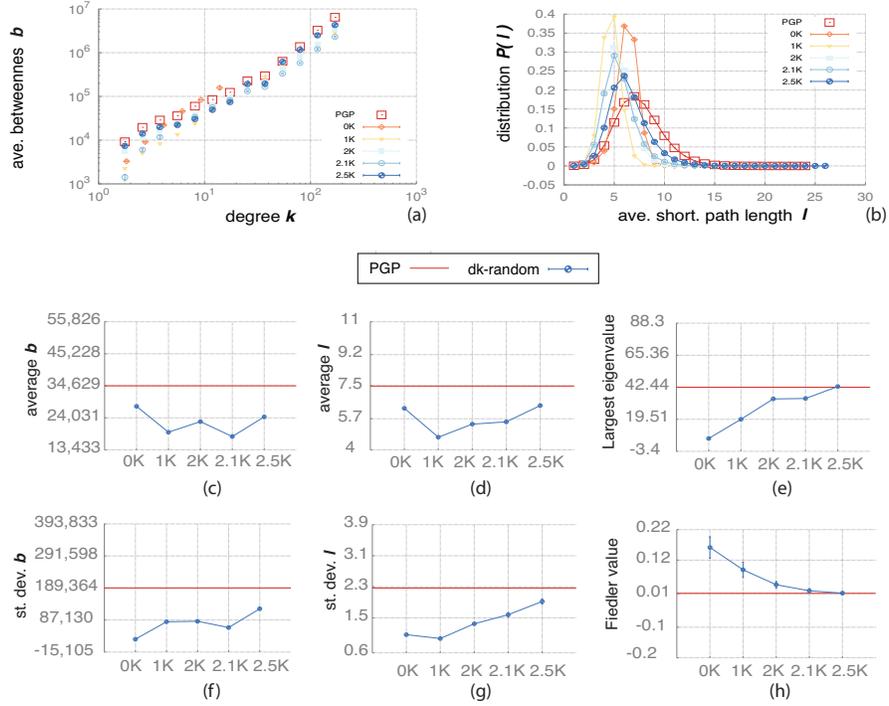}
\caption{\textbf{Macroscopic properties of the PGP network and its $dk$-random graphs.} The figure shows \textbf{(a)} the average betweenness $\bar{b}(k)$ of nodes of degree $k$, \textbf{(b)} the distribution $P(l)$ of hop lengths $l$ of the shortest paths between all pairs of nodes, the \textbf{(c,d)} means and \textbf{(f,g)} standard deviations of the corresponding distributions, \textbf{(e)} the largest eigenvalues of the adjacency matrix $A$, and \textbf{(h)} the Fiedler value, which is the spectral gap (the second largest eigenvalue) of the graph's Laplacian matrix $L=D-A$, where $D$ is the degree matrix, $D_{ij}=\delta_{ij}k_i$, $\delta_{ij}$ the Kronecker delta, and $k_i$ the degree of node $i$. The error bars indicate the standard deviations of the metrics across different graph realizations.}
\label{fig:global}
\end{figure*}

We performed an extensive set of numeric experiments with six real networks---the US air transportation network, an fMRI map of the human brain, the Internet at the level of autonomous systems, a technosocial web of trust among users of the distributed Pretty Good Privacy (PGP) cryptosystem, a human protein interaction map, and an English word adjacency network (Supplementary Note 2 and Supplementary Table 3 present the analysed networks). For each network we compute its average degree, degree distribution, degree correlations, average clustering, averaging clustering of nodes of degree $k$, and based on these $dk$-statistics generate a number of $dk$-random graphs as described above for each $d=0,1,2,2.1,2.5$. Then for each sample we compute a variety of network properties, and report their means and deviations for each combination of the real network, $d$, and the property. Figures~\ref{fig:local}-\ref{fig:global} present the results for the PGP network; Supplementary Note 3, Supplementary Figures 1-10, and Supplementary Tables 1-2 provide the complete set of results for all the considered real networks. The reason why we choose the PGP network as our main example is that this network appears to be ``least random'' among the considered real networks, in the sense that the PGP network requires higher values of $d$ to reproduce its considered properties. The only exception is the brain network. Some of its properties are not reproduced even by $d=2.5$.

Figure~\ref{fig:methods} visualizes the PGP network and its $dk$-randomizations. The figure illustrates the convergence of $dk$-series applied to this network. While the $0k$-random graph has very little in common with the real network, the $1k$-random one is somewhat more similar, even more so for $2k$, and there is very little visual difference between the real PGP network and its $2.5k$-random counterpart. This figure is only an illustration though, and to have a better understanding of how similar the network is to its randomization, we compare their properties.

We split the properties that we compare into the following categories. The {\em microscopic properties} are local properties of individual nodes and subgraphs of small size. These properties can be further subdivided into those that are defined by the $dk$-distributions---the degree distribution, average neighbor degree, clustering, Fig.~\ref{fig:local}---and those that are not fixed by the $dk$-distributions---the concentrations of subgraphs of size $3$ and $4$, Fig.~\ref{fig:motifs}. The {\em mesoscopic properties}---$k$-coreness and $k$-density (the latter is also known as $m$-coreness or edge multiplicity, Supplementary Note 1), Fig.~\ref{fig:decomposition}---depend both on local and global aspects of network organization. Finally, the {\em macroscopic properties} are truly global ones---betweenness, the distribution of hop lengths of shortest paths, and spectral properties, Fig.~\ref{fig:global}. In Supplementary Note 3 we also report some extremal properties, such as the graph diameter (the length of the longest shortest path), and Kolmogorov-Smirnov distances between the distributions of all the considered properties in real networks and their corresponding $dk$-random graphs. The detailed definitions of all the properties that we consider can be found in Supplementary Note 1.

In most cases---henceforth by ``case'' we mean a combination of a real network and one of its considered property---we observe a nice convergence of properties as $d$ increases. In many cases there is no statistically significant difference between the property in the real network and in its $2.5k$-random graphs. In that sense these graphs, that is, random graphs whose degree distribution and degree-dependent clustering $\bar{c}(k)$ are as in the original network, capture many other important properties of the real network.

Some properties always converge. This is certainly true for the microscopic properties in Fig.~\ref{fig:local}, simply confirming that our $dk$-sampling algorithm operates correctly. But many properties that are not fixed by the $dk$-distributions converge as well. Neither the concentration of subgraphs of size $3$ nor the distribution of the number of neighbors common to a pair of nodes are fully fixed by $dk$-distributions with any $d<3$ by definition, yet $2.5k$-random graphs reproduce them well in all the considered networks. Most subgraphs of size $4$ are also captured at $d=2.5$ in most networks, even though $d=3$ would not be enough to exactly reproduce the statistics of these subgraphs. We note that the improvement in subgraph concentrations at $d=2.5$ compared to $d=2.1$ is particularly striking, Fig.~\ref{fig:motifs}. The mesoscopic and especially macroscopic properties converge more slowly as expected. Nevertheless, quite surprisingly, both mesoscopic properties ($k$-coreness and $k$-density) and some macroscopic properties converge nicely in most cases. The $k$-coreness, $k$-density, and the spectral properties, for instance, converge at $d=2.5$ in all the considered cases other than Internet's Fiedler value. In some cases a property, even global one, converges for $d$ lower than $2.5$. Betweenness, for example, a global property, converges at $d=1$ for the Internet and English word network.

Finally, there are ``outlier'' networks and properties of poor or no $dk$-convergence. Many properties of the brain network, for example, exhibit slow or no convergence. We have also experimented with community structure inferred by different algorithms, and in most cases the convergence is either slow or non-existent as one could expect.

\clearpage
\newpage

\clearpage
\newpage

\section*{Discussion}
\label{sec:discussion}

In general, we should not expect non-local properties of networks to be exactly or even closely reproduced by random graphs with local constraints. The considered brain network is a good example of that this expectation is quite reasonable. The human brain consists of two relatively weakly connected parts, and no $dk$-randomization with low $d$ is expected to reproduce this peculiar global feature, which likely has an impact on other global properties. And indeed we observe in Supplementary Note 3 that its two global properties, the shortest path distance and betweenness distributions, differ drastically between the brain and its $dk$-randomizations.

Another good example is community structure, which is not robust with respect to $dk$-randomizations in all the considered networks. In other words, $dk$-randomizations destroy the original peculiar cluster organization in real networks, which is not surprising, as clusters have too many complex non-local features such as variable densities of internal links, boundaries, etc., which $dk$-randomizations, even with high $d$, are expected to affect considerably.

Surprisingly, what happens for the brain and community structure does not appear representative for many other considered combinations of real networks and their properties.
As a possible explanation, one can think of constraint-based modeling as a satisfiability (SAT) problem: find the elements of the adjacency matrix (1/0, True/False) such that all the given constraints in terms of the functions of the marginals (degrees) of this matrix are obeyed.
We then expect that the $3k$-constraints already correspond to an NP-hard SAT problem, such as 3-SAT, with hardness coming from the global nature of the constraints in the problem. However, many real-world networks evolve based mostly on local dynamical rules and thus we would expect them to contain correlations with $d<3$, that is, below the NP-hard barrier. The primate brain, however, has likely evolved through global constraints, as indicated by the dense connectivity
across all functional areas and the existence of a strong core-periphery structure in which the core heavily concentrates on areas
within the associative cortex, with connections to and from all the primary input and subcortical areas~\cite{Markov2013}.

However, in most cases, the considered networks are $dk$-random with $d\leq2.5$, that is, $d\leq2.5$ is enough to reproduce not only basic microscopic (local) properties but also mesoscopic and even macroscopic (global) network properties~\cite{VazquezTopologicalRelationship2004,GuimeraClasses2007,TakemotoStructure2007,FosterClustering2011,ColomerClustering2013}. This finding means that these more sophisticated properties are effectively random in the considered networks, or more precisely, that the observed values of these properties are effective consequences of particular degree distributions and, optionally, degree correlations and clustering that the networks have. This further implies that attempts to find explanations for these complex but effectively random properties should probably be abandoned, and redirected to explanations of why and how degree distributions, correlations, and clustering emerge in real networks, for which there already exists a multitude of
approaches~\cite{DoMeSa01,KlEg02,Vazquez2003Growing,Serrano2005Clustering,PaBoKr11,bianconi2014b}. On the other hand, the features that we found non-random do require separate explanations, or perhaps a different system of null models.

We reiterate that the $dk$-randomization system makes it clear that there is no {\it a priori\/} preferred null model for network randomization. To tell how statistically significant a particular feature is, it is necessary to compare this feature in the real network against the same feature in an ensemble of random graphs, a null model. But one is free to choose any random graph model. In particular, any $d$ defines a random graph ensemble, and we find that many properties, most notably the frequencies of small subgraphs that define motifs~\cite{MiSh02-motifs}, strongly depend on $d$ for many considered networks. Therefore choosing any specific value of $d$, or more generally, any specific null model to study the statistical significance of a particular structural network feature, requires some non-trivial justification before this feature can be claimed important for any network function.

Yet another implication of our results is that if one looks for network topology generators that would veraciously reproduce certain properties of a given real network---a task that often comes up in as diverse disciplines as biology~\cite{KuBa06} and computer science~\cite{MeLaMaBy01-phys}---one should first check how $dk$-random these properties are. If they are $dk$-random with low $d$, then one may not need any sophisticated mission-specific topology generators. The $dk$-random graph generation algorithms discussed here can be used for that purpose in this case. We note that there exists an extension of a subset of these algorithm for networks with arbitrary annotations of links and nodes~\cite{DiKr09}---directed or colored (multilayer) networks, for instance.

The main caveat of our approach is that we have no proof that our $dk$-random graph generation algorithms for $d=2.1$ and $d=2.5$ sample graphs uniformly at random from the ensemble. The random graph ensembles and edge rewiring processes employed here are known to suffer from problems such as degeneracy and hysteresis~\cite{Foster2010Hysteresis,Coolen2014Loops,Horvat2014Degeneracy}. Ideally, we would wish to calculate analytically the exact expected value of a given property in an ensemble. This is currently possible only for very simple properties in soft ensembles with $d=0,1,2$~\cite{GarlaschelliLikelihood2011,Squartini2015Sampling}. Some mathematically rigorous results are available for $d=0,1$ and for some exponential random graph models~\cite{ChatterjeeDegreeSequences2011,ChatterjeeERGM2013}. Many of these results rely on graphons~\cite{LovaszBook2012} that are applicable to dense graphs only, while virtually all real networks are sparse~\cite{BasslerSparse2011}. Some rigorous approaches to sparse networks are beginning to emerge~\cite{Bollobas2011Sparse,Borgs2014LpI}, but the rigorous treatment of global properties, which tend to be highly non-trivial functions of adjacency matrices, in random graph ensembles with $d>2$ constraints, appear to be well beyond the reach in the near future. Yet if we ever want to fully understand the relationship between the structure, function, and dynamics of real networks, this future research direction appears to be of a paramount importance.

\section*{Acknowledgements}

We acknowledge financial support by NSF Grants No.\ CNS-1039646, CNS-1345286, CNS-0722070, CNS-0964236, CNS-1441828, CNS-1344289, CNS-1442999, CCF-1212778, and DMR-1206839; by AFOSR and DARPA Grants No.\ HR0011-12-1-0012 and FA9550-12-1-0405; by DTRA Grant No.\ HDTRA-1-09-1-0039; by Cisco Systems; by the Ministry of Education, Science, and Technological Development of the Republic of Serbia under Project No.\ ON171017; by the ICREA Academia Prize, funded by the {\it Generalitat de Catalunya}; by the Spanish MINECO Project No.\ FIS2013-47282-C2-1-P; by the {\it Generalitat de Catalunya} Grant No.\ 2014SGR608; and by European Commission Multiplex FP7 Project No. 317532.

\section*{Contributions}
All authors contributed to the development and/or implementation of the concept, discussed and analysed the results.
C.O., M.M.D., and P.C.S.\ implemented the software for generating $dk$-graphs and analyzed their properties.
D.K. wrote the manuscript, incorporating comments and contributions from all authors.

\section*{Competing financial interests}
The authors declare no competing financial interests.

\section*{Correspondence}
Correspondence and requests for materials should be addressed to C.O.\ (chiara@caida.org) and D.K.\ (dima@neu.edu).

\clearpage
\newpage

\renewcommand{\figurename}{Supplementary Figure}
\renewcommand{\tablename}{Supplementary Table}
\renewcommand{\algorithmcfname}{Supplementary Algorithm}

\section*{Supplementary Figures}

\begin{figure*}[!h]
\centering
\subfloat[AIR]{%
  \includegraphics[width=0.5\textwidth]{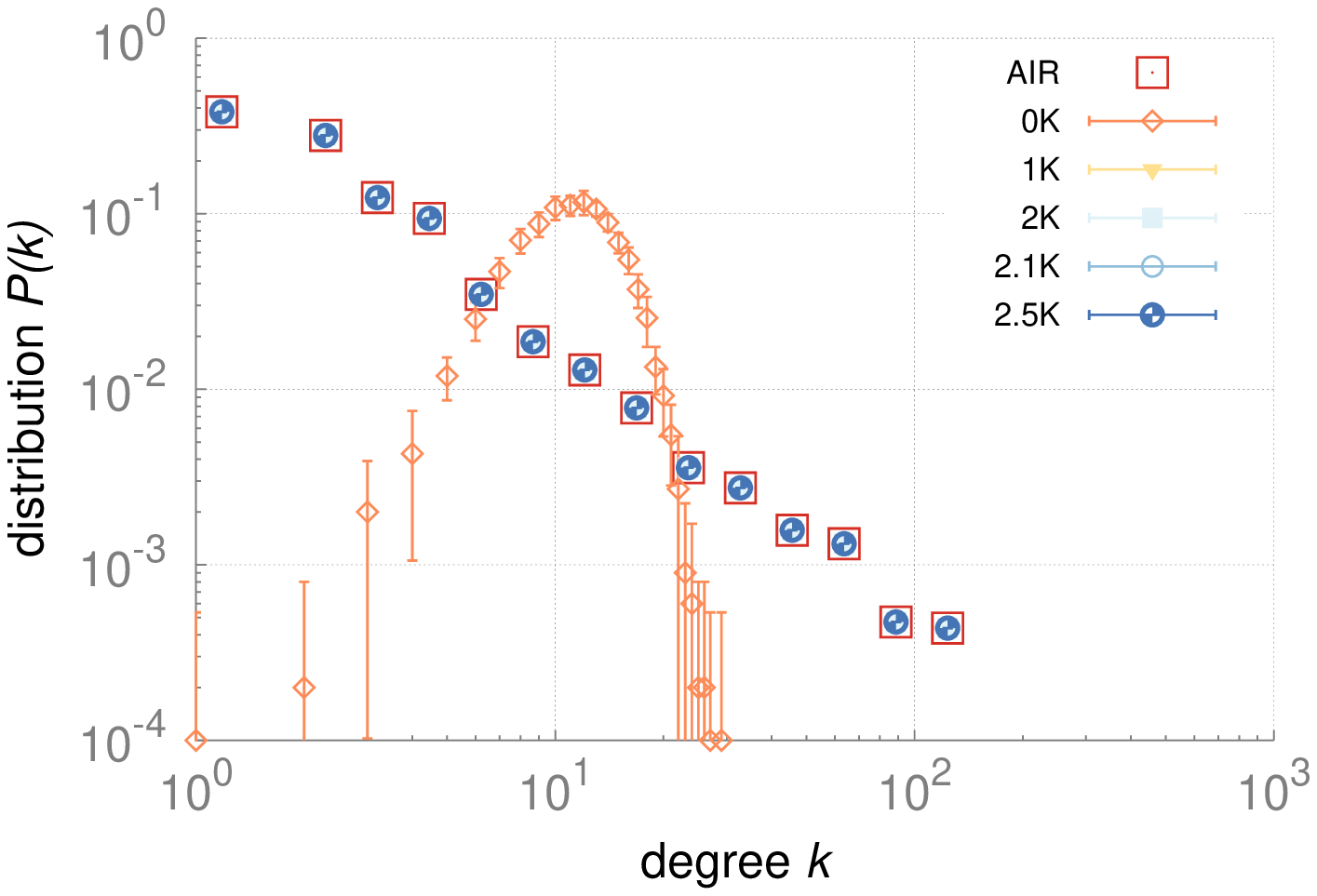}
}
\subfloat[BRAIN]{%
  \includegraphics[width=0.5\textwidth]{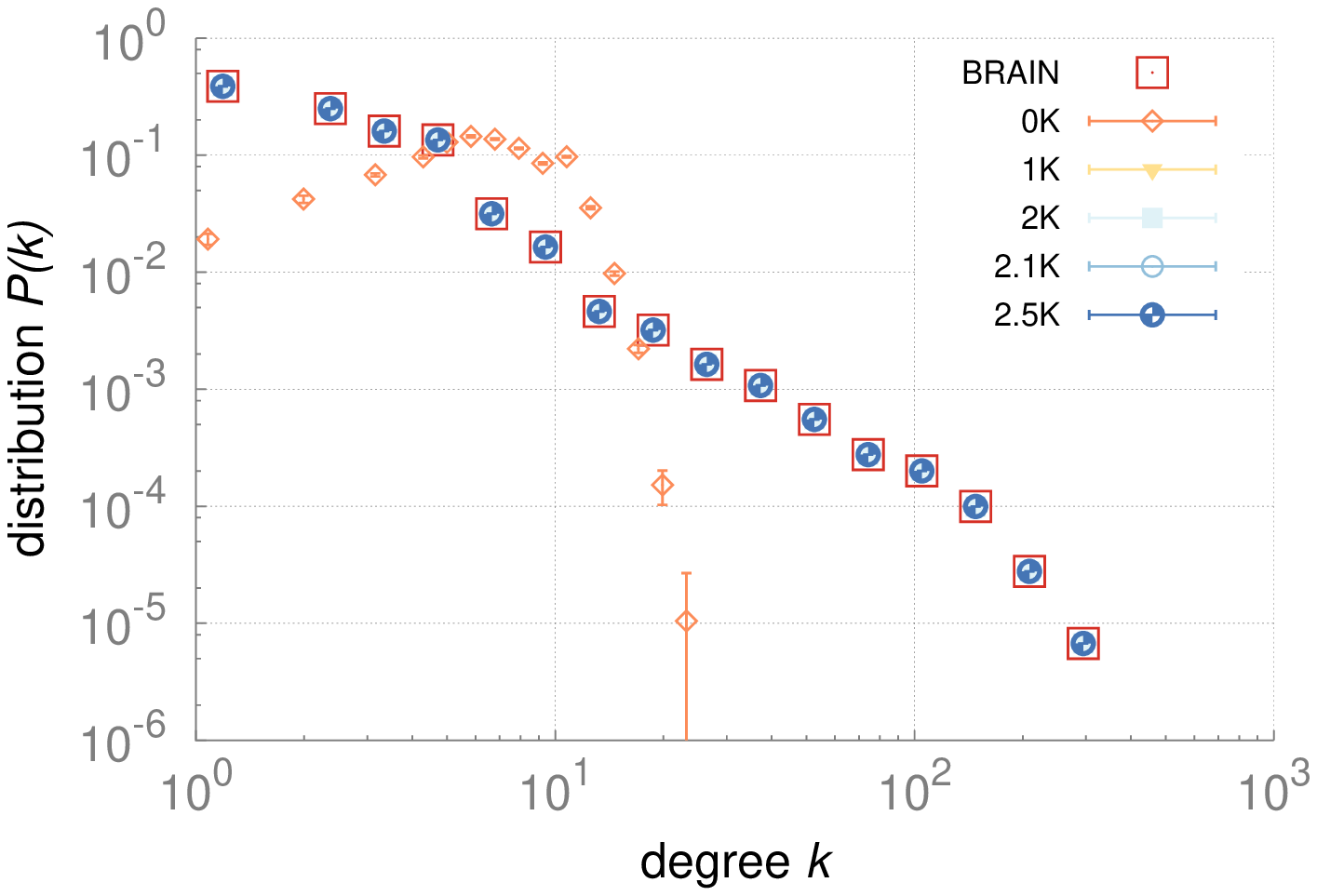}
}\\
\subfloat[WORDS]{%
  \includegraphics[width=0.5\textwidth]{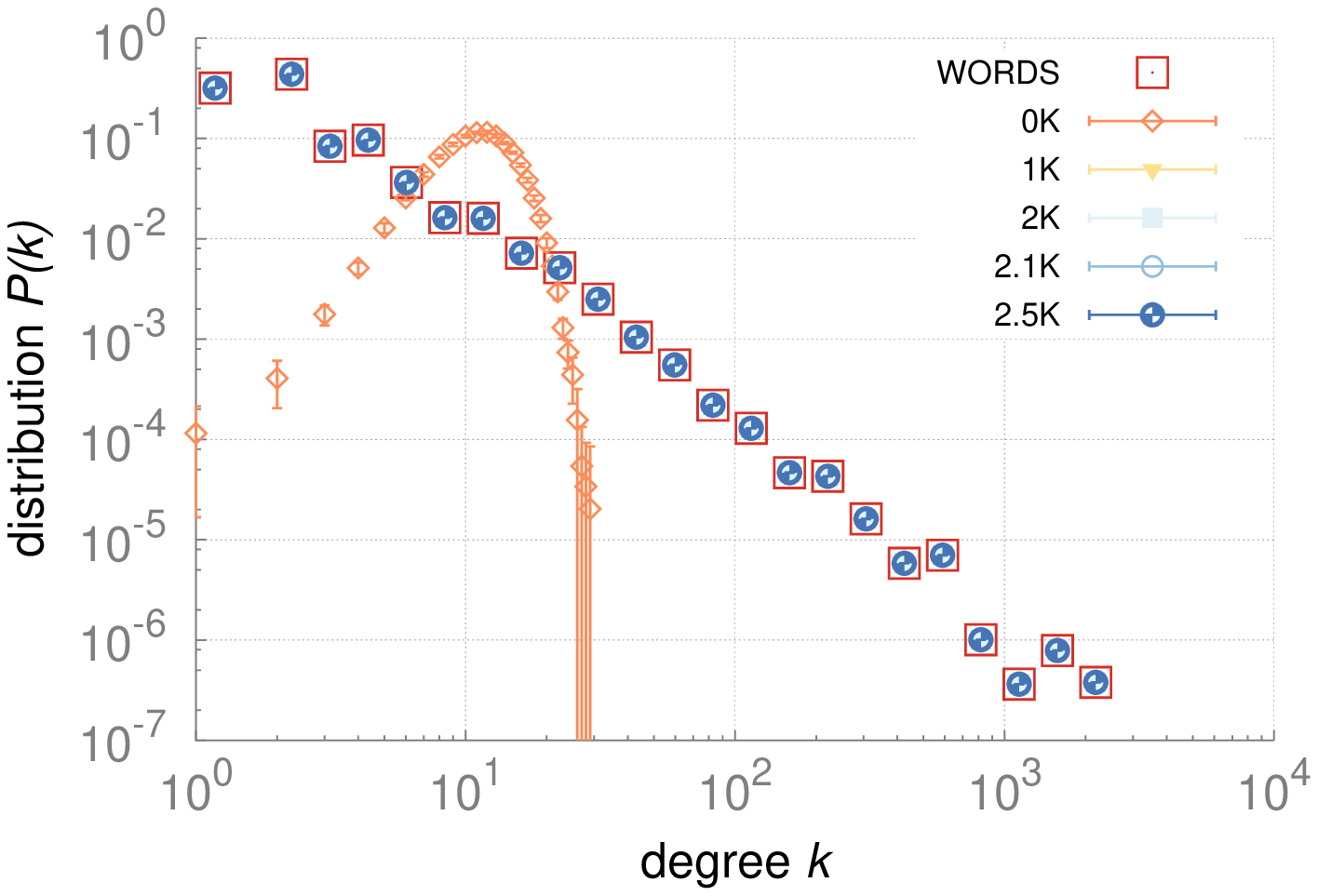}
}
\subfloat[INTERNET]{%
  \includegraphics[width=0.5\textwidth]{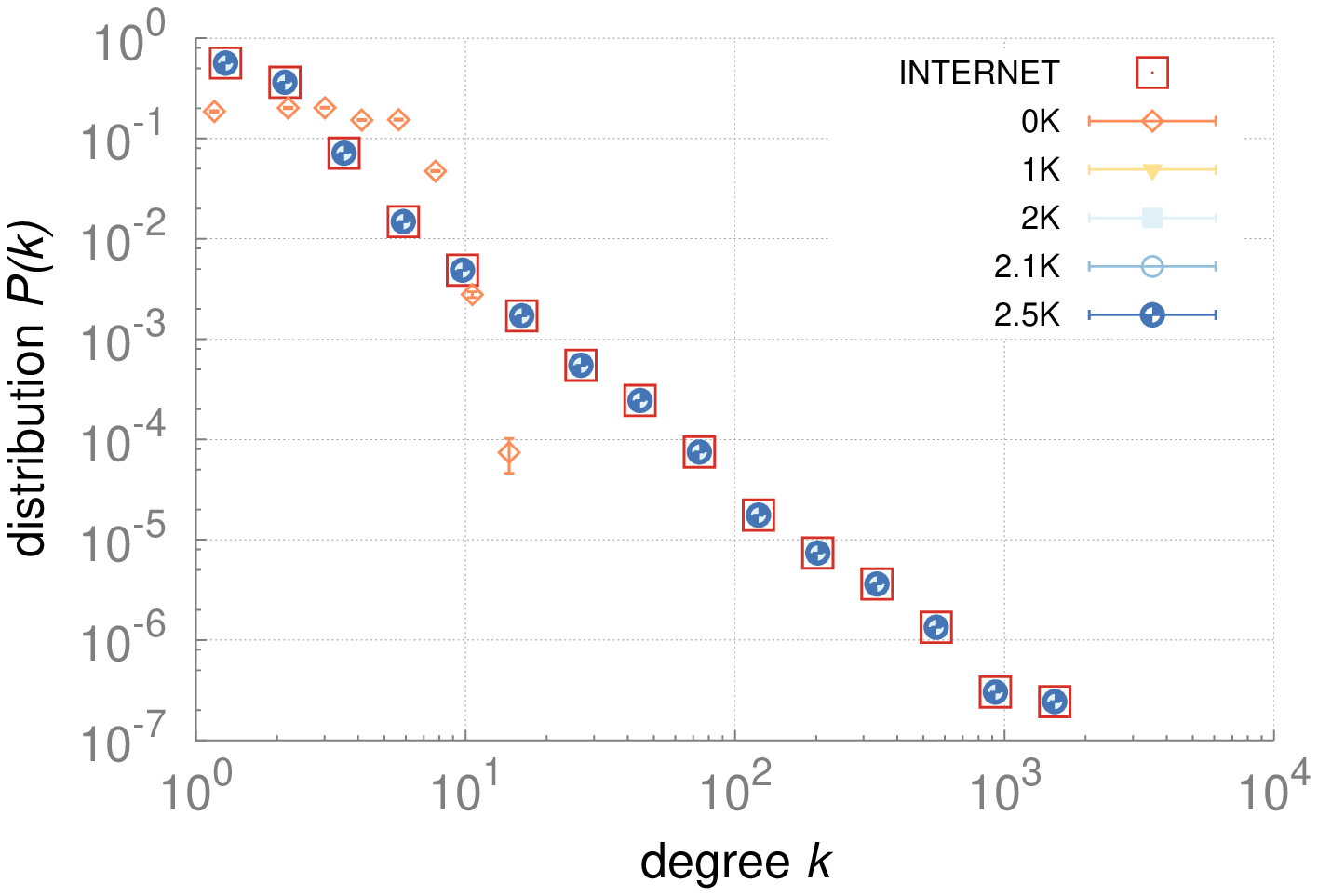}
}\\
\subfloat[PGP]{%
  \includegraphics[width=0.5\textwidth]{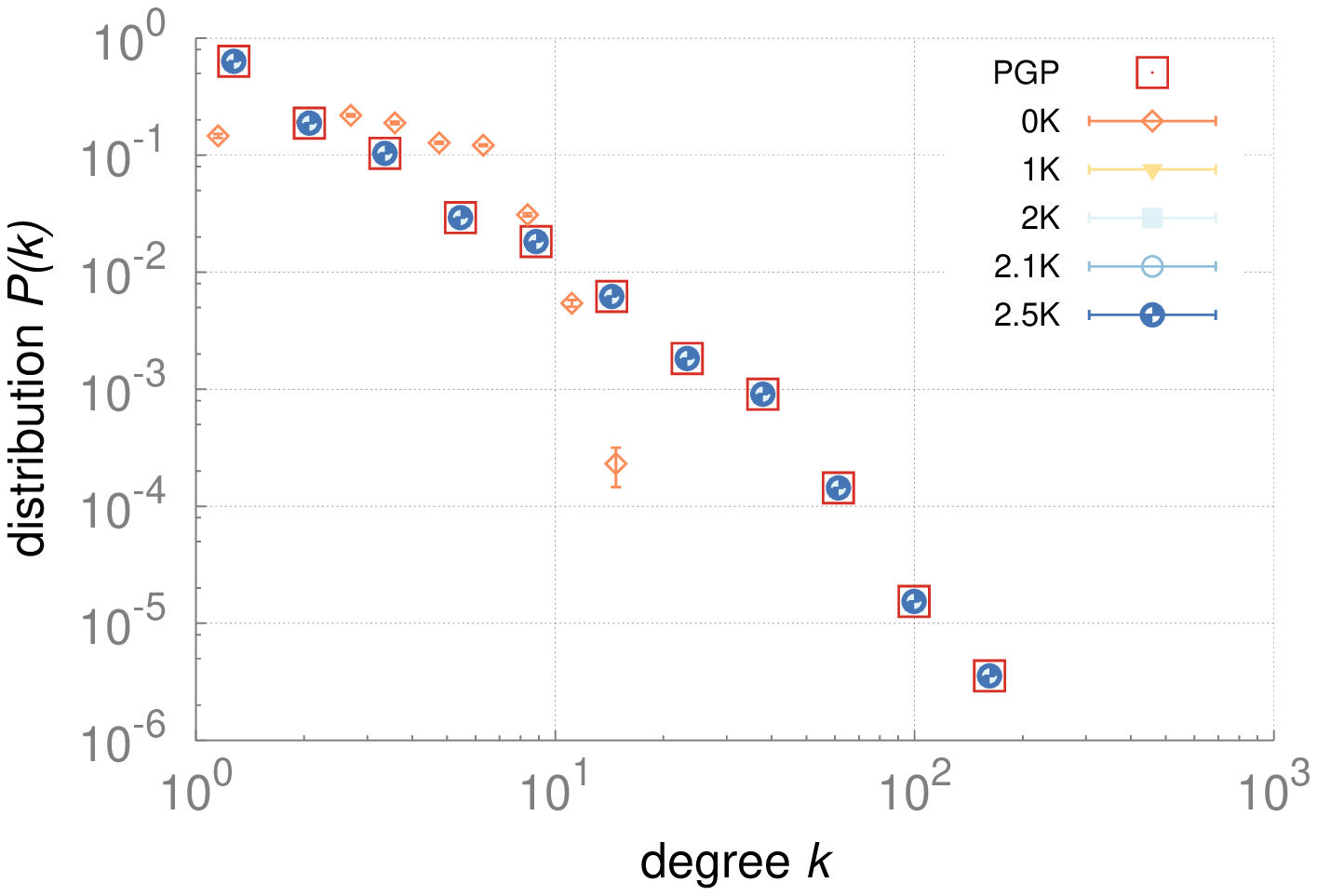}
}
\subfloat[PPI]{%
  \includegraphics[width=0.5\textwidth]{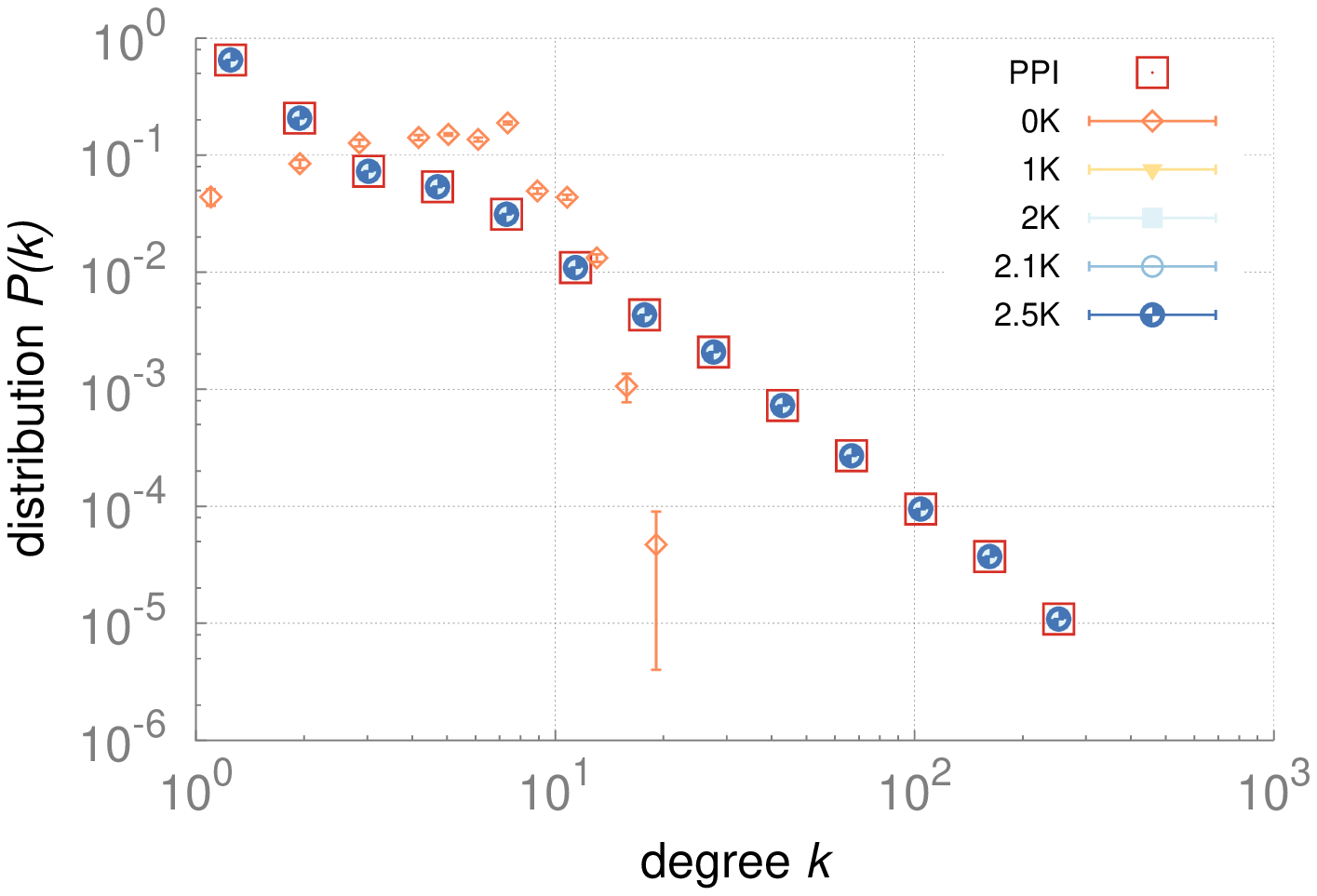}
}\\
\caption{\textbf{Degree distributions in real networks and their $dk$-randomizations.}}
\label{fig:k}
\vspace*{-5cm}
\end{figure*}

\begin{figure*}
\centering
\subfloat[AIR]{%
  \includegraphics[width=0.5\textwidth]{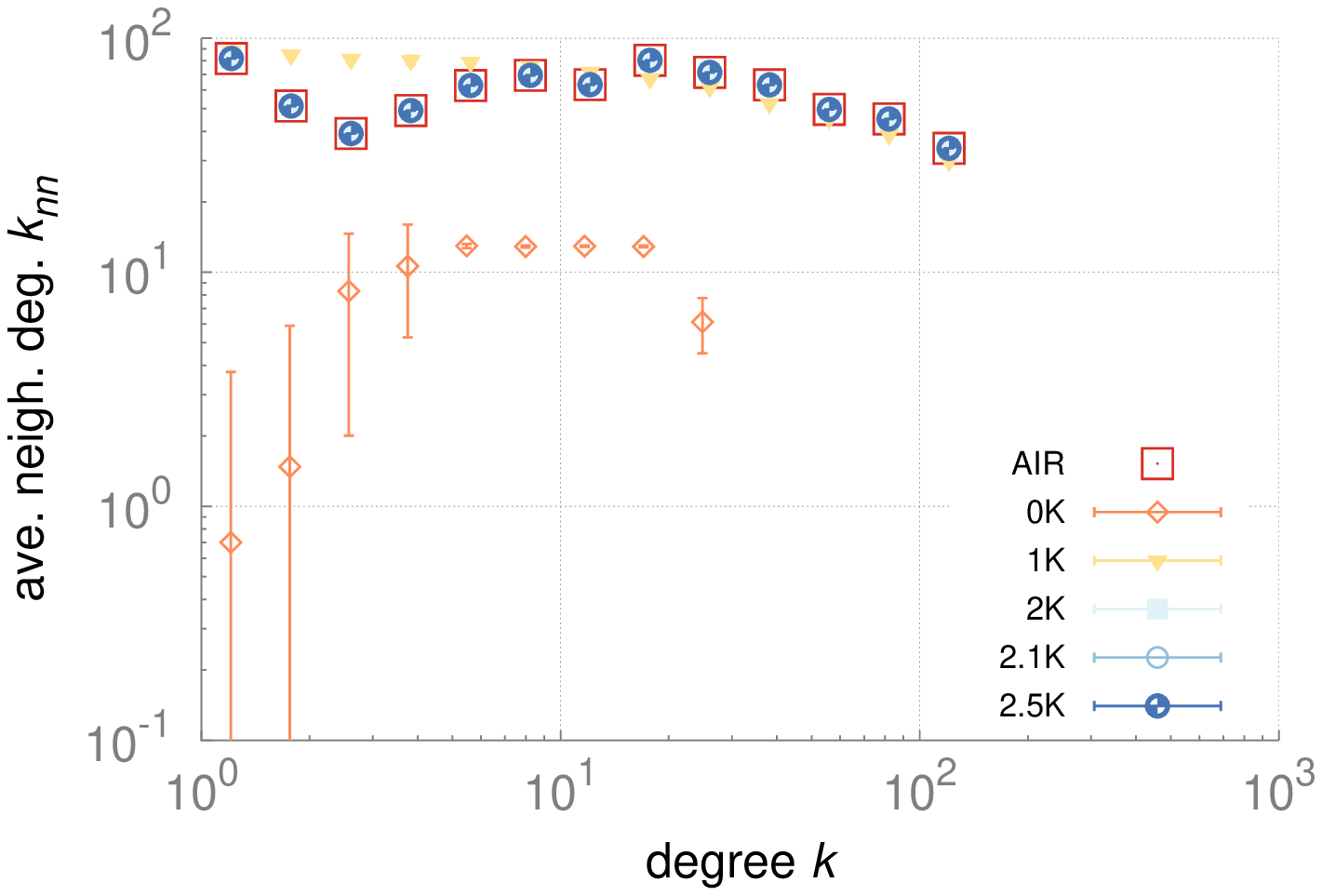}
}
\subfloat[BRAIN]{%
  \includegraphics[width=0.5\textwidth]{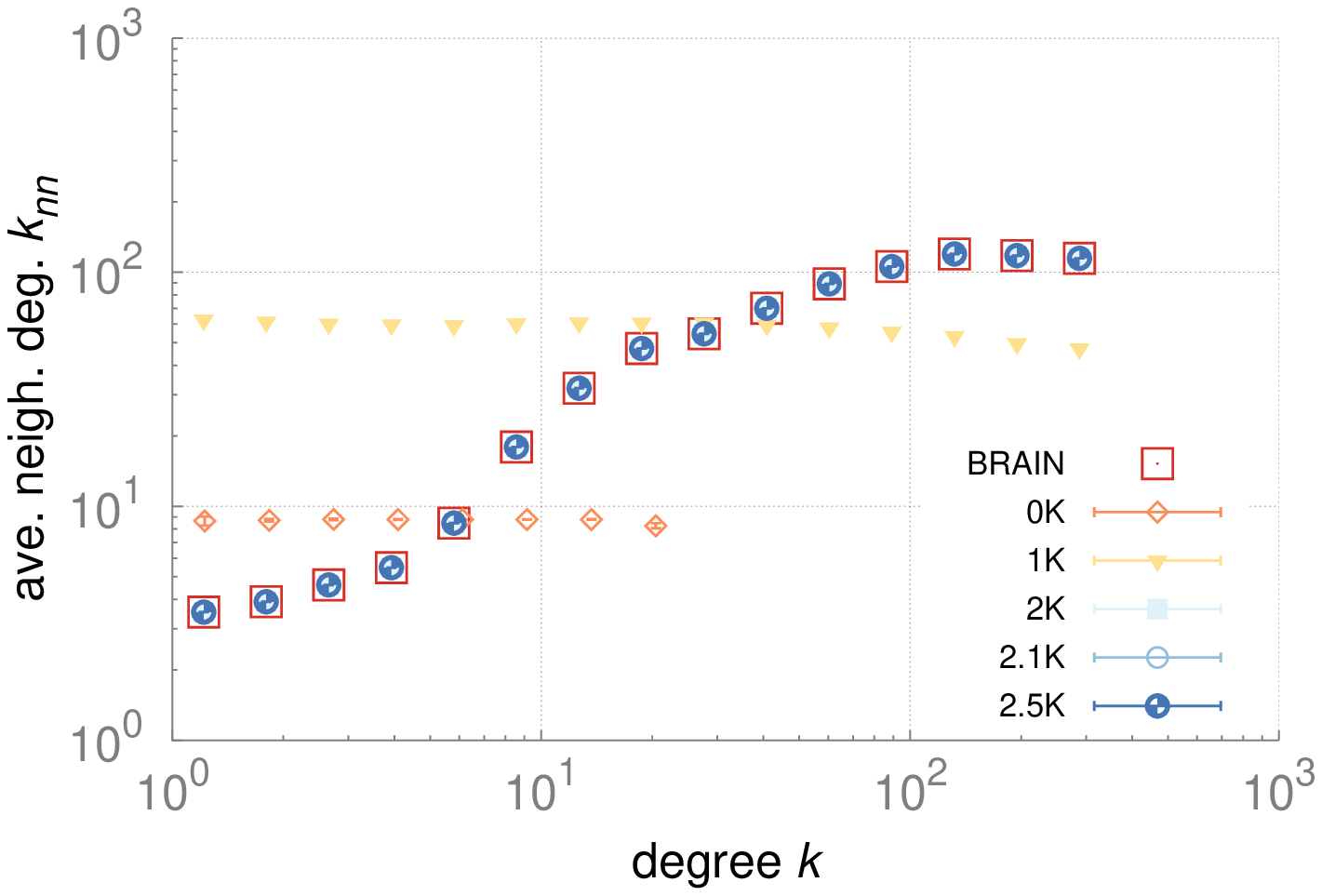}
}\\
\subfloat[WORDS]{%
  \includegraphics[width=0.5\textwidth]{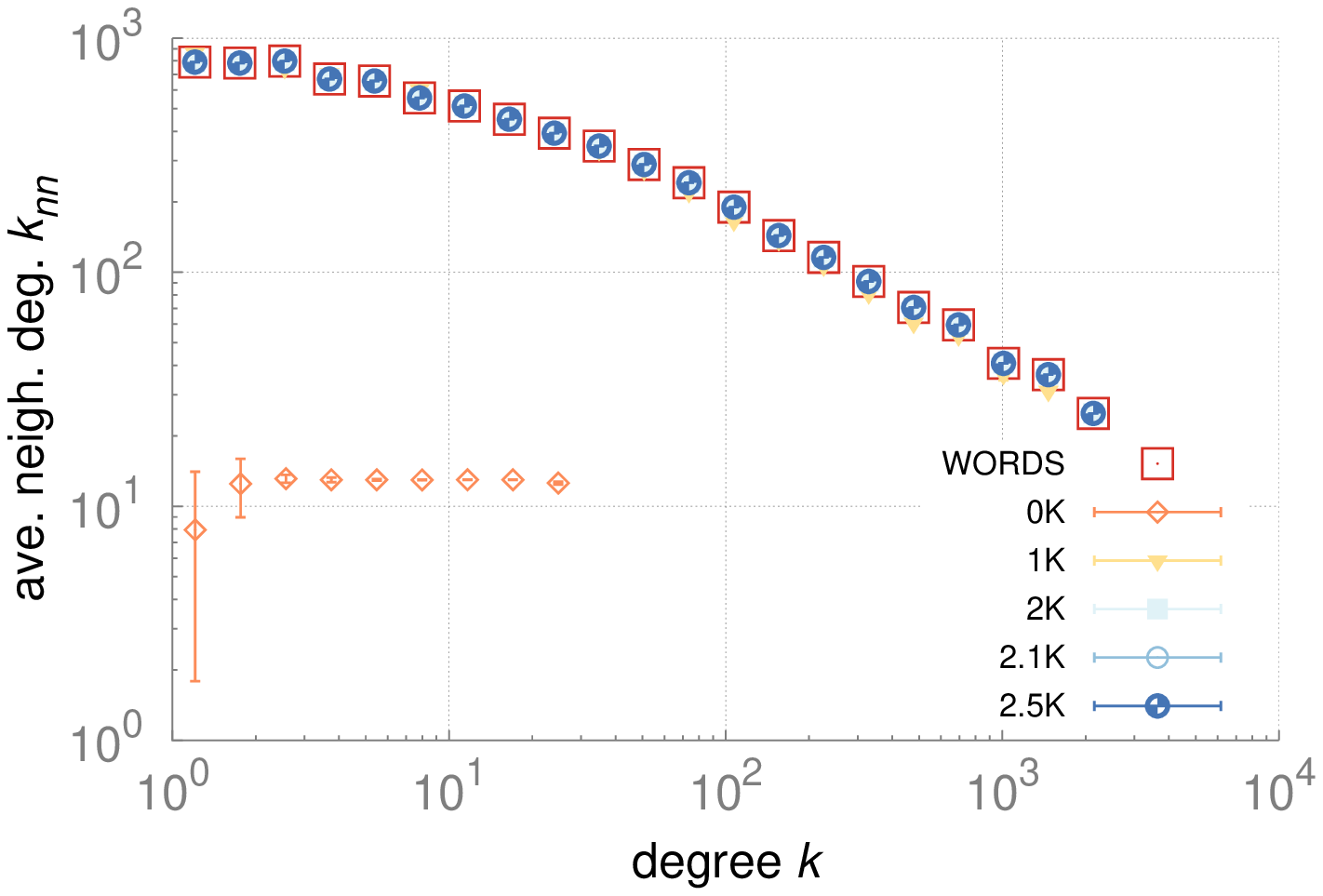}
}
\subfloat[INTERNET]{%
  \includegraphics[width=0.5\textwidth]{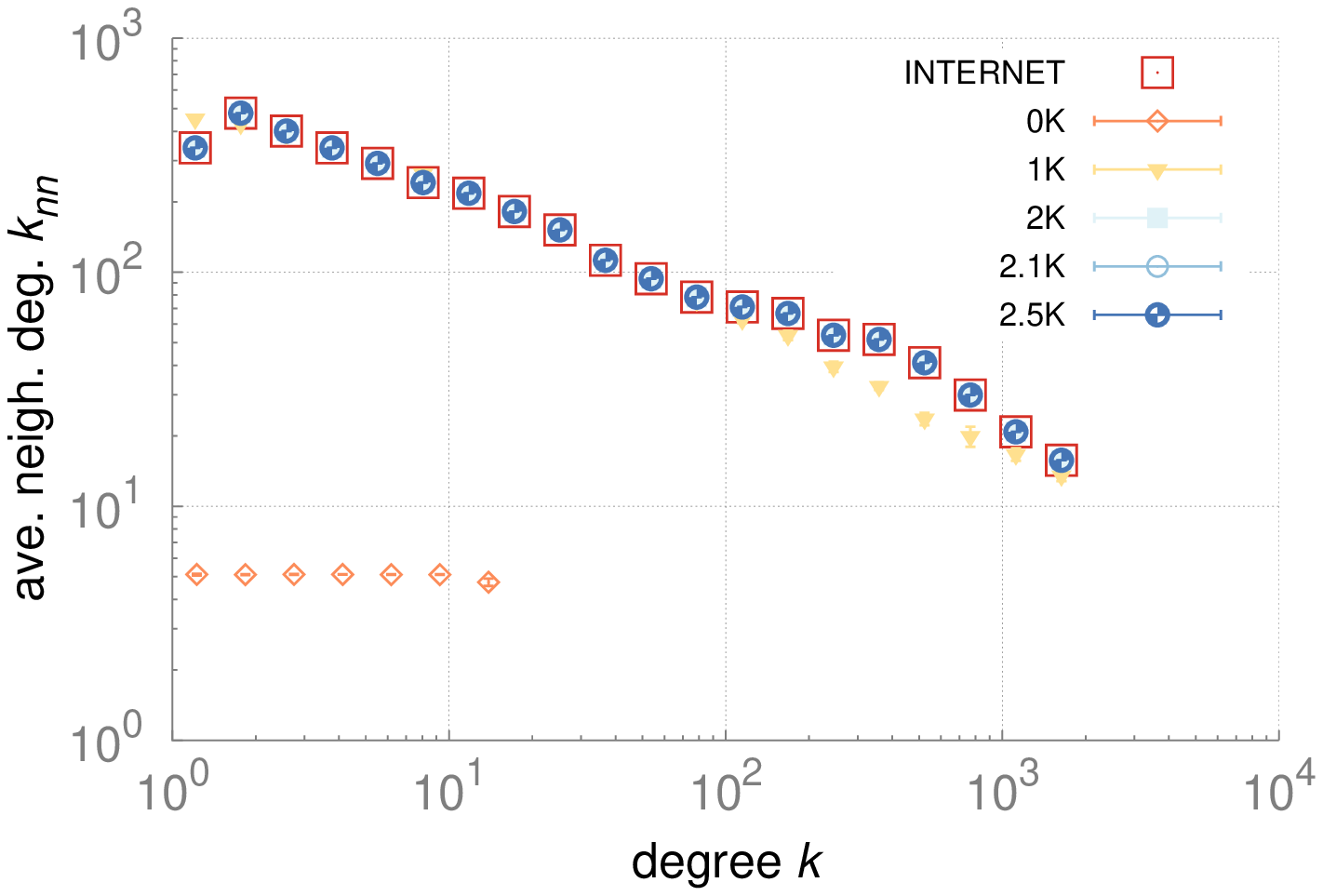}
}\\
\subfloat[PGP]{%
  \includegraphics[width=0.5\textwidth]{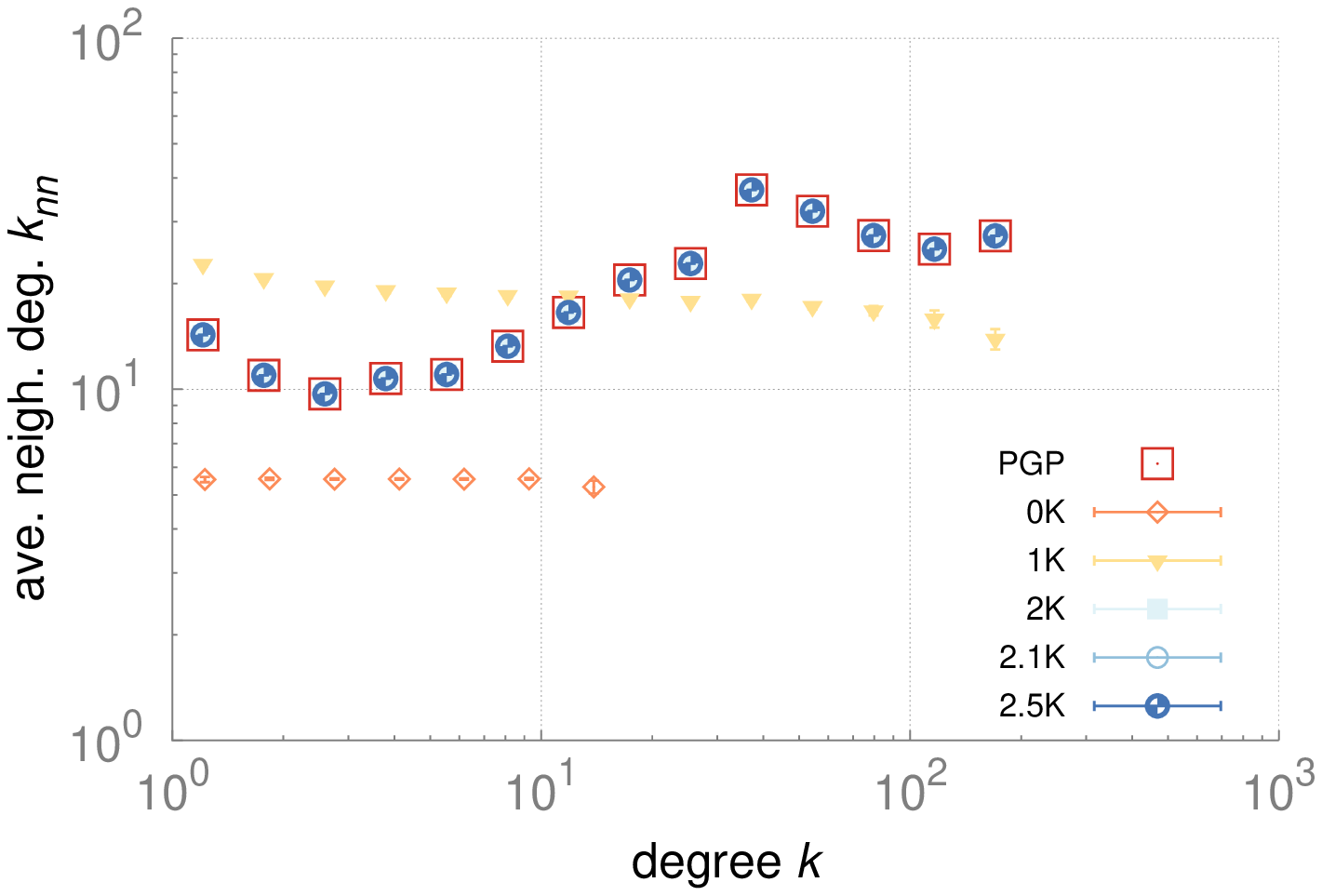}
}
\subfloat[PPI]{%
  \includegraphics[width=0.5\textwidth]{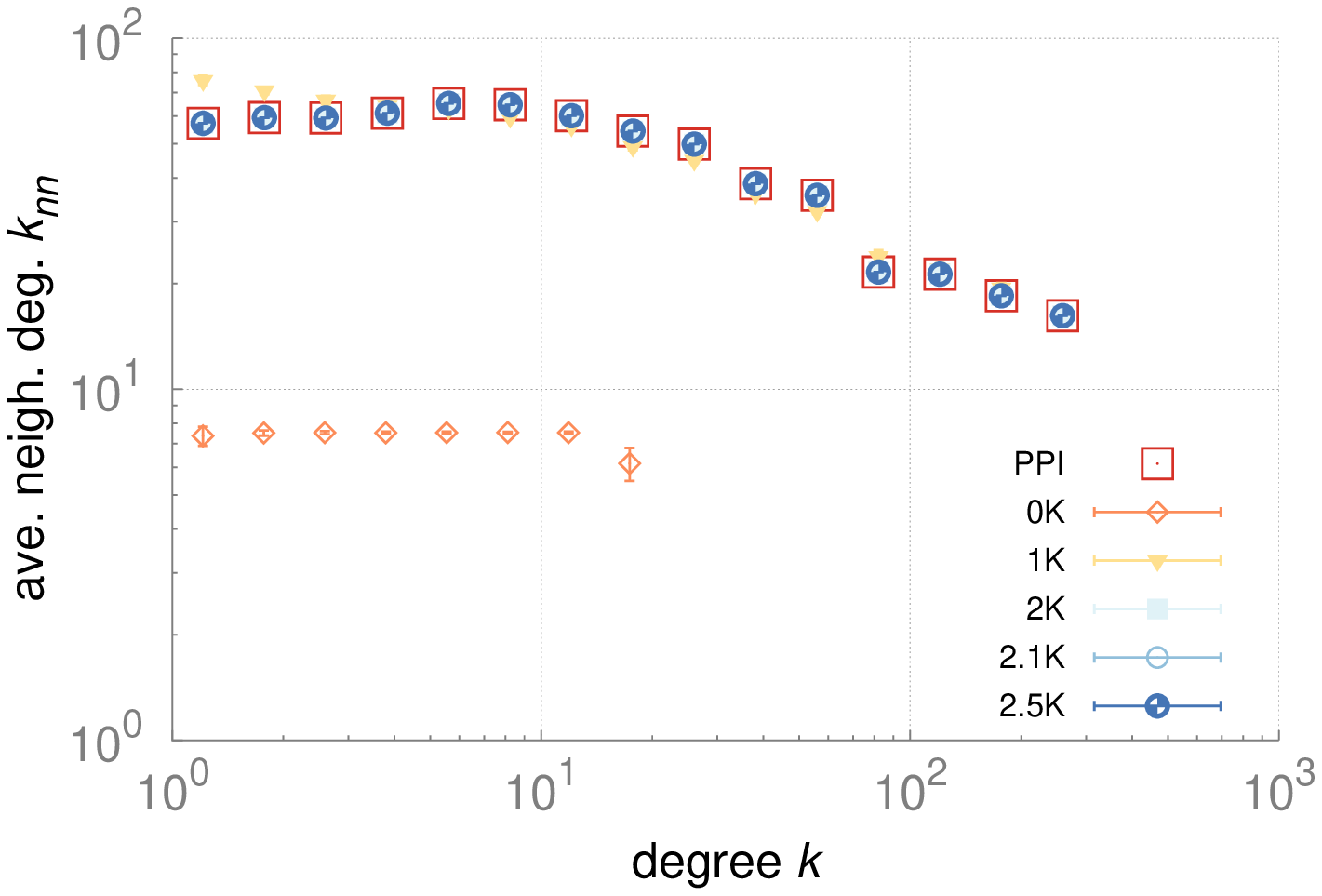}
}\\
\caption{\textbf{Average nearest neighbor degrees (ANNDs) of nodes of a given degree in real networks and their $dk$-randomizations.}}
\label{fig:knn}
\end{figure*}

\begin{figure*}
\centering
\subfloat[AIR]{%
  \includegraphics[width=0.5\textwidth]{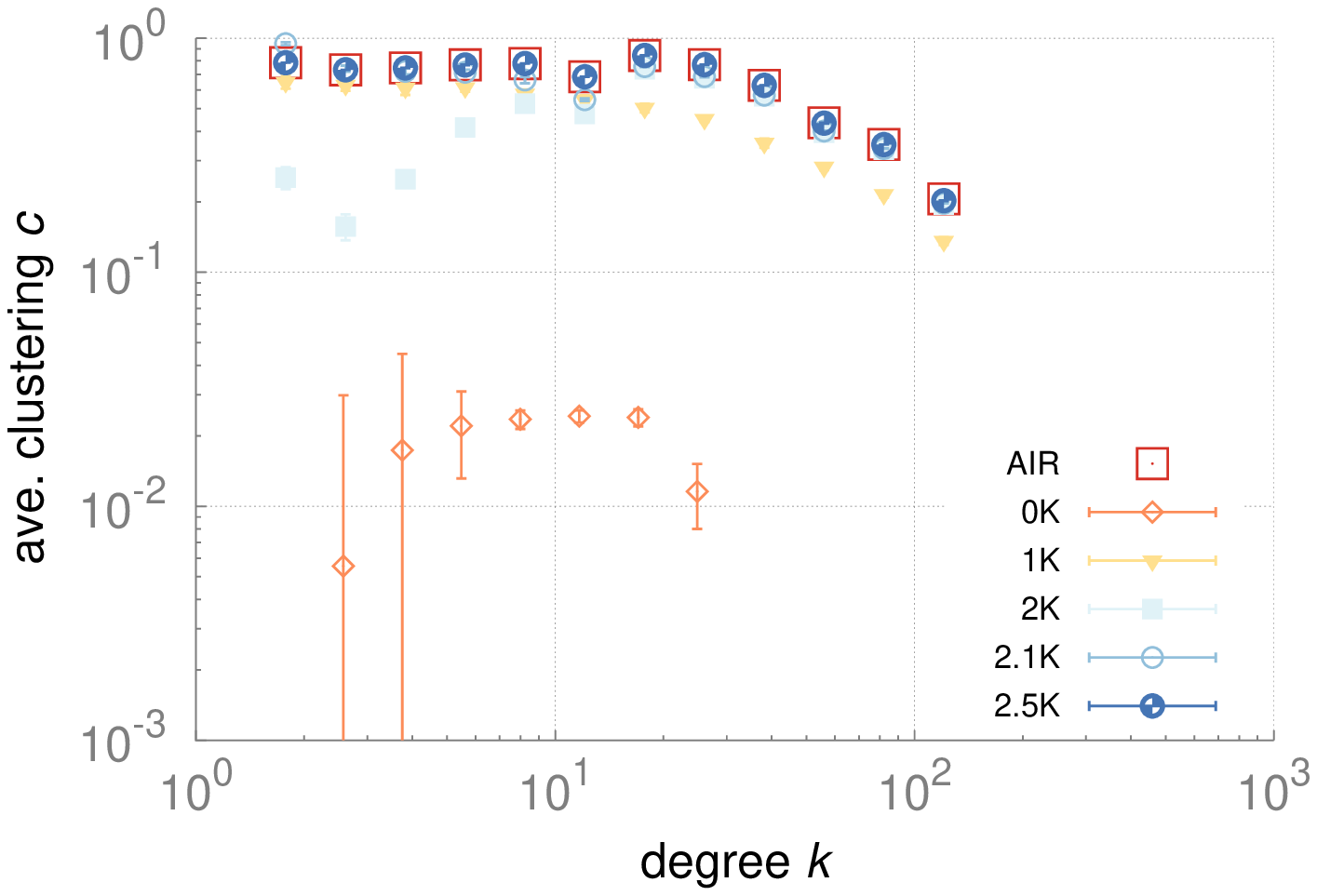}
}
\subfloat[BRAIN]{%
  \includegraphics[width=0.5\textwidth]{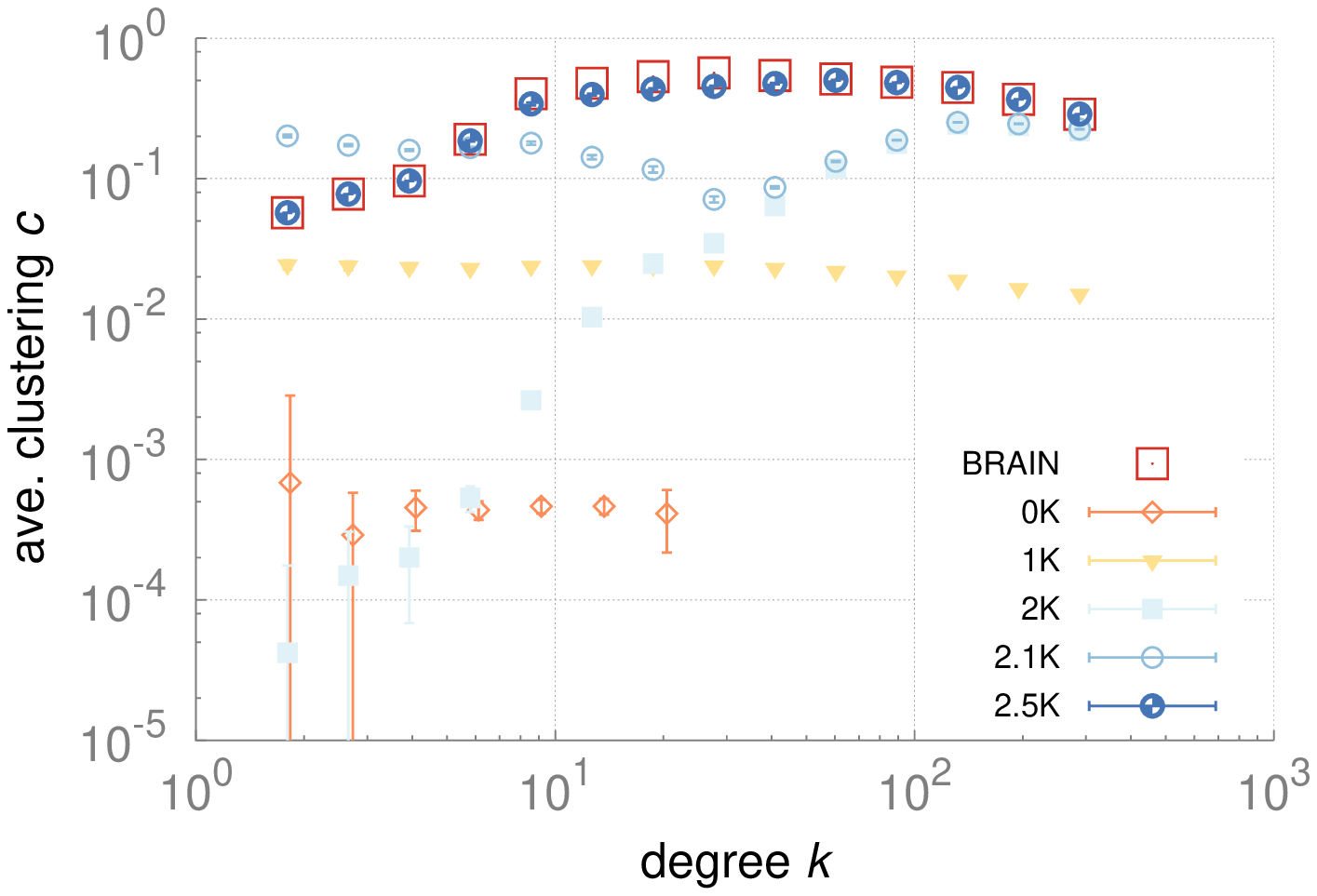}
}\\
\subfloat[WORDS]{%
  \includegraphics[width=0.5\textwidth]{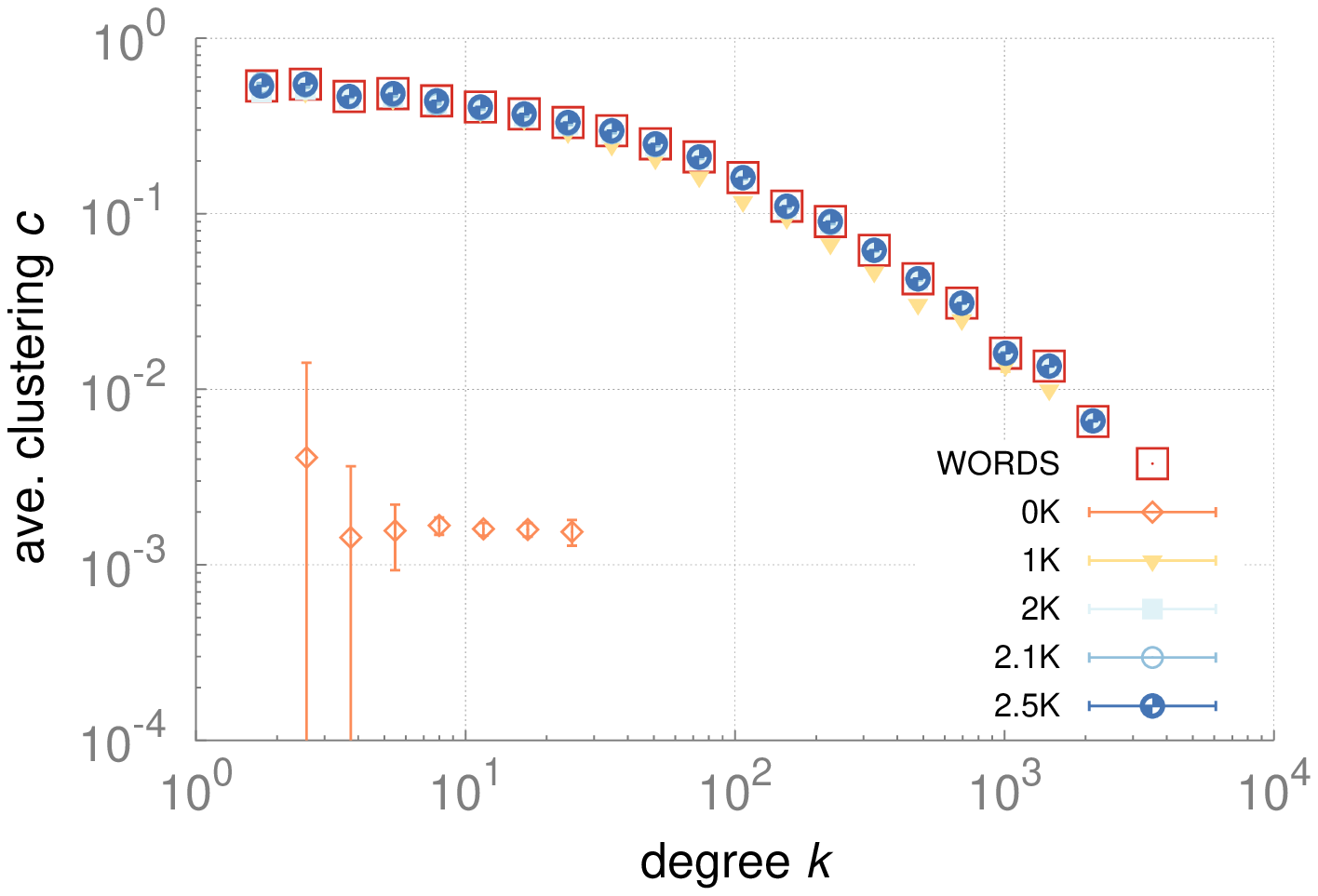}
}
\subfloat[INTERNET]{%
  \includegraphics[width=0.5\textwidth]{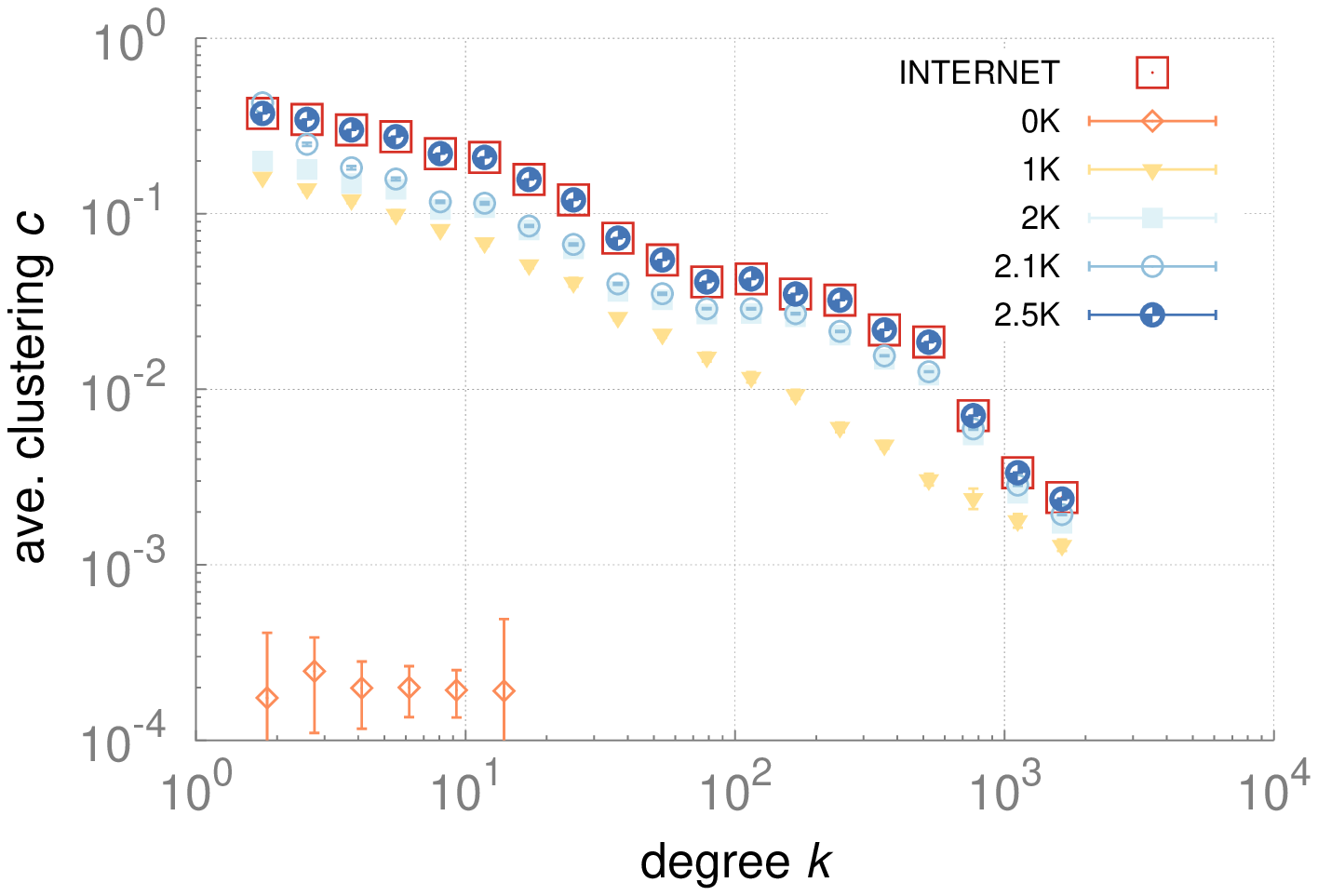}
}\\
\subfloat[PGP]{%
  \includegraphics[width=0.5\textwidth]{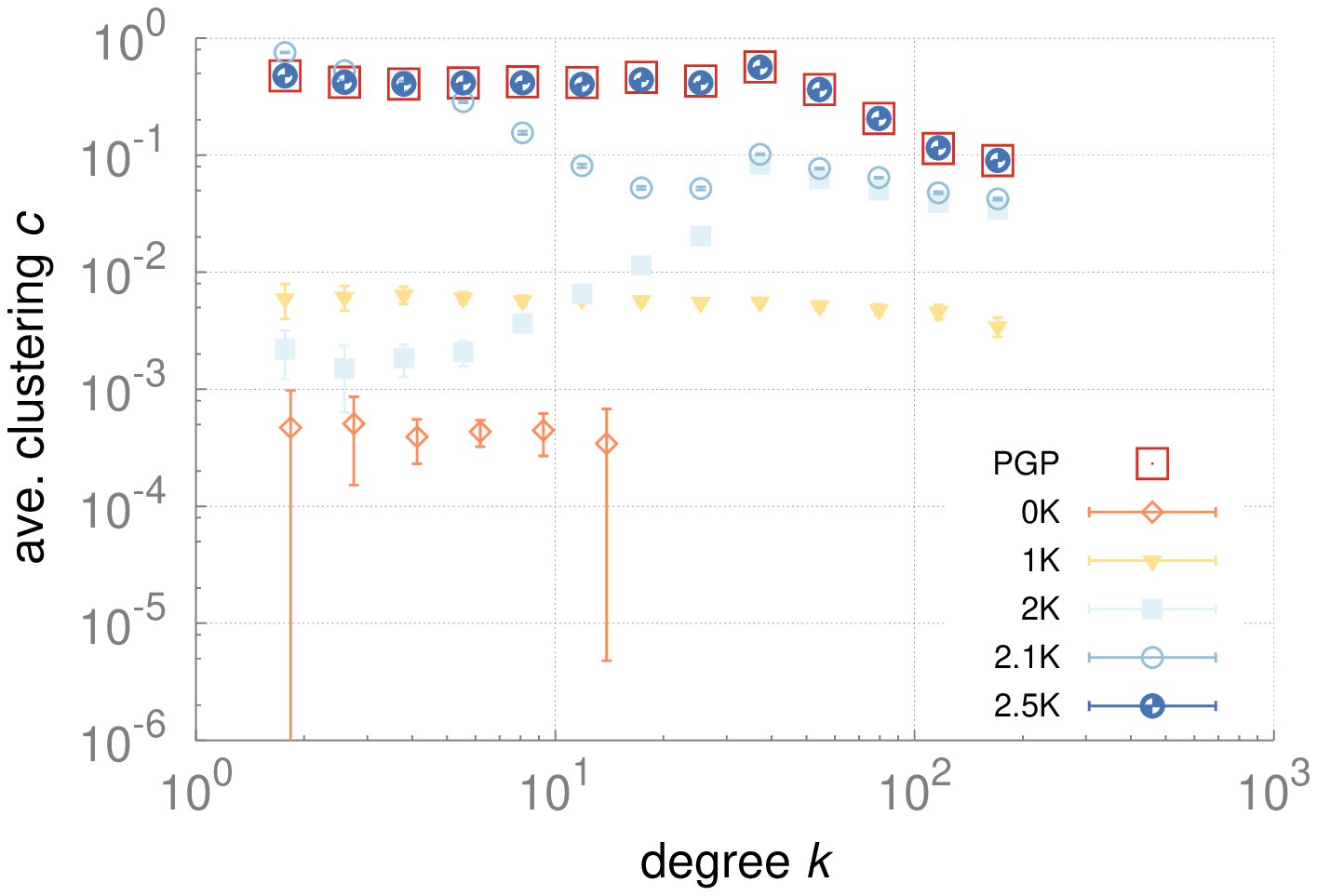}
}
\subfloat[PPI]{%
  \includegraphics[width=0.5\textwidth]{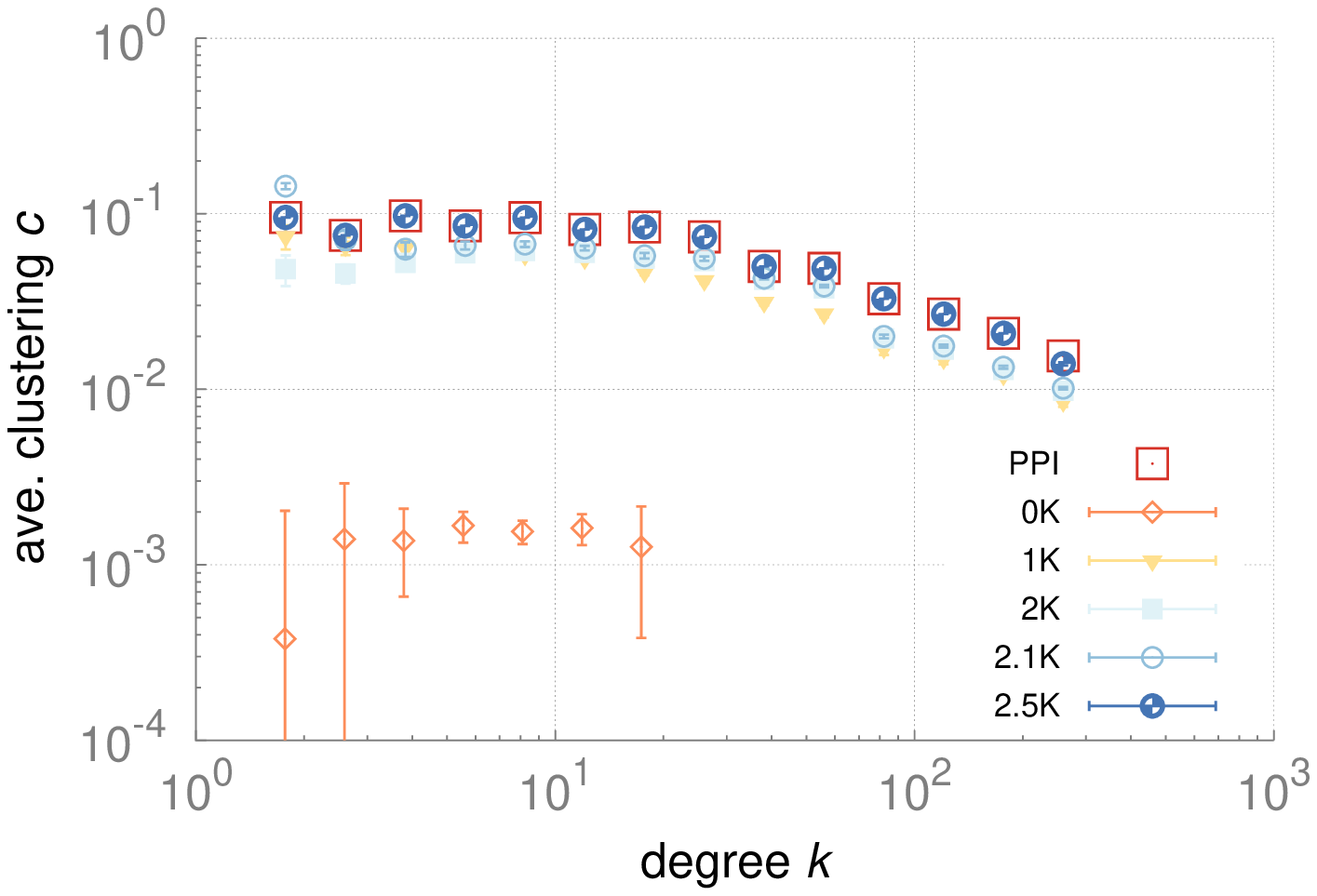}
}\\
\caption{\textbf{Average clustering of nodes of a given degree in real networks and their $dk$-randomizations.}}
\label{fig:cc}
\end{figure*}

\begin{figure*}
\centering
\makebox[\textwidth][c]{\includegraphics[trim= 0cm 0cm 5cm 0cm,
  clip=true, width=1.2\textwidth]{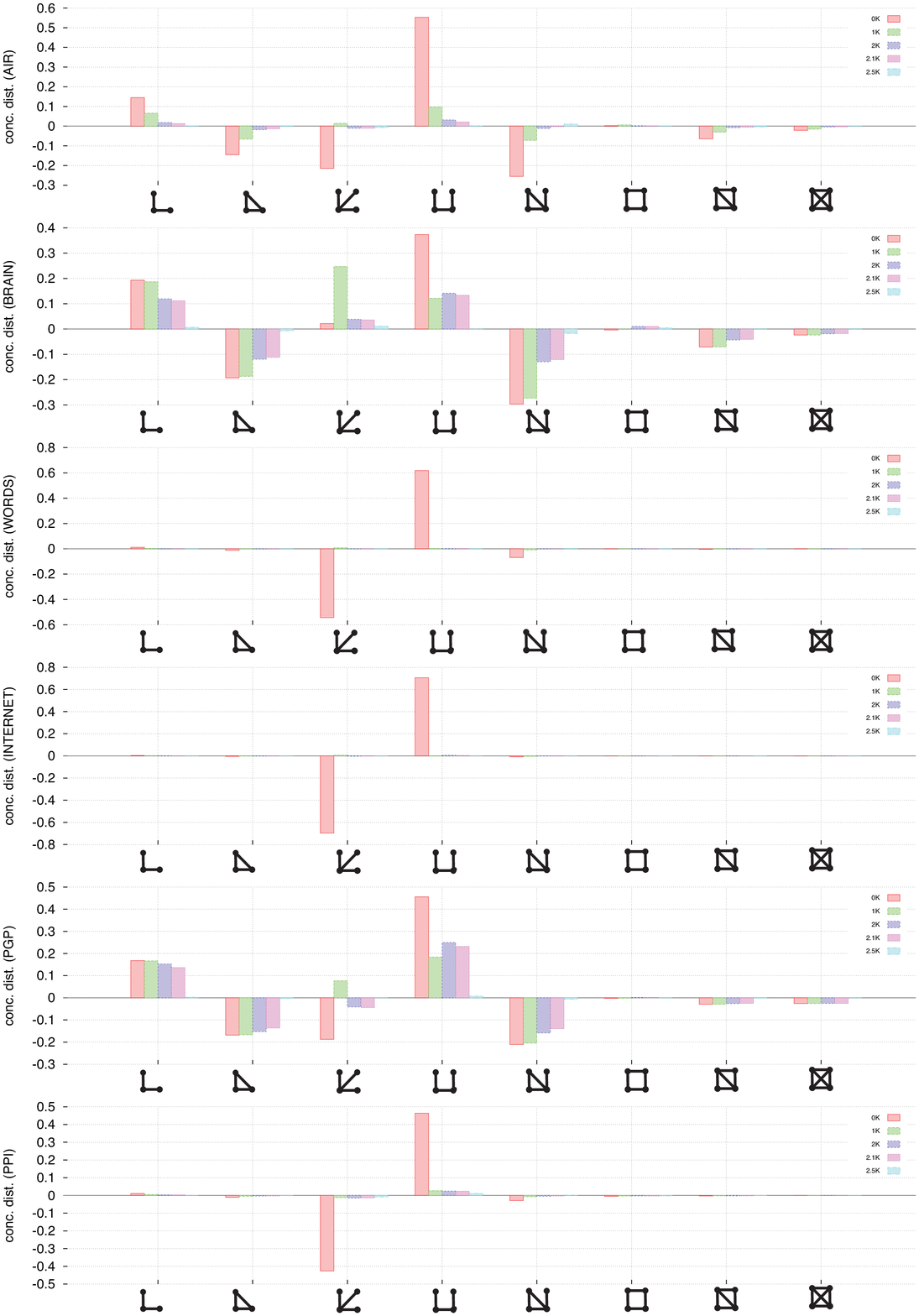}}%
\caption{\textbf{Subgraph concentration differences betweesn
  $dk$-randomizations and real networks.}}
\label{fig:motifssup}
\end{figure*}

\begin{figure*}
\centering
\subfloat[AIR]{%
  \includegraphics[width=0.5\textwidth]{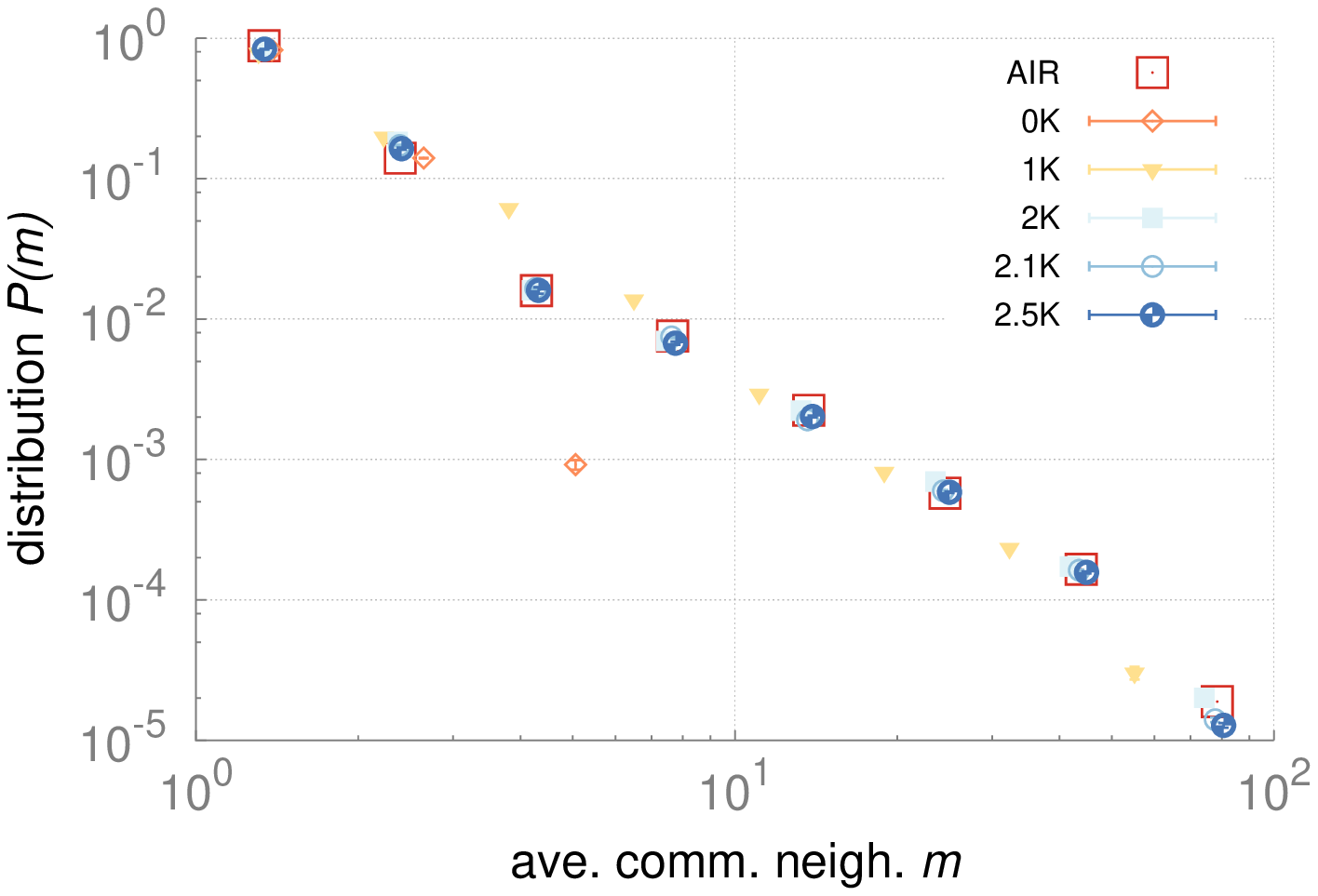}
}
\subfloat[BRAIN]{%
  \includegraphics[width=0.5\textwidth]{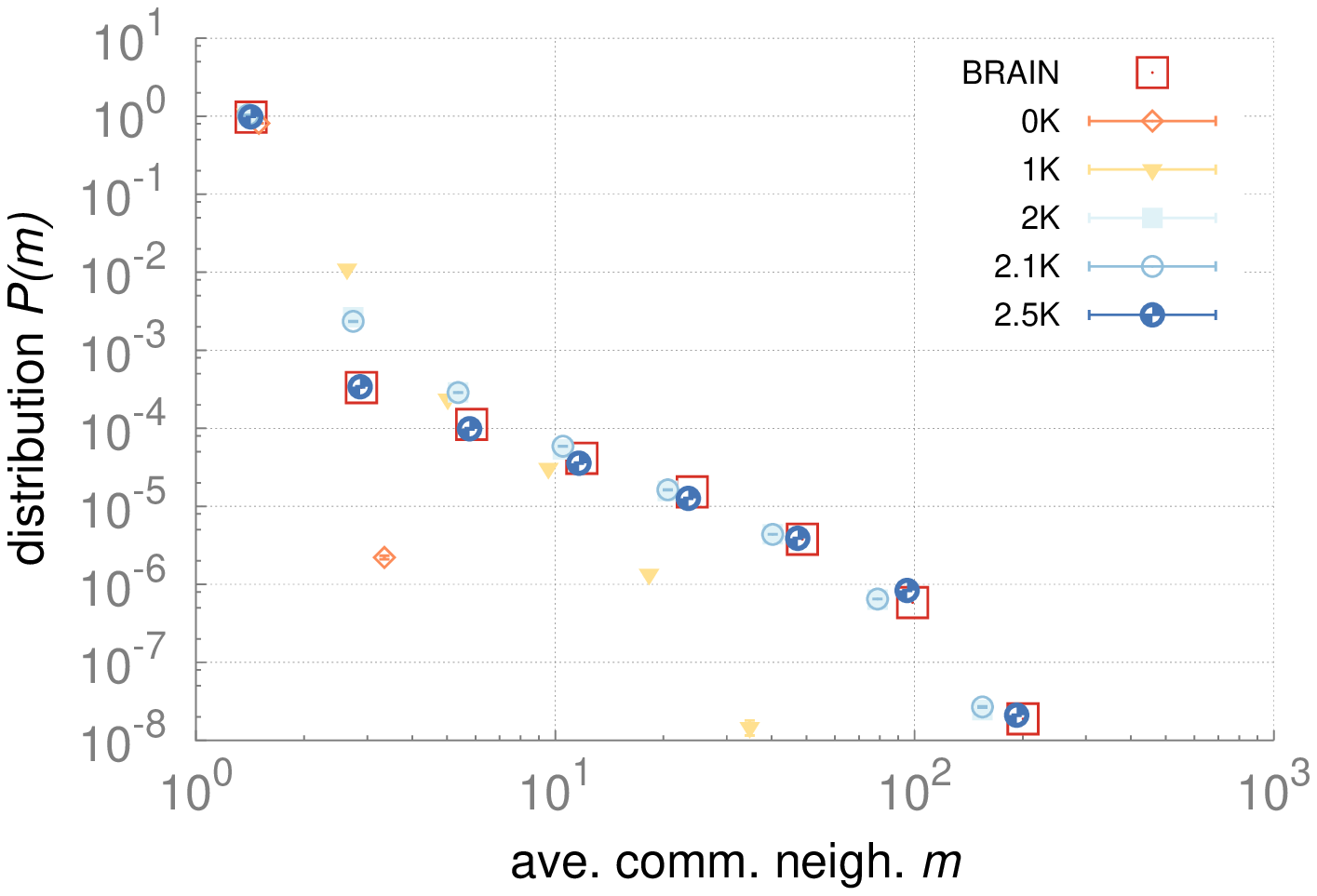}
}\\
\subfloat[WORDS]{%
  \includegraphics[width=0.5\textwidth]{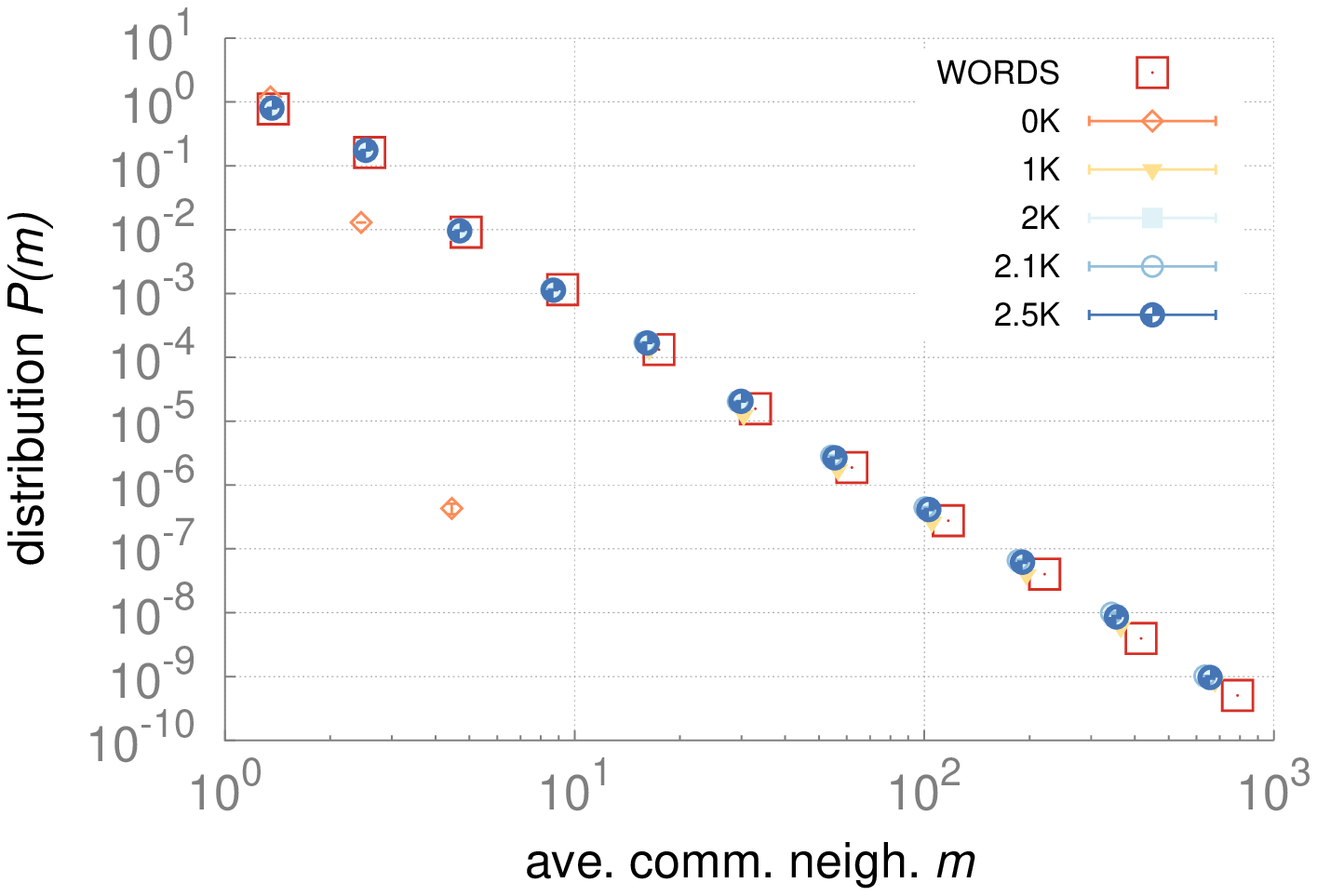}
}
\subfloat[INTERNET]{%
  \includegraphics[width=0.5\textwidth]{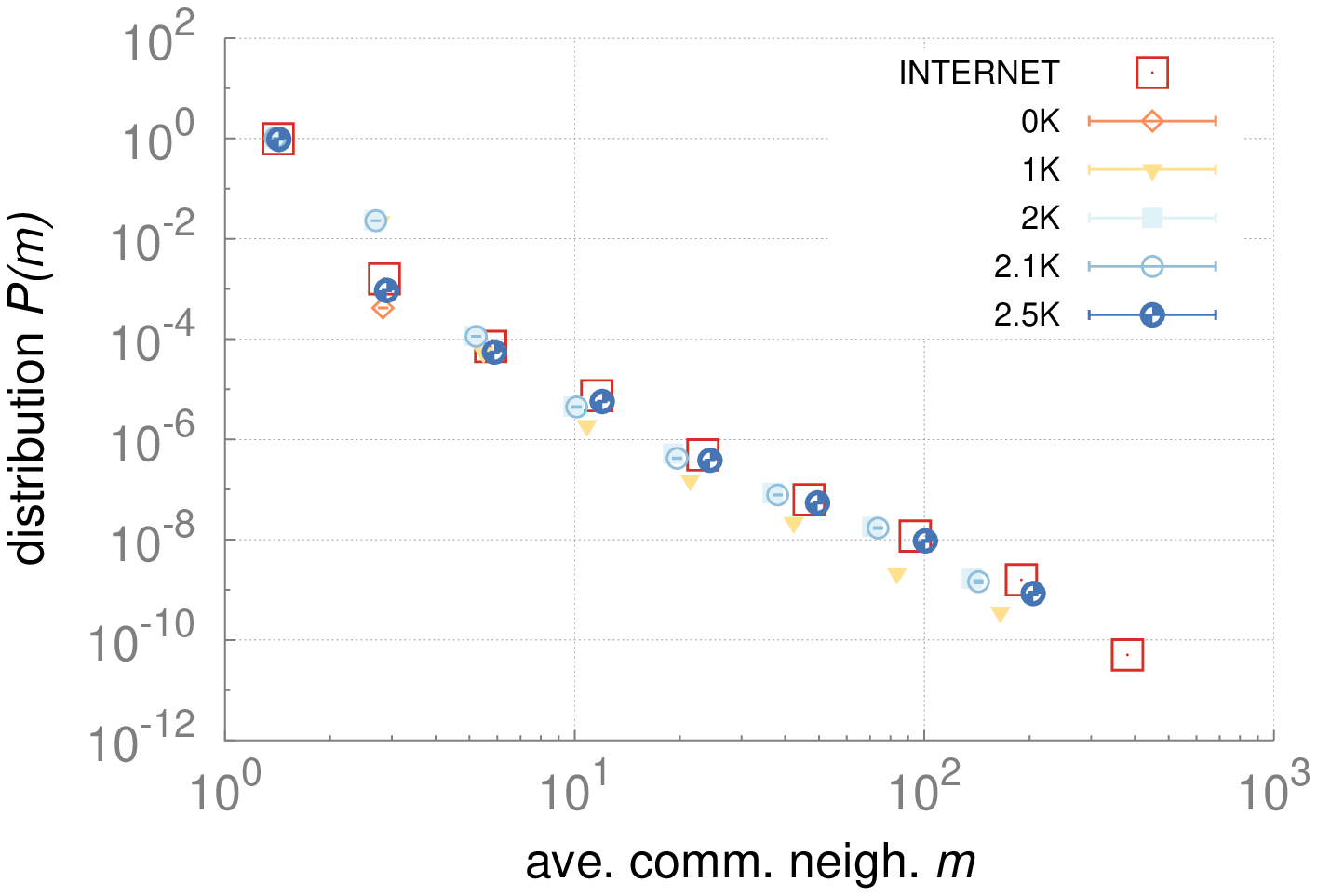}
}\\
\subfloat[PGP]{%
  \includegraphics[width=0.5\textwidth]{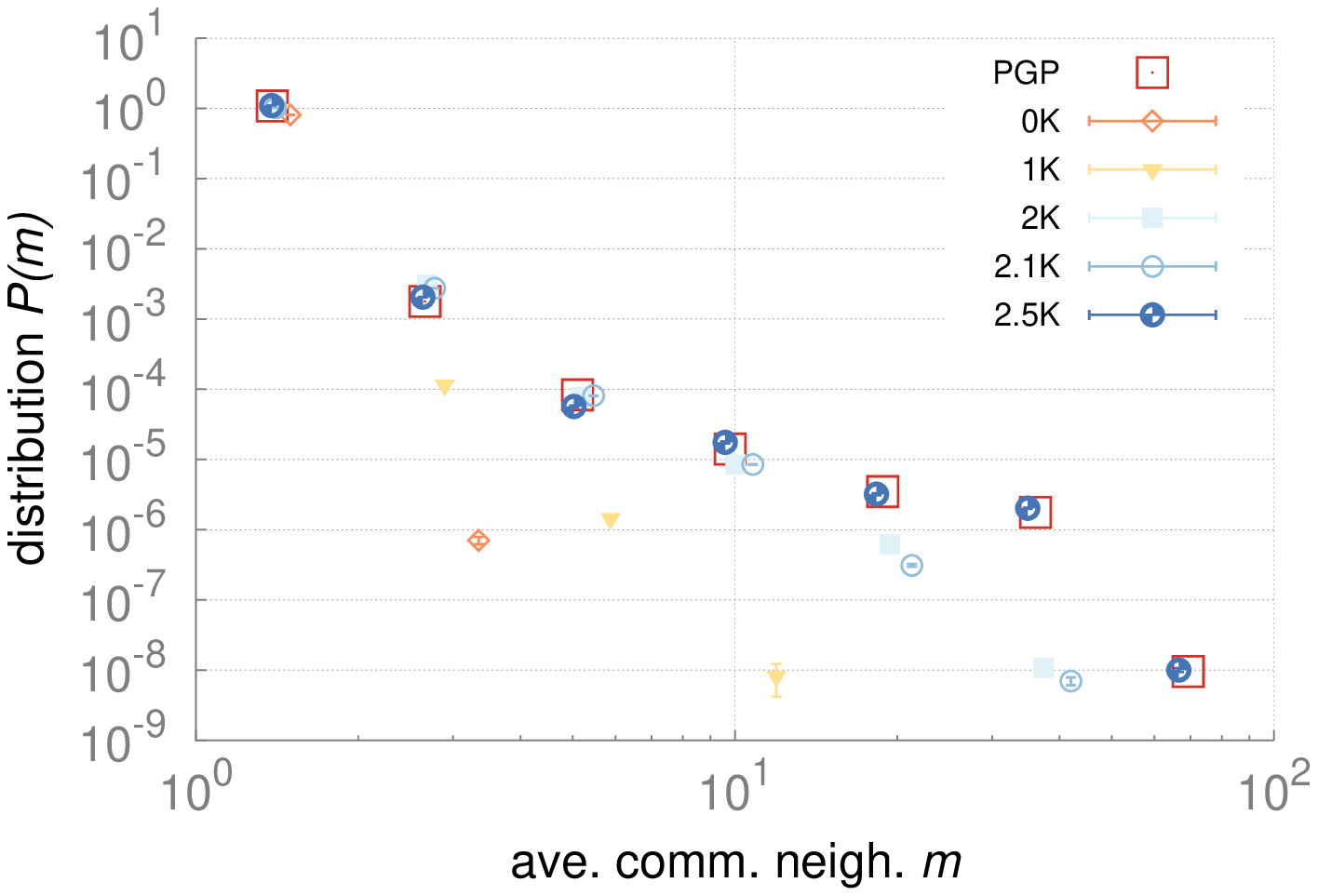}
}
\subfloat[PPI]{%
  \includegraphics[width=0.5\textwidth]{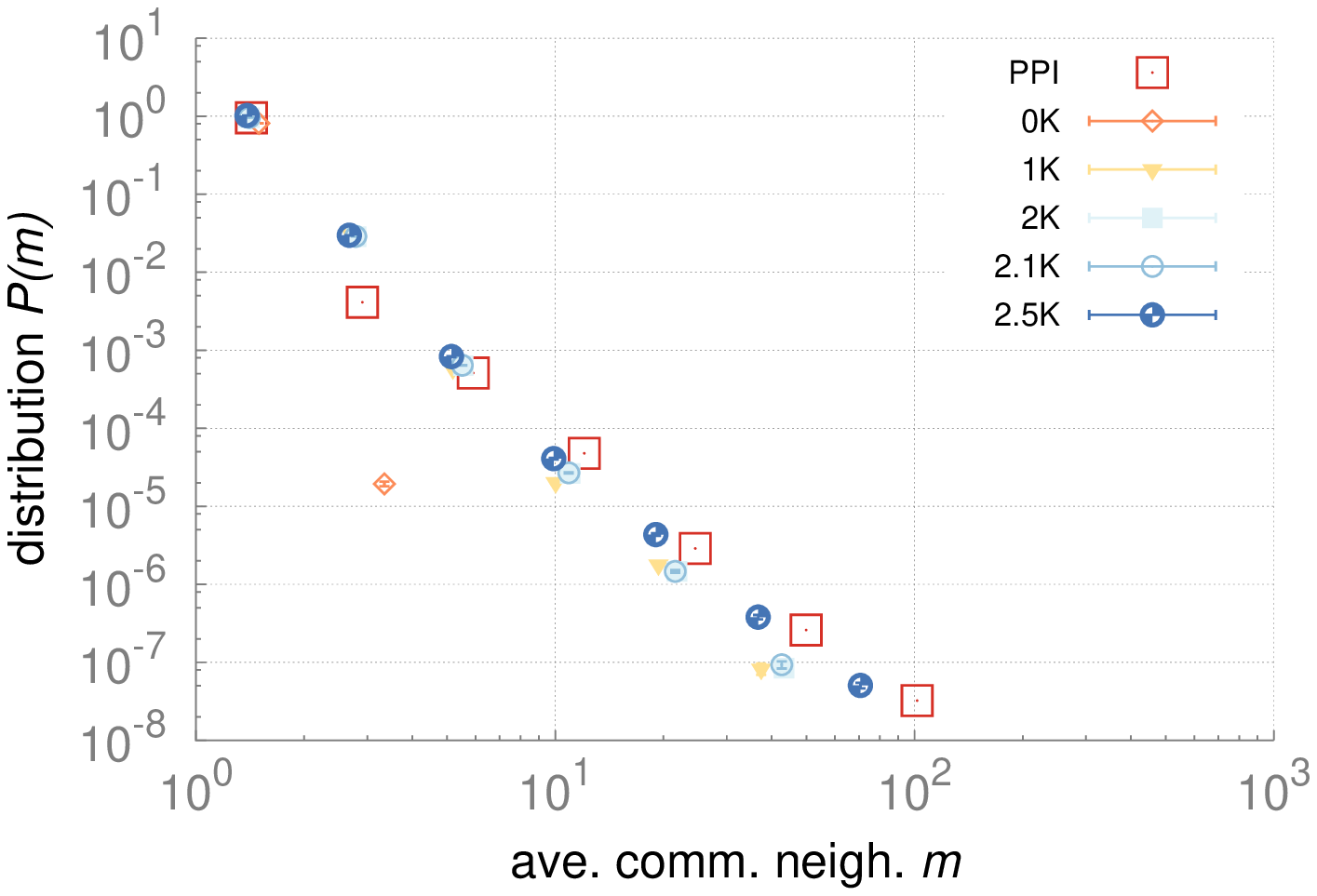}
}\\
\caption{\textbf{Common neighbor distributions in real networks and their $dk$-randomizations.}}
\label{fig:cn}
\end{figure*}

\begin{figure*}
\centering
\subfloat[AIR]{%
  \includegraphics[width=0.5\textwidth]{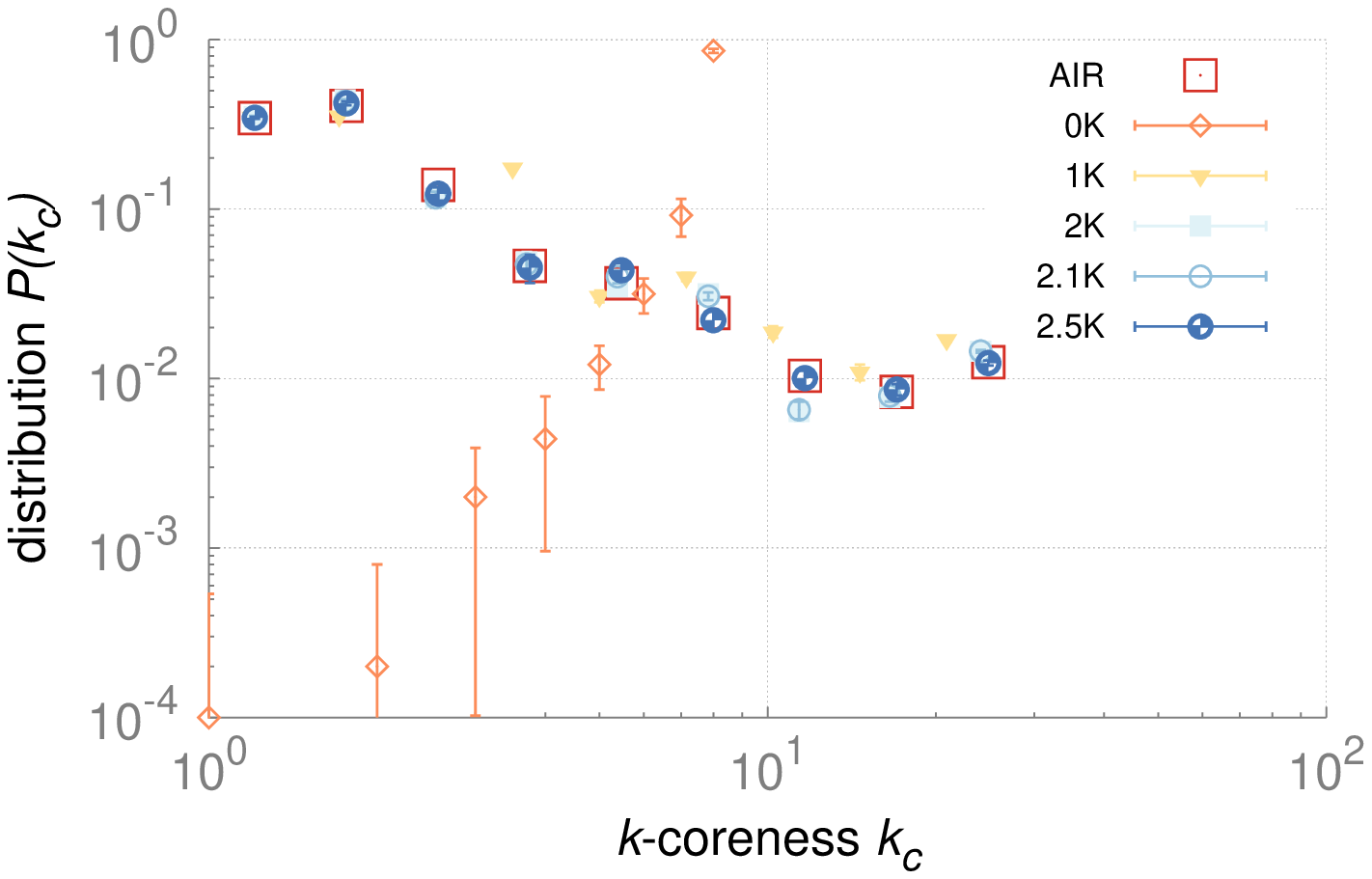}
}
\subfloat[BRAIN]{%
  \includegraphics[width=0.5\textwidth]{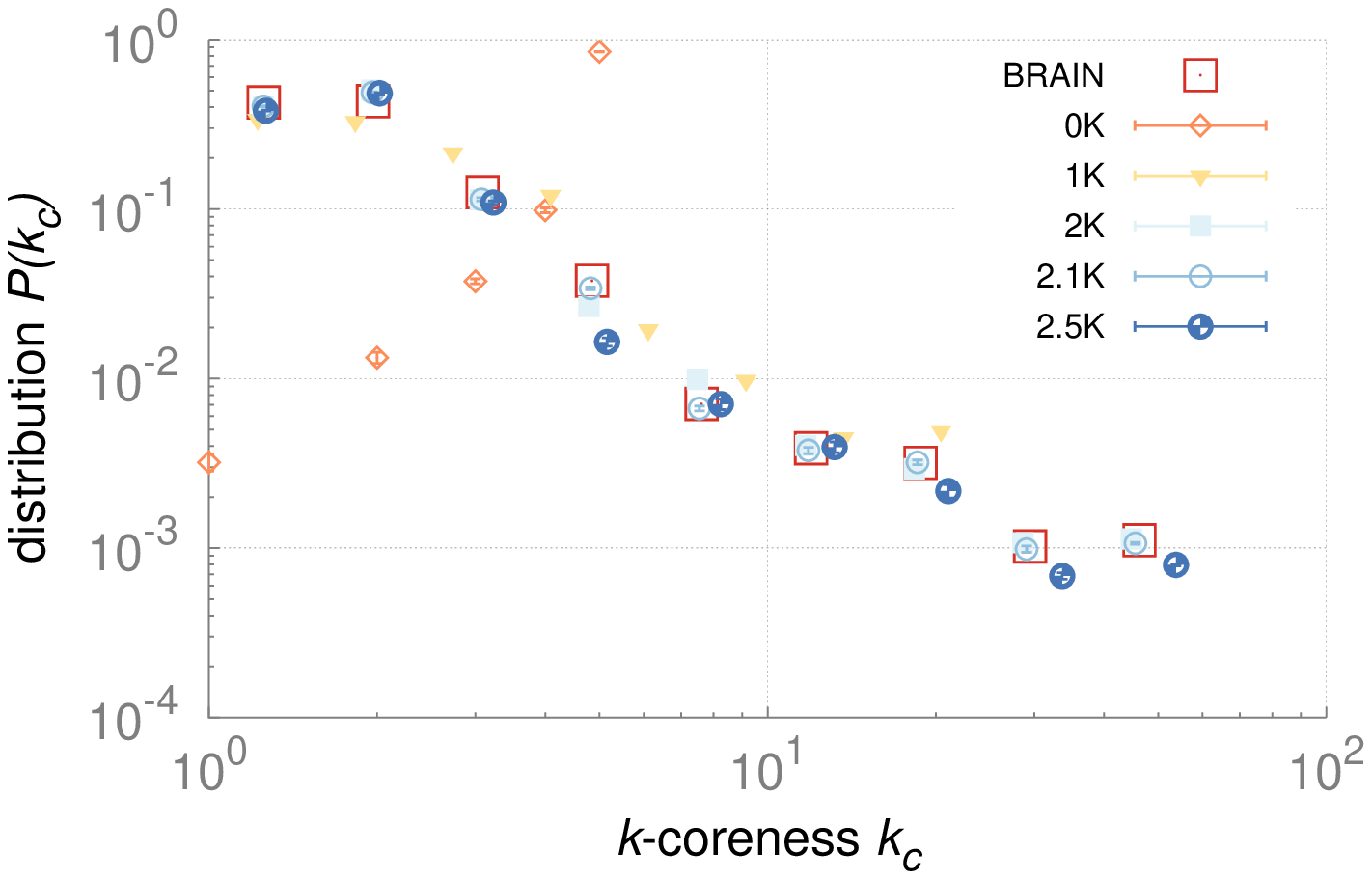}
}\\
\subfloat[WORDS]{%
  \includegraphics[width=0.5\textwidth]{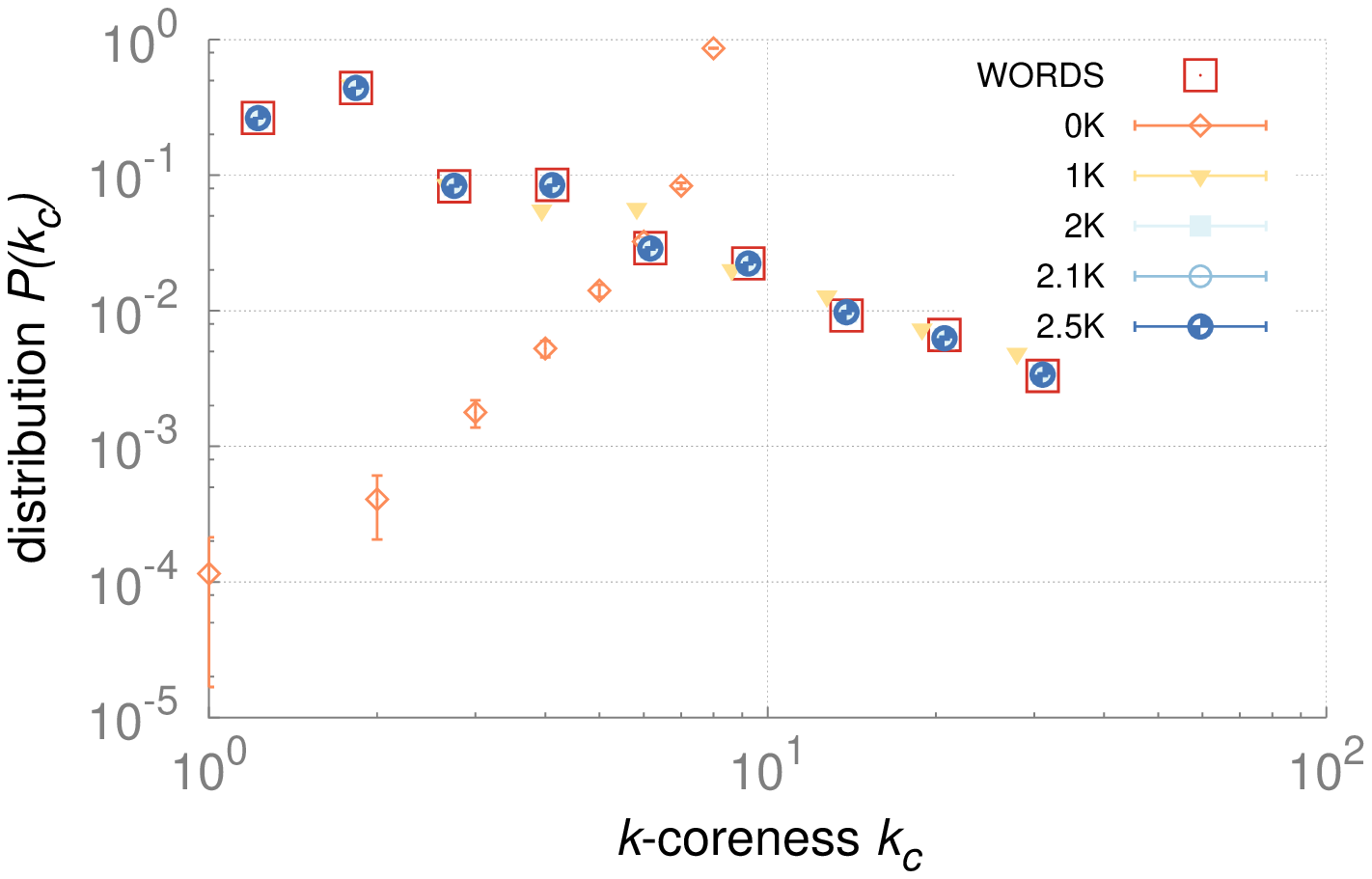}
}
\subfloat[INTERNET]{%
  \includegraphics[width=0.5\textwidth]{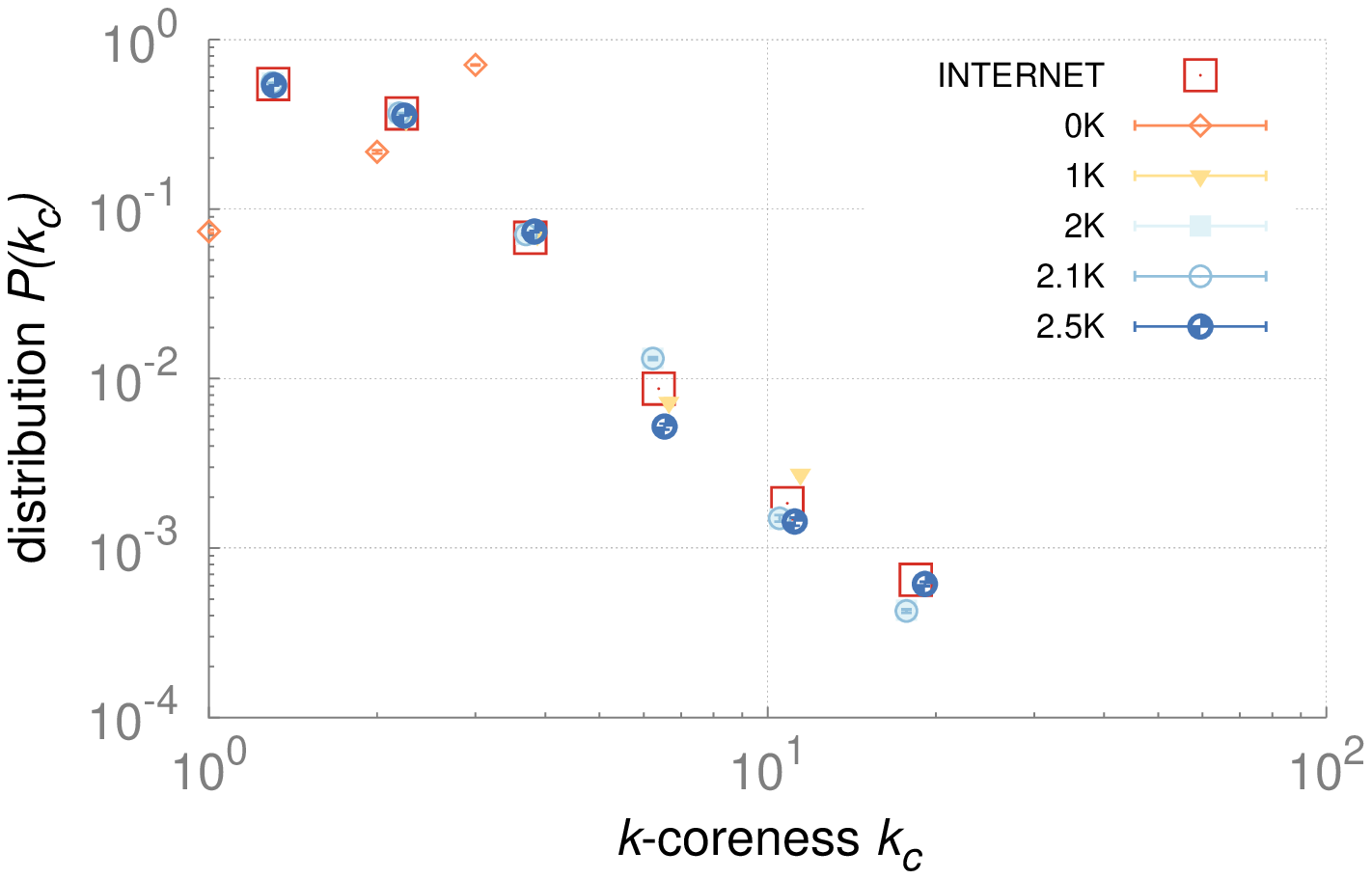}
}\\
\subfloat[PGP]{%
  \includegraphics[width=0.5\textwidth]{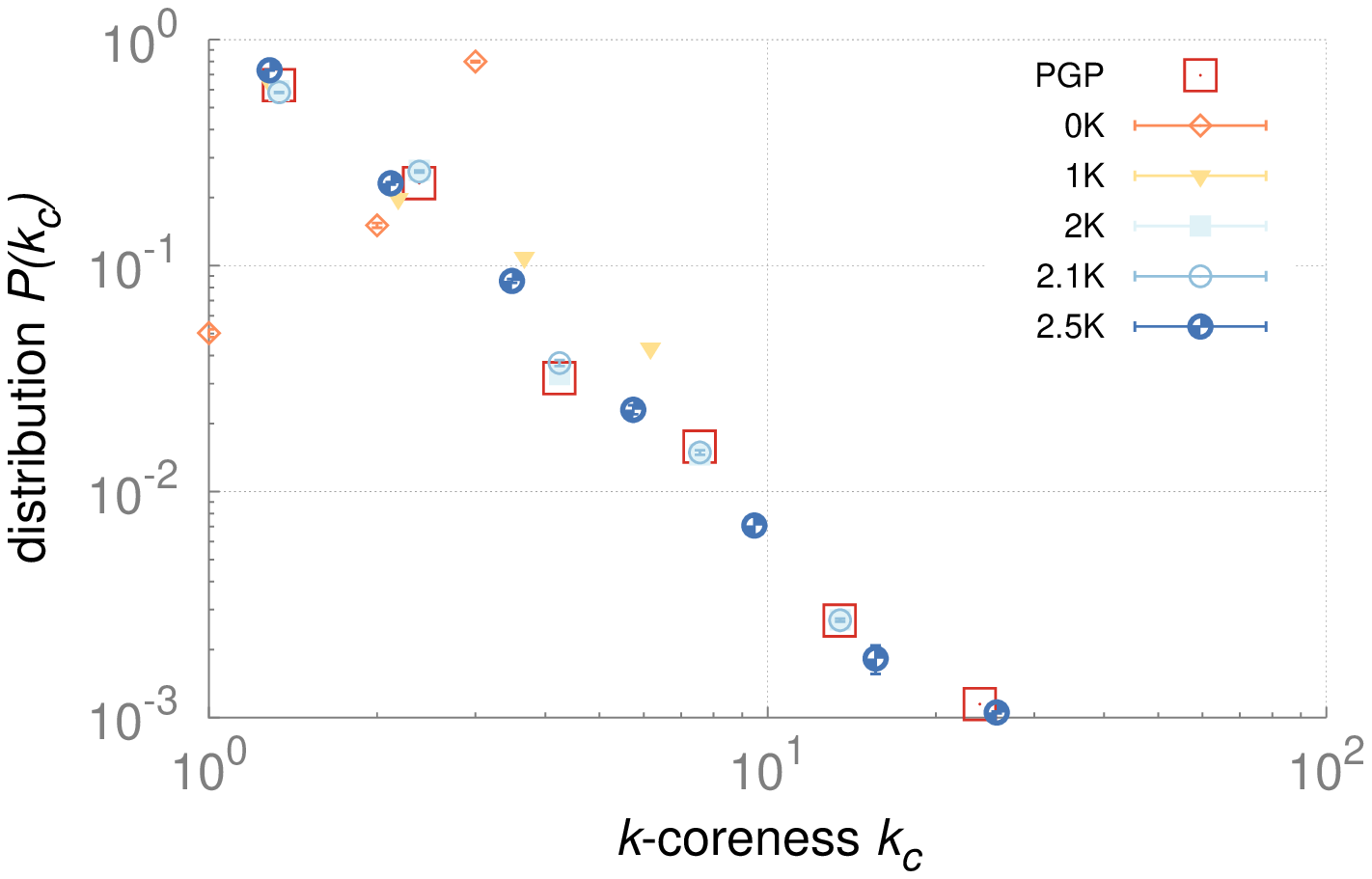}
}
\subfloat[PPI]{%
  \includegraphics[width=0.5\textwidth]{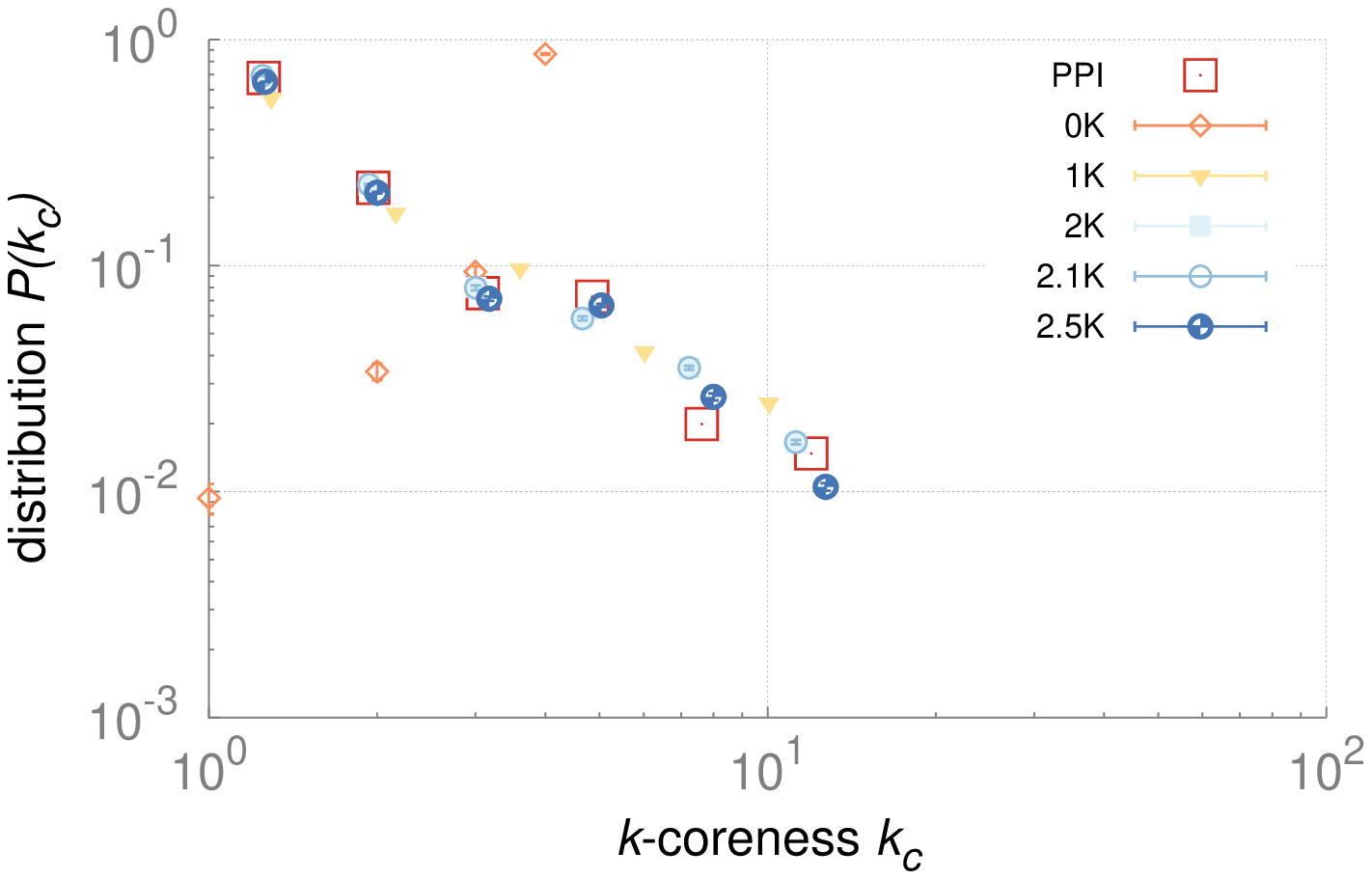}
}\\
\caption{\textbf{$k$-coreness distributions in real networks and their $dk$-randomizations.}}
\label{fig:kcore}
\end{figure*}

\begin{figure*}
\centering
\subfloat[AIR]{%
  \includegraphics[width=0.5\textwidth]{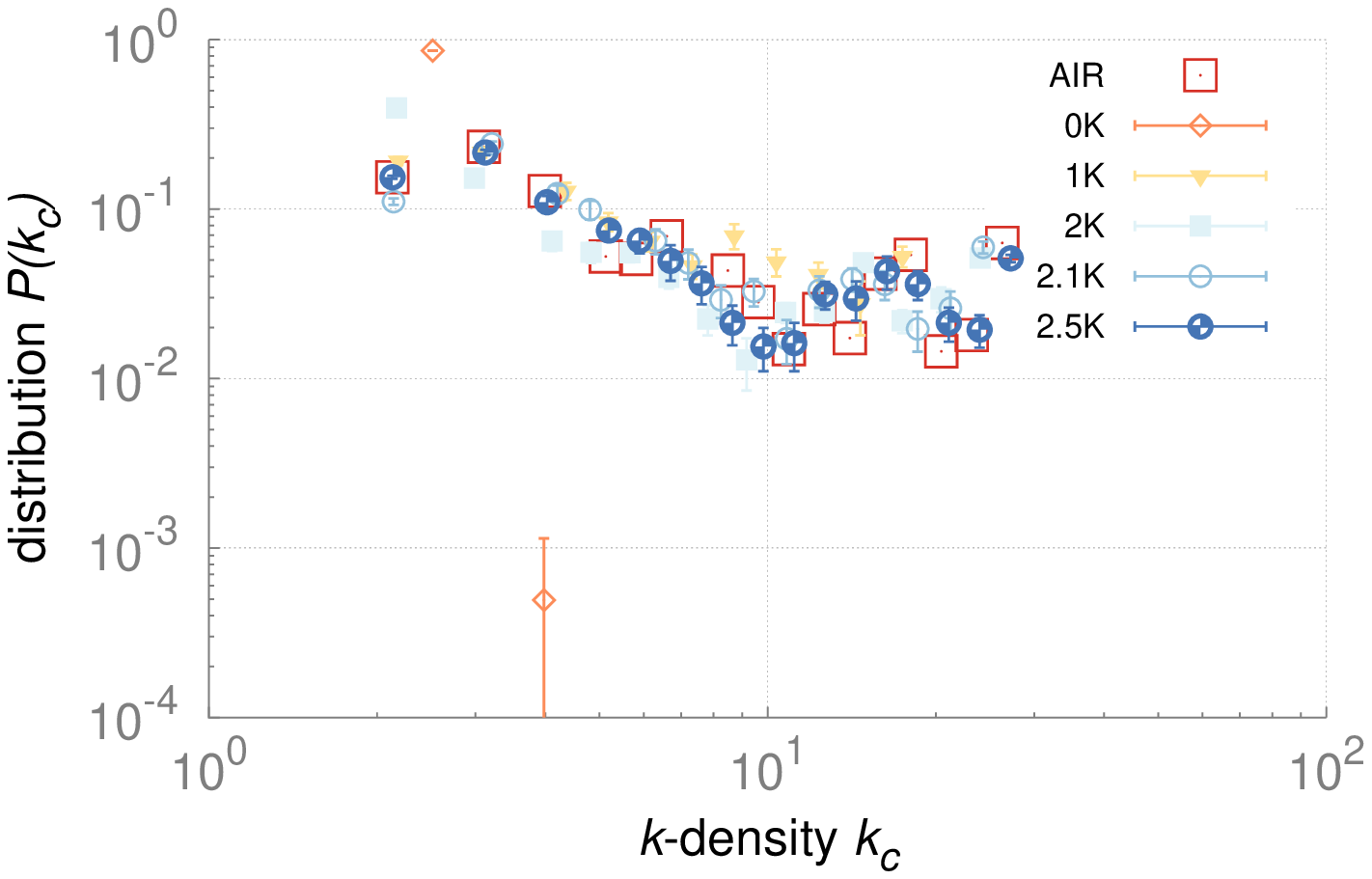}
}
\subfloat[BRAIN]{%
  \includegraphics[width=0.5\textwidth]{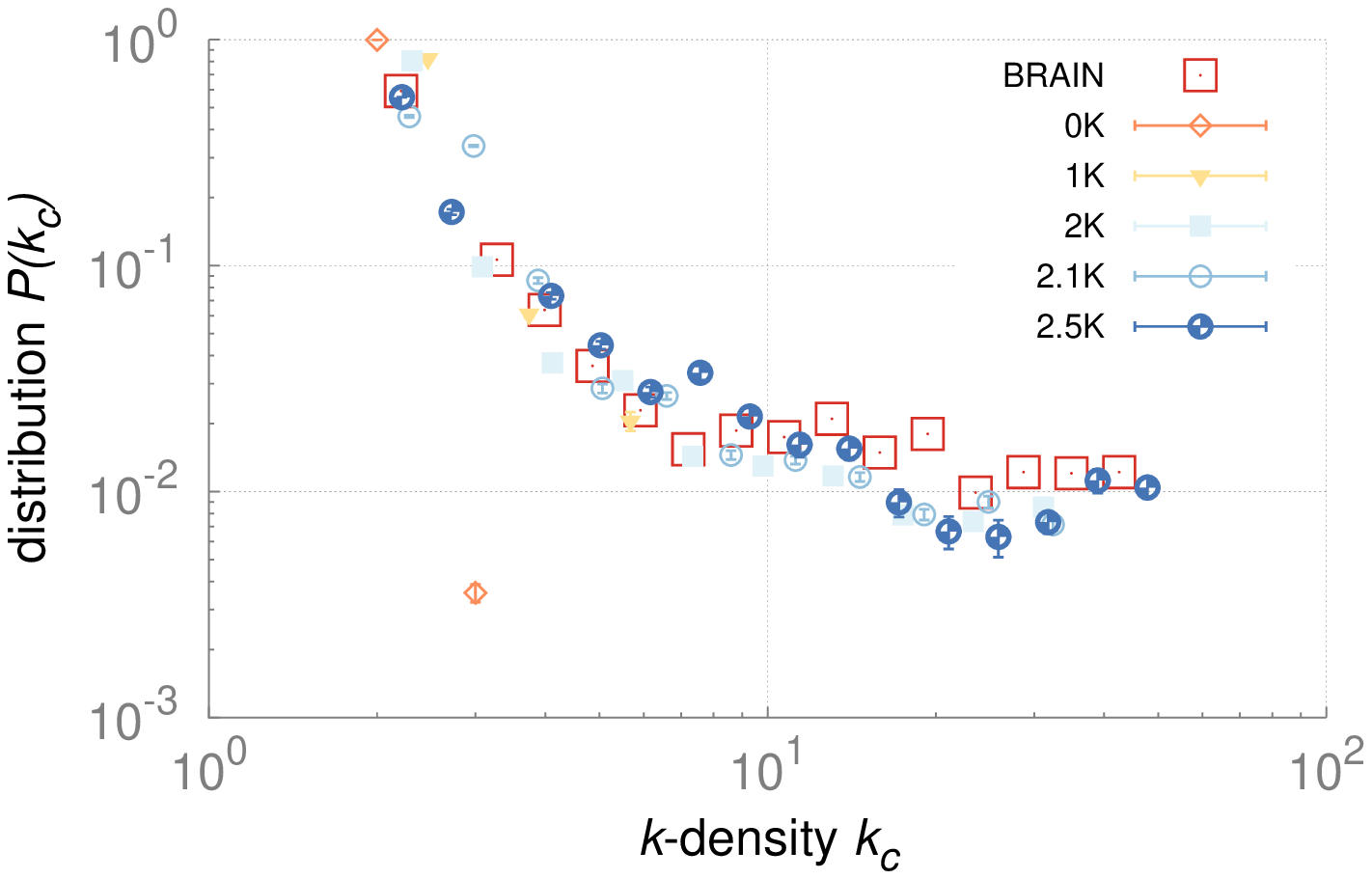}
}\\
\subfloat[WORDS]{%
  \includegraphics[width=0.5\textwidth]{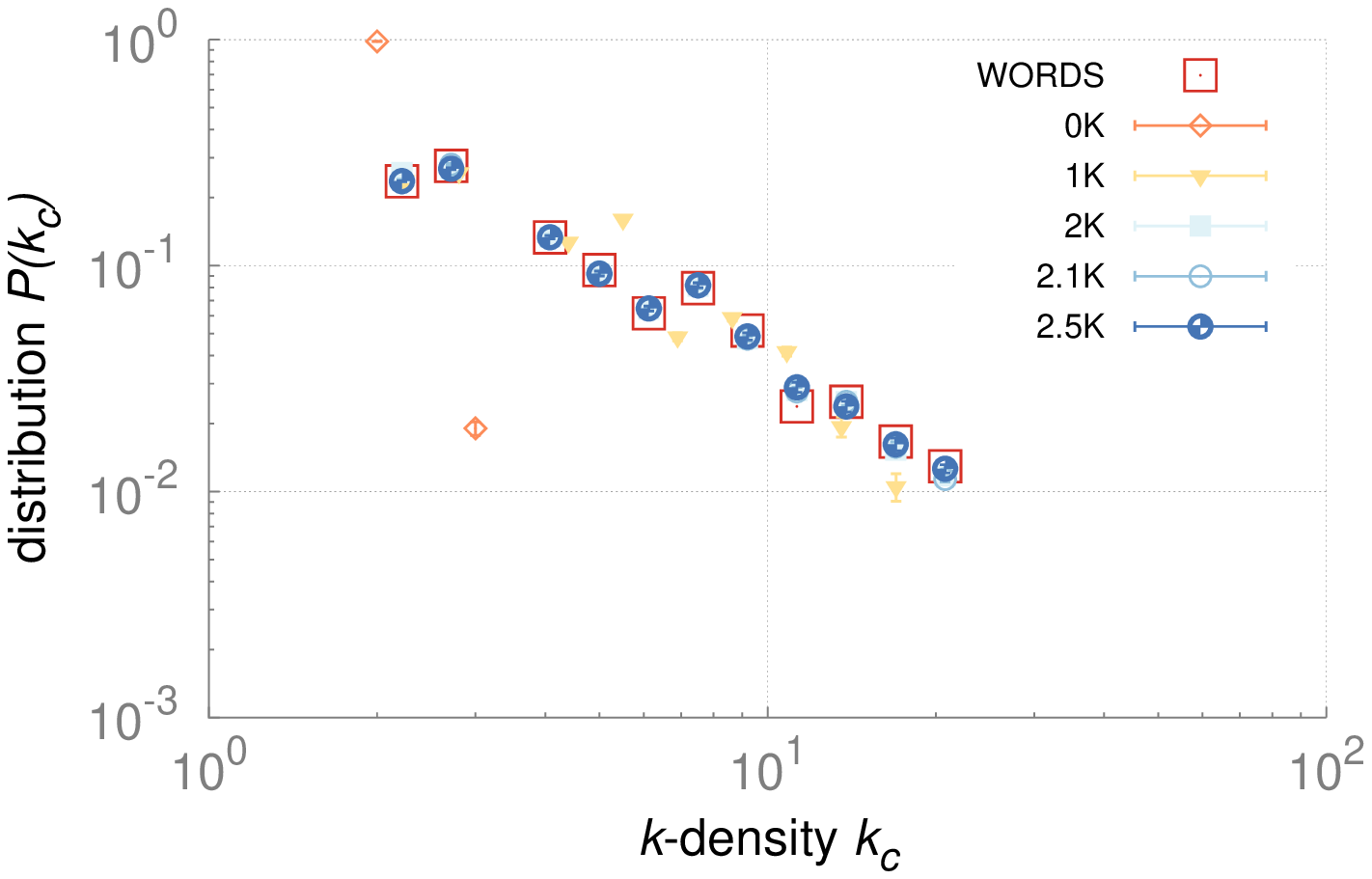}
}
\subfloat[INTERNET]{%
  \includegraphics[width=0.5\textwidth]{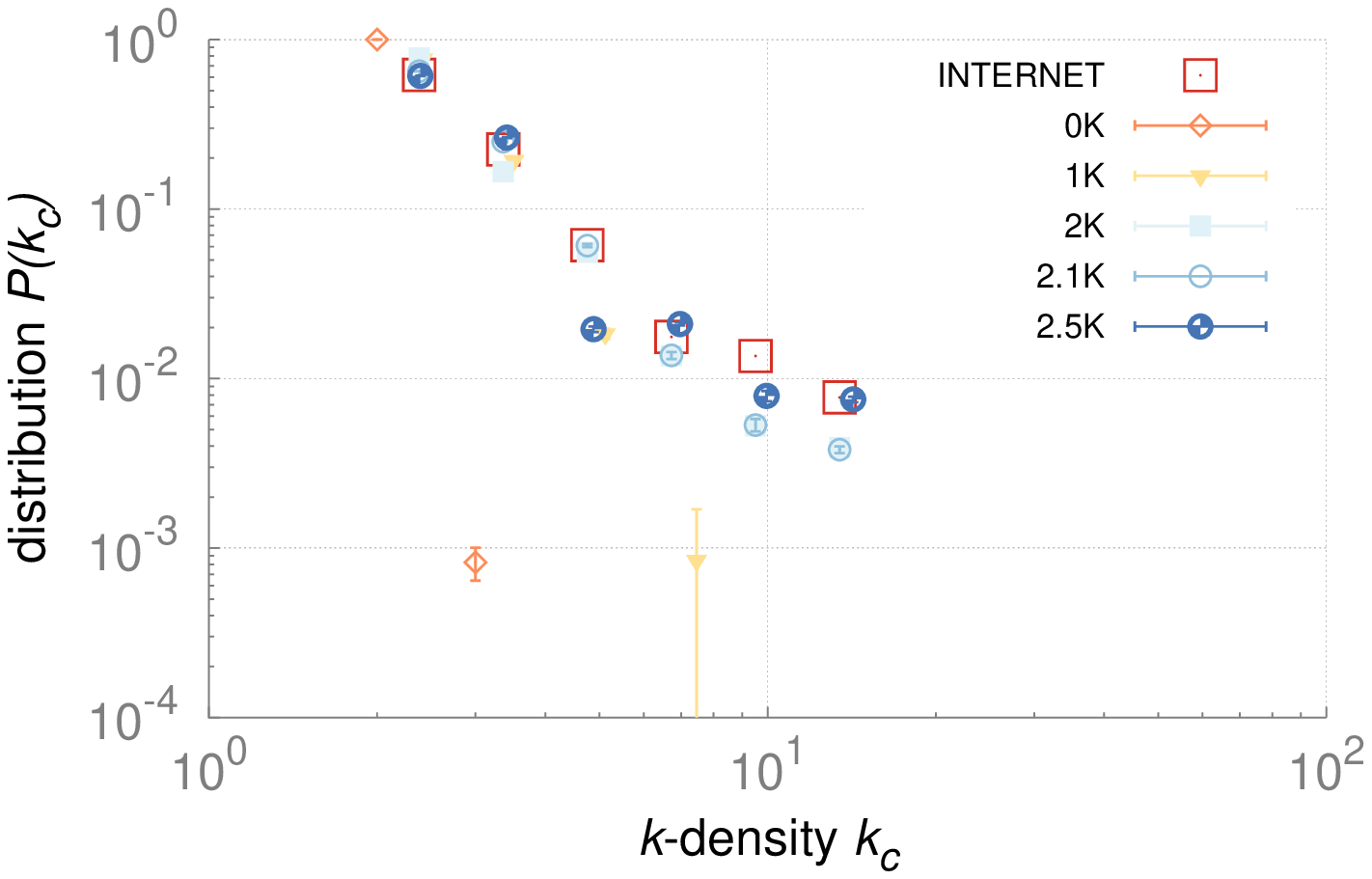}
}\\
\subfloat[PGP]{%
  \includegraphics[width=0.5\textwidth]{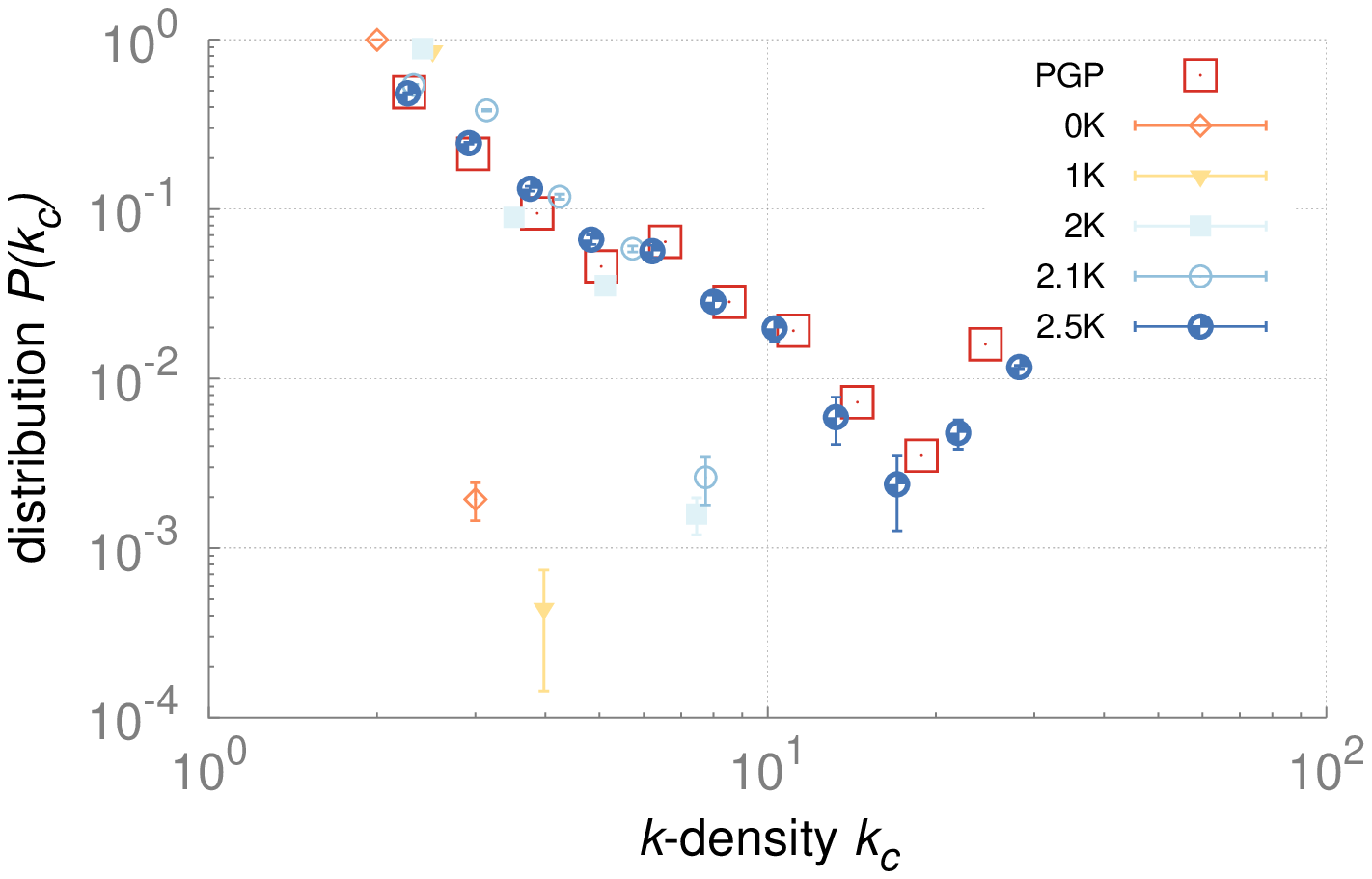}
}
\subfloat[PPI]{%
  \includegraphics[width=0.5\textwidth]{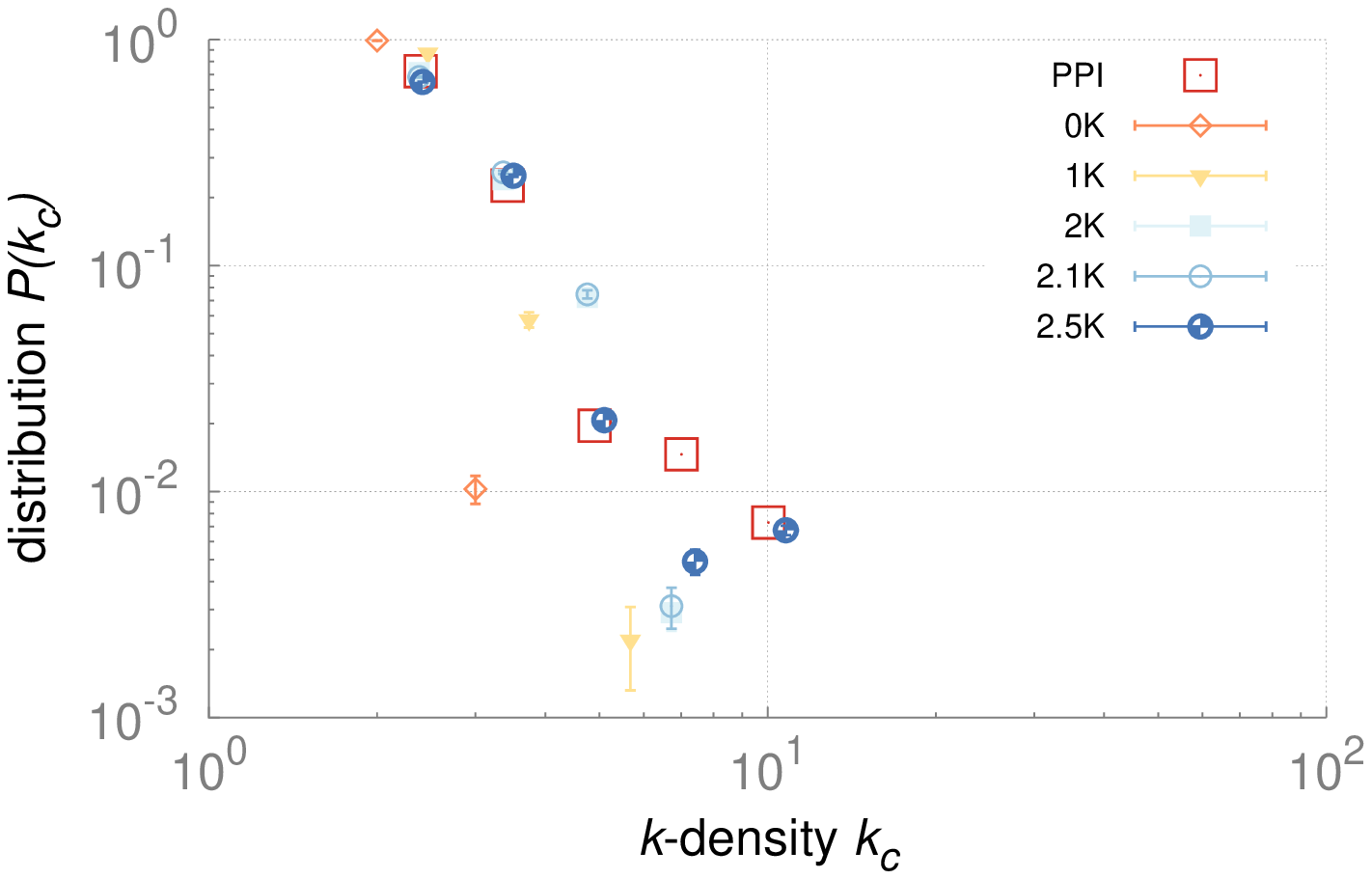}
}\\
\caption{\textbf{$k$-denseness distributions in real networks and their $dk$-randomizations.}}
\label{fig:kdense}
\end{figure*}

\begin{figure*}
\centering
\subfloat[AIR]{%
  \includegraphics[width=0.5\textwidth]{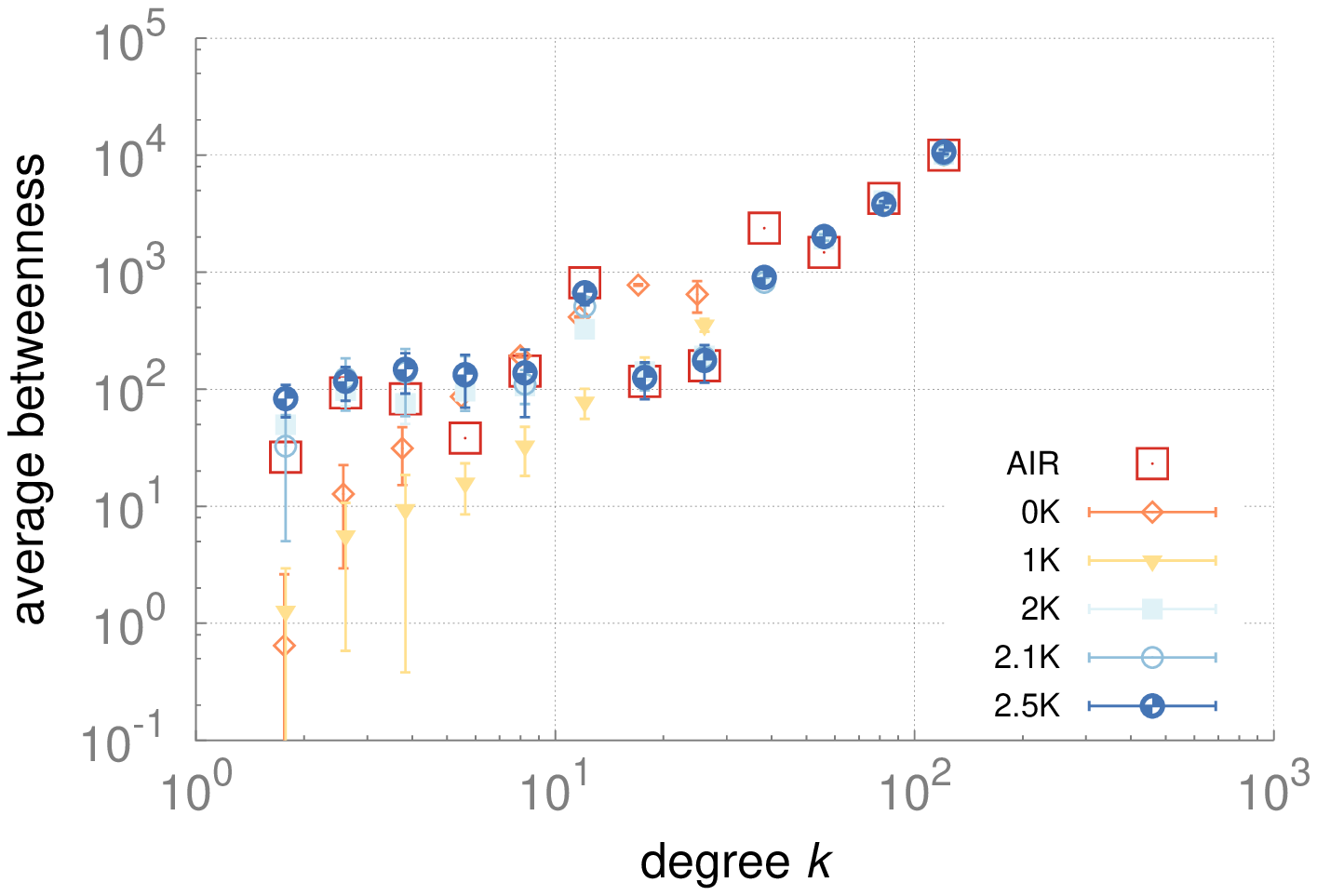}
}
\subfloat[BRAIN]{%
  \includegraphics[width=0.5\textwidth]{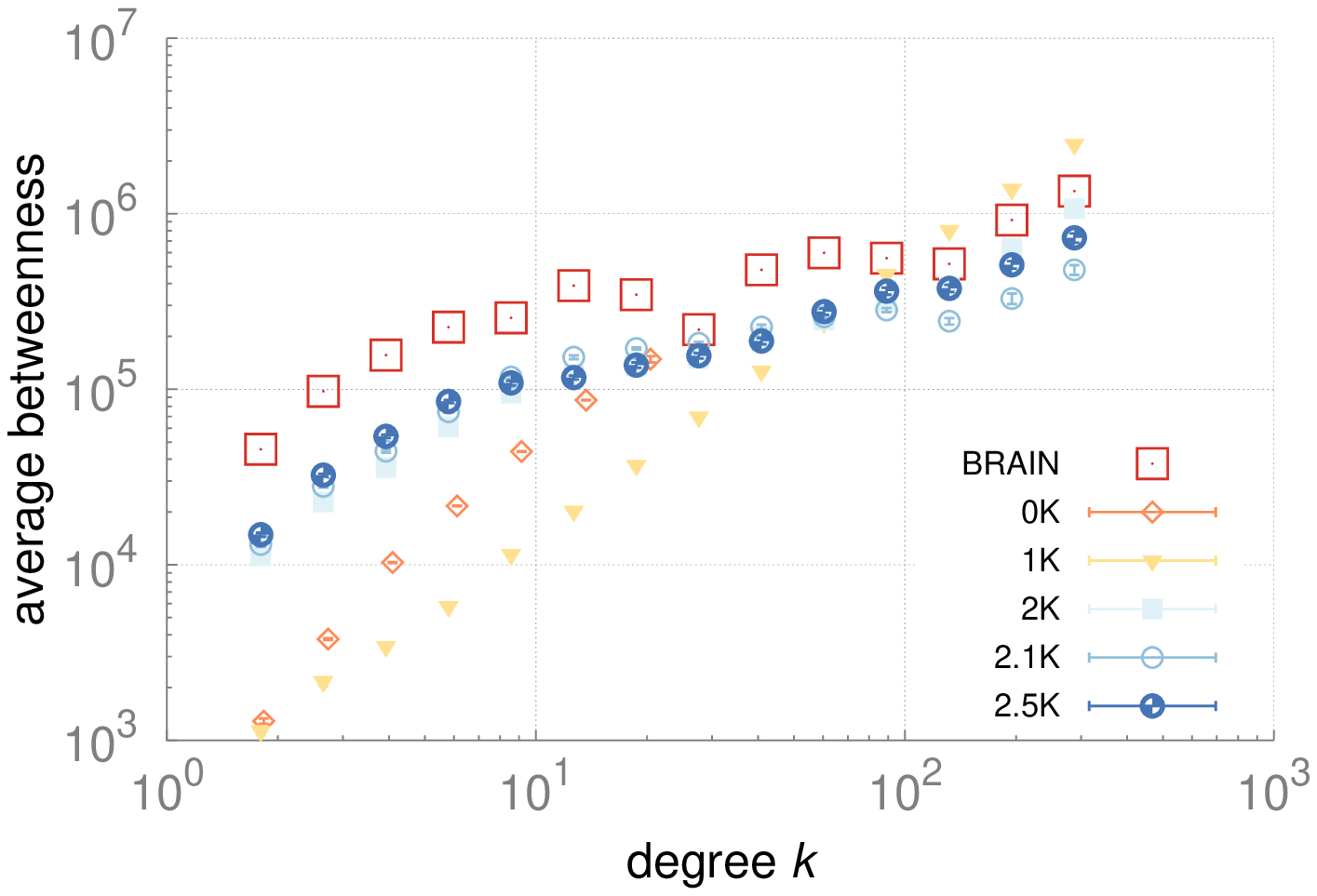}
}\\
\subfloat[WORDS]{%
  \includegraphics[width=0.5\textwidth]{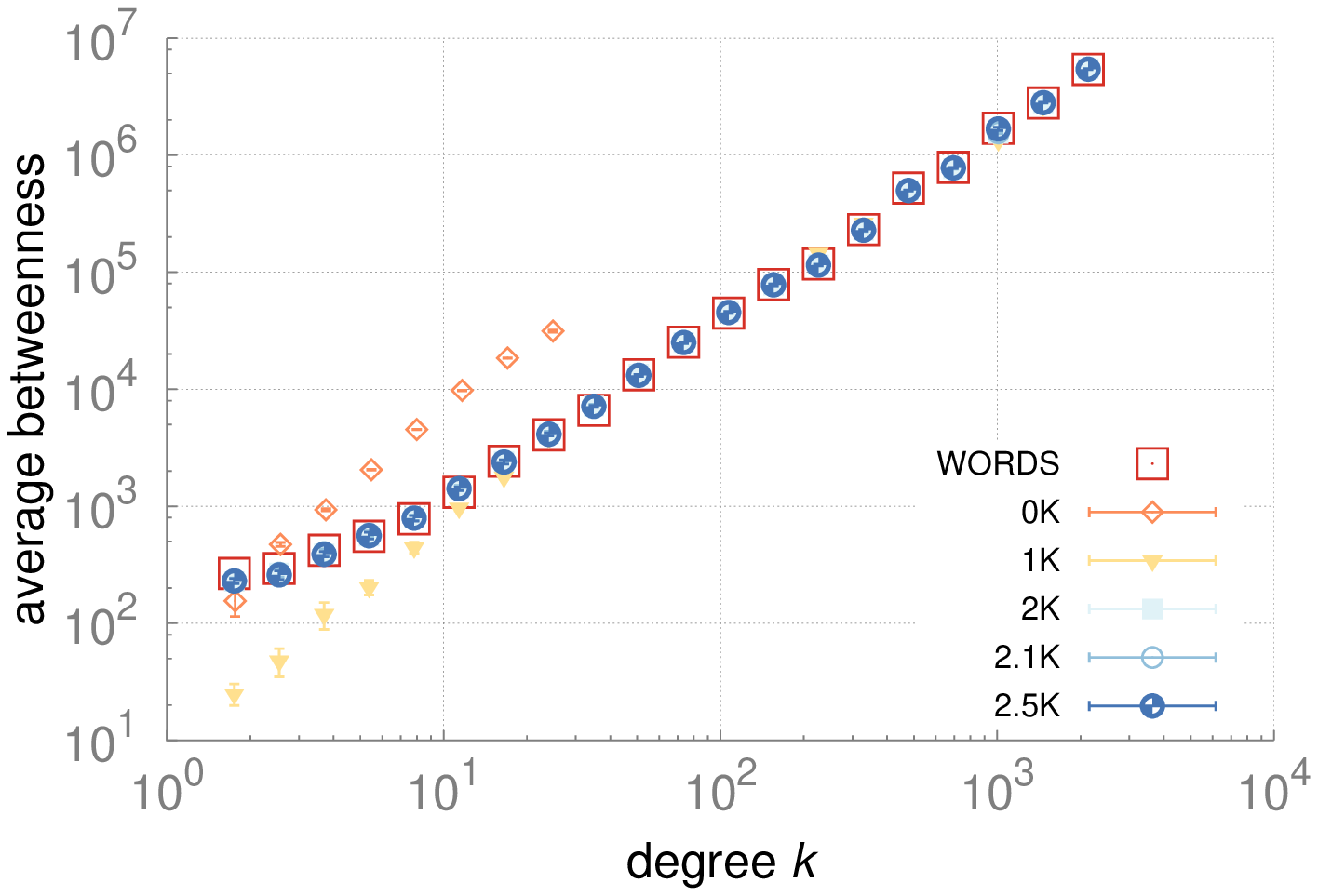}
}
\subfloat[INTERNET]{%
  \includegraphics[width=0.5\textwidth]{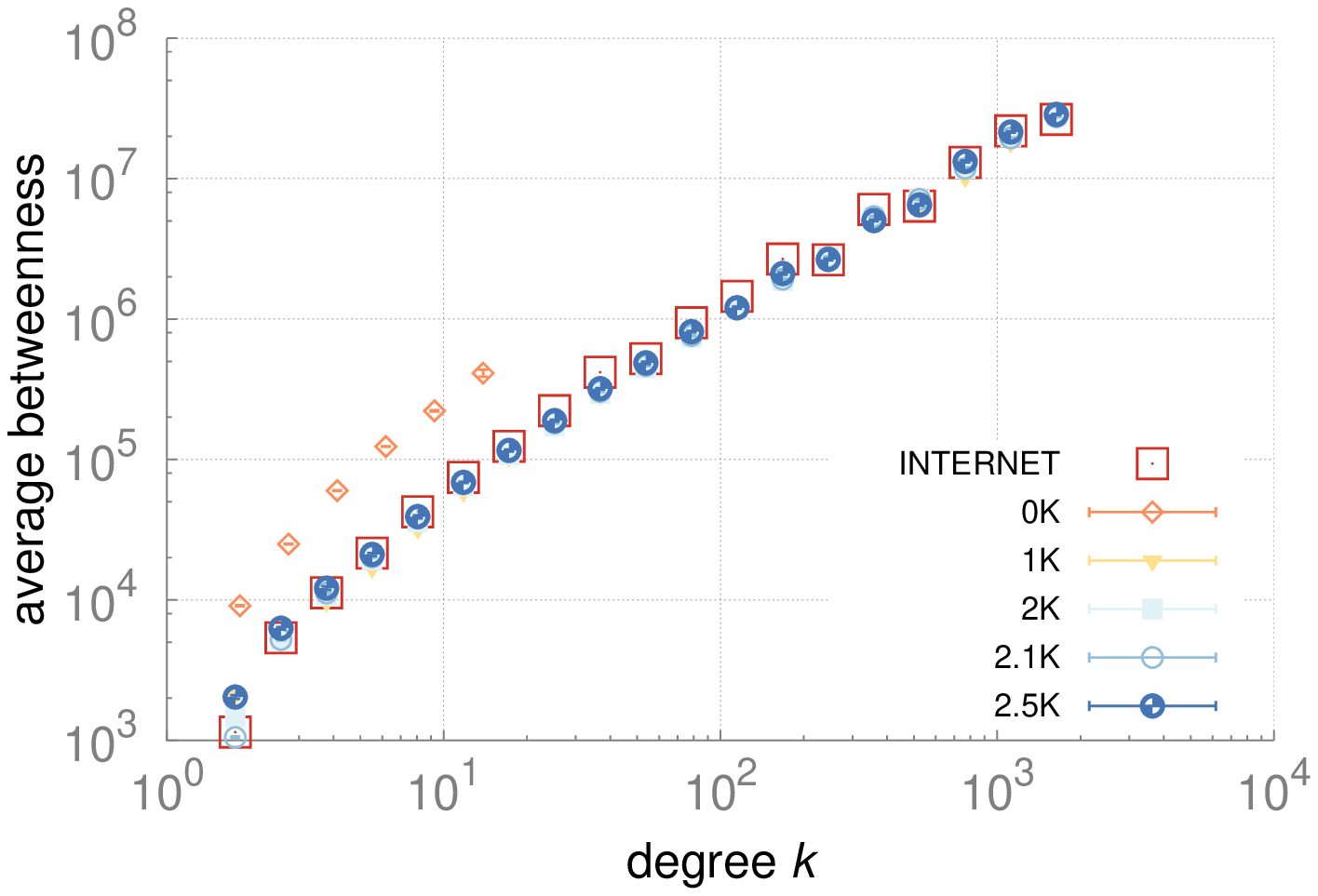}
}\\
\subfloat[PGP]{%
  \includegraphics[width=0.5\textwidth]{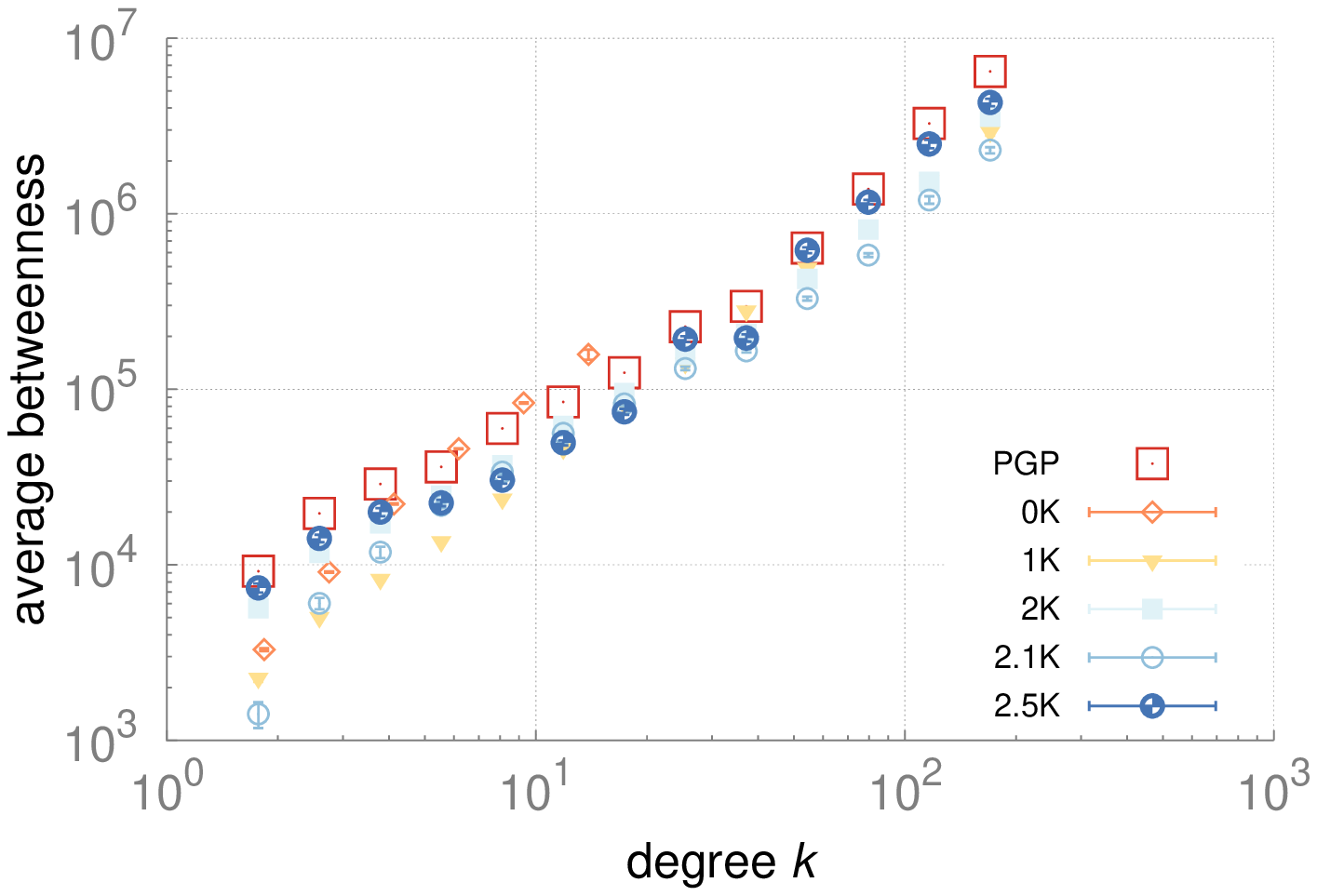}
}
\subfloat[PPI]{%
  \includegraphics[width=0.5\textwidth]{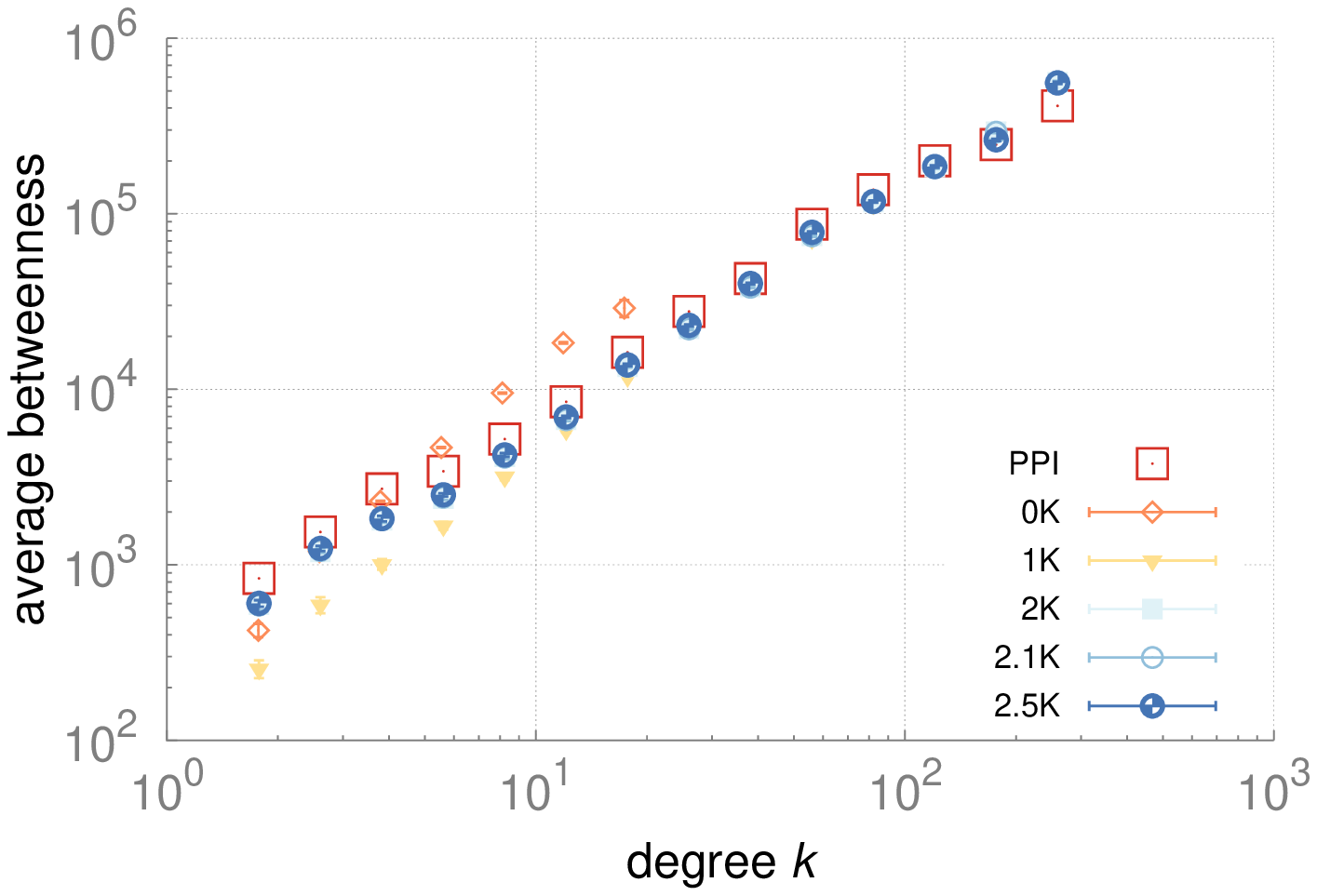}
}\\
\caption{\textbf{Average betweenness of nodes of a given degree in real networks and their $dk$-randomizations.}}
\label{fig:bet}
\end{figure*}

\begin{figure*}
\centering
\subfloat[AIR]{%
  \includegraphics[width=0.5\textwidth]{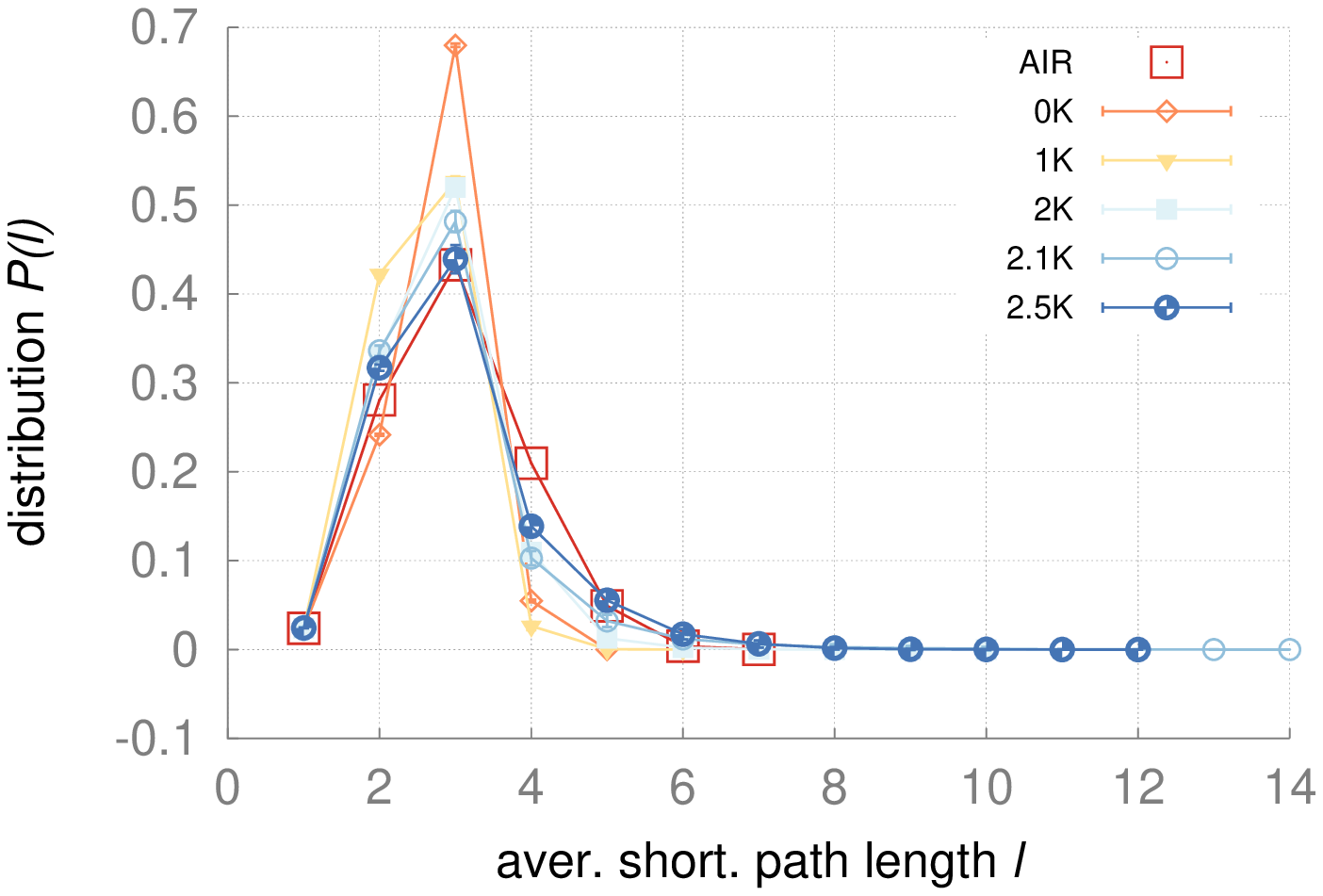}
}
\subfloat[BRAIN]{%
  \includegraphics[width=0.5\textwidth]{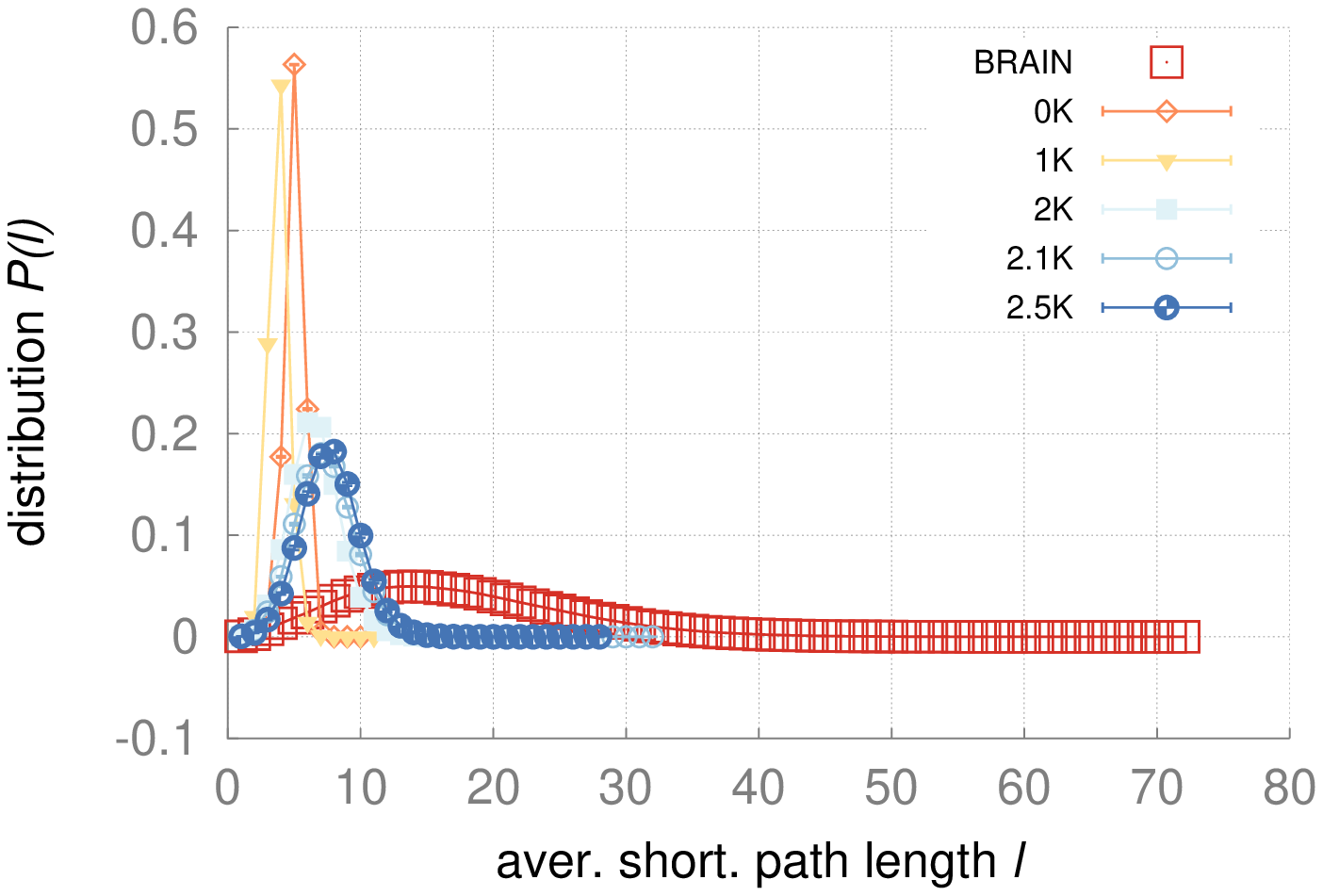}
}\\
\subfloat[WORDS]{%
  \includegraphics[width=0.5\textwidth]{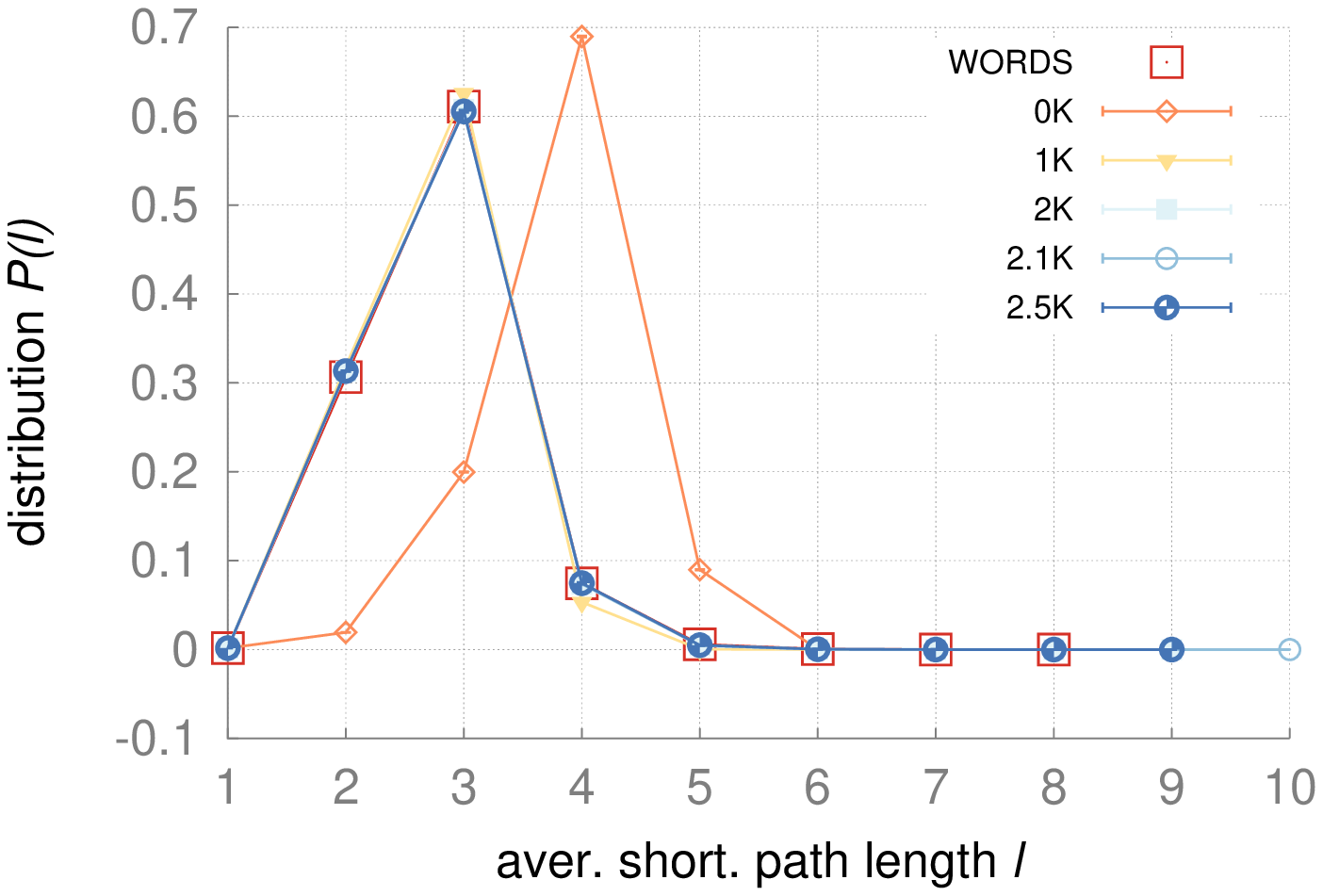}
}
\subfloat[INTERNET]{%
  \includegraphics[width=0.5\textwidth]{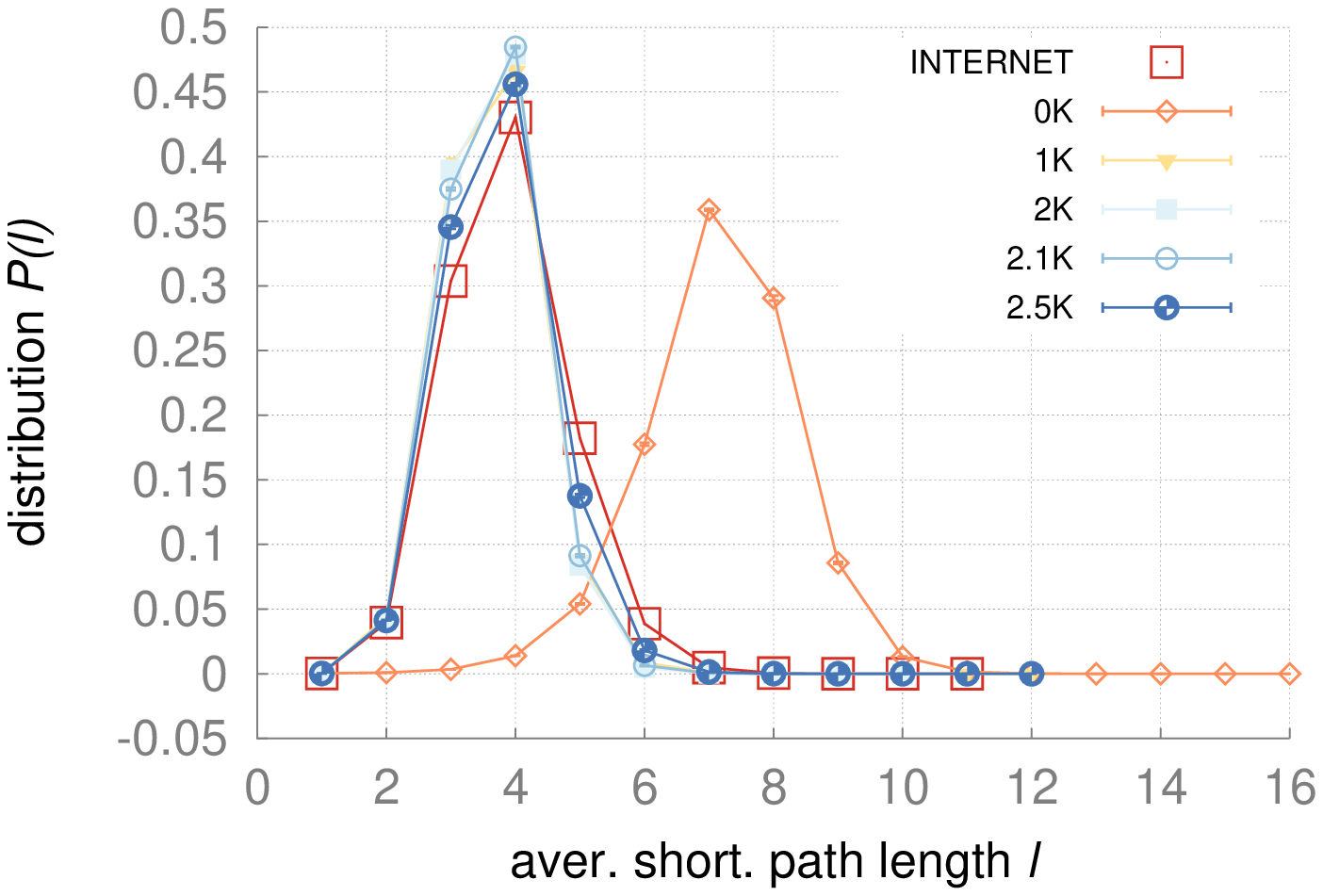}
}\\
\subfloat[PGP]{%
  \includegraphics[width=0.5\textwidth]{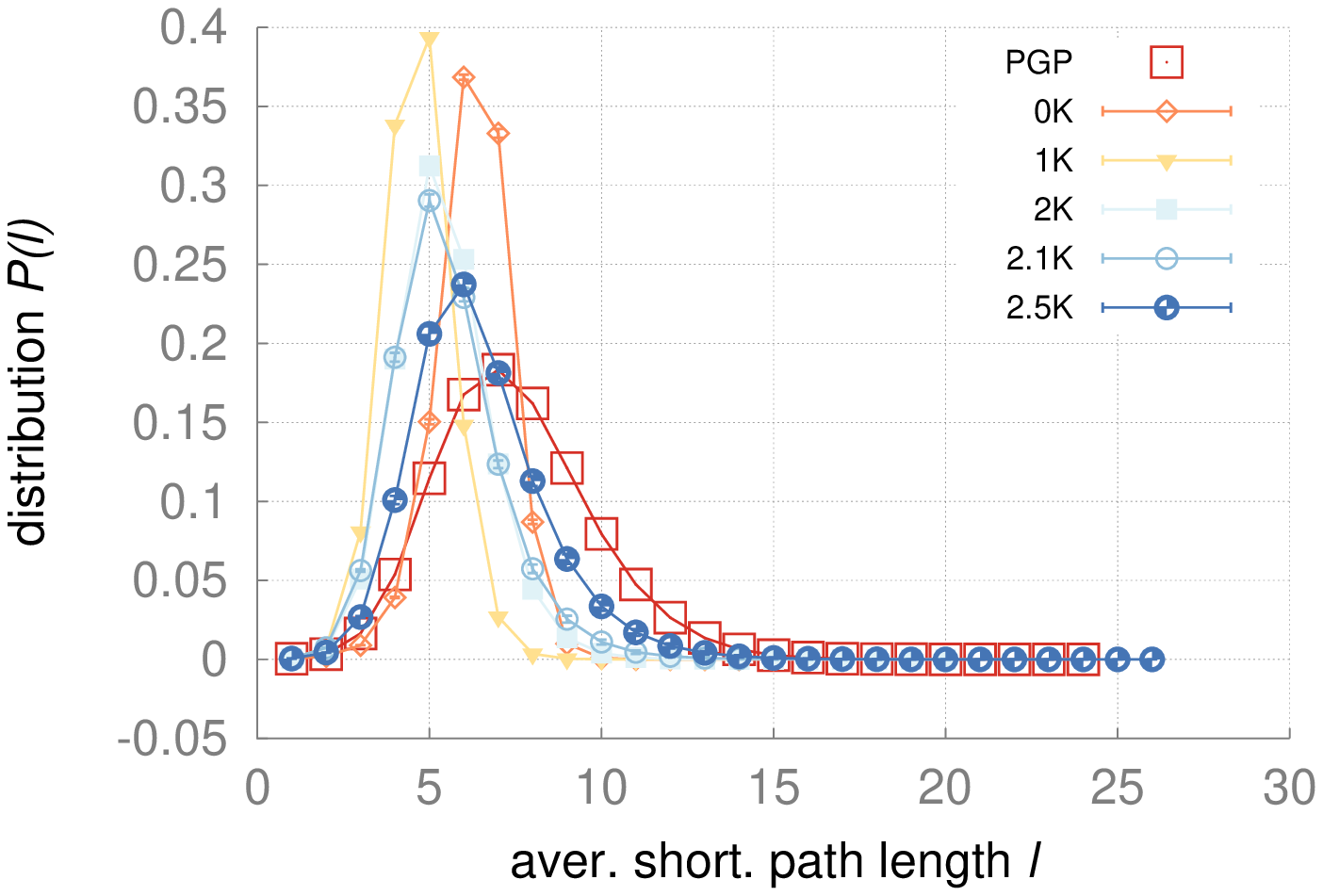}
}
\subfloat[PPI]{%
  \includegraphics[width=0.5\textwidth]{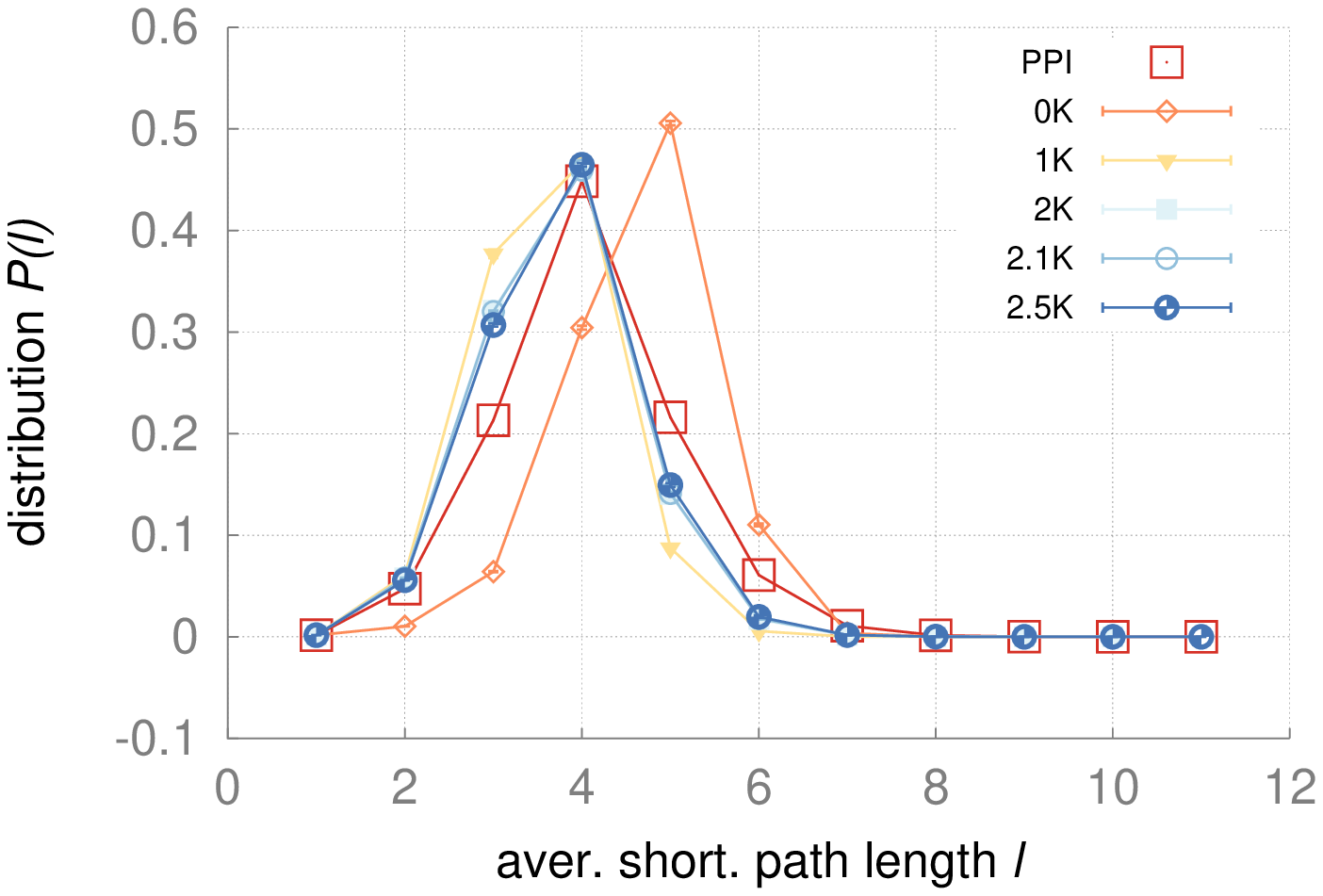}
}\\
\caption{\textbf{Shortest path distance distributions in real networks and their $dk$-randomizations.}}
\label{fig:distance}
\end{figure*}

\begin{figure*}
\centering
\makebox[\textwidth][c]{\includegraphics[width=1.2\textwidth]{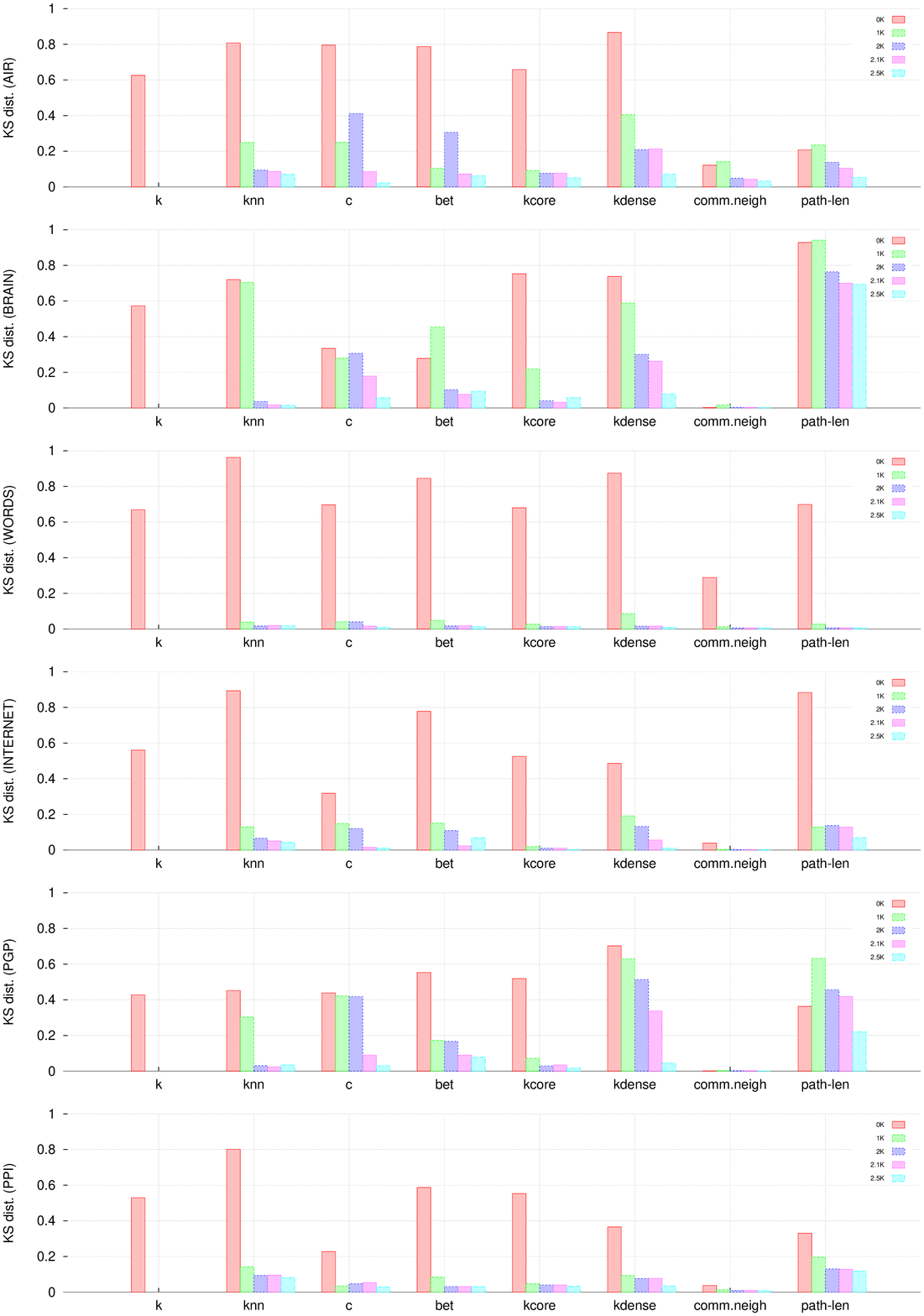}}%
\caption{\textbf{Kolmogorov-Smirnov distances between real networks and their $dk$-randomizations.}}
\label{fig:ks}
\end{figure*}

\newpage

\section*{Supplementary Tables}

\begin{table*}[ht]
\centering
\caption{\textbf{Largest eigenvalues, averaged across different realizations for each $d$, and their standard deviations in parentheses.}}
\label{tab:largeigen}
\makebox[\textwidth][c]{
\begin{tabular}{|c||c||ccccc|}
\hline
  &	Original 		&	$0k$	&	$1k$	& $2k$   & $2.1k$ & $2.5k$ \\
\hline
\hline			
AIR & 48.07 & 12.97(0.08) & 42.41(0.24) & 47.46(0.01) & 47.51(0.02) & 47.82(0.03) \\
BRAIN & 119.66 & 8.91(0.01) & 54.89(0.26) & 113.41(0.02) & 114.09(0.06) & 122.27(0.20) \\
WORDS & 109.44 & 13.06(0.02) & 104.12(0.28) & 108.82(0.03) & 108.80(0.04) & 108.92(0.02) \\
INTERNET & 67.17 & 5.36(0.01) & 56.02(0.33) & 61.15(0.03) & 61.32(0.06) & 65.34(0.10) \\
PGP & 42.44 & 5.77(0.02) & 19.50(0.24) & 34.08(0.03) & 34.40(0.05) & 42.95(0.12) \\
PPI & 38.56 & 8.05(0.05) & 32.47(0.17) & 34.07(0.04) & 34.05(0.04) & 35.56(0.10) \\\hline
\end{tabular}}
\end{table*}

\begin{table*}[ht]
\centering
\caption{\textbf{Spectral gaps, averaged across different realizations for each $d$, and their standard deviations in parentheses.}}
\label{tab:spgap}
\makebox[\textwidth][c]{
\begin{tabular}{|c||c||ccccc|}
\hline
  &	Original 		&	$0K$	&	$1K$	& $2K$   & $2.1K$ & $2.5K$ \\
\hline
\hline			
AIR & 29.34 & 6.04(0.09) & 32.61(0.46) & 37.86(0.25) & 37.21(0.21) & 30.93(0.29) \\
BRAIN & 40.97 & 2.90(0.06) & 35.52(0.31) & 77.53(0.11) & 76.59(0.27) & 42.71(0.35) \\
WORDS & 65.31 & 5.86(0.02) & 65.28(0.51) & 68.53(0.14) & 68.47(0.12) & 68.21(0.15) \\
INTERNET & 17.56 & 0.70(0.05) & 14.94(0.53) & 18.83(0.07) & 18.55(0.11) & 19.53(0.25) \\
PGP & 4.25 & 0.98(0.04) & 5.51(0.31) & 18.01(0.18) & 17.55(0.21) & 4.71(0.19) \\
PPI & 11.69 & 2.25(0.07) & 15.75(0.27) & 16.44(0.19) & 16.28(0.20) & 10.76(0.17) \\
\hline
\end{tabular}}
\end{table*}

\begin{table*}[ht]
\centering
\caption{\textbf{The considered networks, their abbreviations, and the numbers of nodes and links in them.}}
\label{tab:net-abbr}
\vspace{0.5cm}
\begin{tabular}{|l|c|c|c|c}
\hline
Network & Abbr. & $N$ & $M$ \\
\hline \hline
US air transportation network~\cite{CoPaSaVe07}  & AIR           & 500       & 2,980 \\
Brain network~\cite{EgCh05}                                  & BRAIN       & 17,455  & 67,895 \\
English word network~\cite{MiItKaLeSoAyShAl04} & WORDS     & 7,377    & 44,205 \\
Internet AS-level~\cite{MaKrFo06}                         & INTERNET & 20,906  & 42,994 \\
PGP web of trust~\cite{BoPa04}                              & PGP           & 10,680  & 24,316 \\
Protein interaction network~\cite{Vidal2014HI}     & PPI            & 4,099    & 13,355 \\
\hline
\end{tabular}
\end{table*}

\begin{table*}[ht]
\centering
\caption{\textbf{Parameters used for the $dk$-randomization (left) and $2.1k$/$2.5k$-targeting $2k$-preserving (right) rewiring processes ($M$ the number of edges in the real network, $\bar{c}$ average clustering, $\bar{c}(k)$ average clustering of nodes of degree~$k$).}}
\label{tab:targeting-parameters}
\makebox[\textwidth][c]{
\begin{tabular}{|c||cc||cccccc|}
\hline
\multirow{2}{*}{$G_T$} & \multicolumn{2}{c||}{$dk$-randomization} & \multicolumn{6}{c|}{$p$-targeting $dk$-preserving rewiring} \\
\hhline{~--------}	
			&	$R$ 		&	$d$	&	$p_{G_T}$		&	$d$	&	$R$		&	$\beta_0$	&	$\beta_{factor}$	&	$\alpha$ 	\\
\hline
\hline			
AIR			&	$100M$ & $1/2$ 	& $\bar{c}$, $\bar{c}(k)$ 	& 	2 	&	$10M$ 	&	$10^{-2}$	 		&		$1.4$			&	$5 \cdot 10^{-4}$ \\ 
BRAIN		&	$100M$ & $1/2$ 	& $\bar{c}$, $\bar{c}(k)$ 	& 	2 	&	$10M$   & 	$10^{-2}$			&		$1.1$			&	$5 \cdot 10^{-5}$ \\ 
WORDS		&	$100M$ & $1/2$ 	& $\bar{c}$, $\bar{c}(k)$ 	& 	2 	&	$10M$ 	&	$10^{-2}$	 		&		$1.4$			&	$5 \cdot 10^{-4}$ \\ 
 INTERNET	& 	$100M$ & $1/2$ 	& $\bar{c}$, $\bar{c}(k)$ 	& 	2 	&	$10M$ 	&	$10^{-2}$	 		&		$1.4$			&	$5 \cdot 10^{-4}$ \\ 
PGP			& 	$100M$ & $1/2$ 	& $\bar{c}$, $\bar{c}(k)$ 	& 	2 	&	$100M$ 	&	$10^{-2}$	 		&		$1.1$			&	$5 \cdot 10^{-7}$ \\ 
PPI			& 	$100M$ & $1/2$ 	& $\bar{c}$, $\bar{c}(k)$ 	& 	2 	&	$200M$ 	&	$10^{-2}$	 		&		$1.1$			&	$5 \cdot 10^{-7}$ \\ 
\hline
\end{tabular}}
\end{table*}

\begin{table*}
\centering
\caption{\textbf{$dk$-series vs.\ $d$-series}}
\label{tab:dk.vs.d}
\begin{tabular}{|c||c|c|}
\hline $d$ & $dk$-statistics & $d$-statistics
\\ \hline
   \hline $0$ & $\bar{k}$ & -
\\ \hline $1$ & $N(k)$    & $N$
\\ \hline $2$ & $N(k,k')$ & $M$
\\ \hline \multirow{2}{*}{$3$} & $N_\wedge(k,k',k'')$    & $W$
\\                             & $N_\triangle(k,k',k'')$ & $T$
\\ \hline
\end{tabular}
\end{table*}

\clearpage
\newpage

\section*{Supplementary Notes}

\renewcommand{\thesubsection}{Supplementary Note \arabic{subsection}:}
\renewcommand{\thesubsubsection}{\arabic{subsection}.\arabic{subsubsection}}
\subsection{Network properties}
\renewcommand{\thesubsection}{\arabic{subsection}}
\label{sec:metrics}

Here we describe all the network properties measured and discussed in
Supplementary Note~3 
 and, where meaningful, their relations to $dk$-series. 

\subsubsection{Degree distribution}
The distribution $P(k)$ of node degrees $k$, i.e., the $1k$-distribution, is:
\begin{equation}
P(k)=\frac{N(k)}{N},
\end{equation}
where $N(k)$ is the number of nodes of degree $k$ in the network, and $N$ is the
total number of nodes in it, so that $P(k)$ is normalized, $\sum_kP(k)=1$.
The $1k$-distribution fully defines the $0k$-distribution, i.e., the average degree
$\bar{k}$ in the network, by
\begin{equation}\label{eq:bar_k}
\bar{k}=\sum_kkP(k),
\end{equation}
but not {\it vice versa}.

\subsubsection{Average nearest neighbor degree (ANND)}

The average degree $\bar{k}_{nn}(k)$ of nearest neighbors of nodes of degree $k$ is a
commonly used projection of the joint degree distribution (JDD) $P(k,k')$, i.e.,
the $2k$-distribution. The JDD is defined as
\begin{equation}
P(k,k')=\mu(k,k')\frac{N(k,k')}{2M},
\end{equation}
where $N(k,k')=N(k',k)$ is the number of links between nodes of degrees $k$
and $k'$ in the network, $M$ is the total number of links in it, and
\begin{equation}
\mu(k,k')=
\begin{cases}
2&\text{if $k=k'$,}\\
1&\text{otherwise,}
\end{cases}
\end{equation}
so that $P(k,k')$ is normalized, $\sum_{k,k'}P(k,k')=1$. The $2k$-distribution
fully defines the $1k$-distribution by
\begin{equation}
P(k)=\frac{\bar{k}}{k}\sum_{k'}P(k,k'),
\end{equation}
but not {\it vice versa}. The average neighbor degree $\bar{k}_{nn}(k)$ is a
projection of the $2k$-distribution $P(k,k')$ via
\begin{equation}
\bar{k}_{nn}(k)=\frac{\bar{k}}{kP(k)}\sum_{k'}k'P(k,k')=
\frac{\sum_{k'}k'P(k,k')}{\sum_{k'}P(k,k')}.
\end{equation}

\subsubsection{Clustering}

Clustering of node $i$ is the number of triangles $\triangle_i$ it belongs to,
or equivalently the number of links among its neighbors,
divided by the maximum such number, which is $k(k-1)/2$,
where $k$ is $i$'s degree, $\deg(i)=k$.
The average clustering coefficient of the network is
\begin{equation}\label{eq:clustering-avg}
\bar{c} = \frac{1}{N} \sum_{i} \frac{\triangle_i}{k_i(k_i-1)/2}
\end{equation}
Averaging over all nodes of degree $k$, the degree-dependent clustering is
\begin{equation}\label{eq:clustering-definition}
\bar{c}(k) = \frac{2\triangle(k)}{k(k-1)N(k)},\;\text{where}\;
\triangle(k) = \sum_{i:\,\deg(i)=k}\triangle_i.
\end{equation}
The degree-dependent clustering is a commonly used projection
of the $3k$-distribution. (See~\cite{SeBo06a,SeBo06b} for an alternative formalism involving three point correlations.)
The $3k$-distribution is actually two
distributions characterizing the concentrations of the two
non-isomorphic degree-labeled subgraphs of size~$3$, wedges and triangles:\\
\centerline{\includegraphics[width=2in]{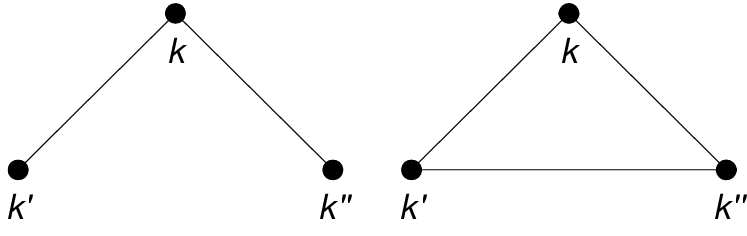}}.
Let $N_\wedge(k',k,k'')=N_\wedge(k'',k,k')$ be the number wedges
involving nodes of degrees $k$, $k'$, and $k''$, where $k$ is the central node degree,
and let $N_\triangle(k,k',k'')$ be the number of triangles consisting of nodes of degrees
$k$, $k'$, and $k''$, where $N_\triangle(k,k',k'')$ is assumed to be symmetric with respect
to all permutations of its arguments. Then the two components of the $3K$-distribution are
\begin{eqnarray}
P_\wedge(k',k,k'') &=& \mu(k',k'')\frac{N_\wedge(k',k,k'')}{2W},\\
P_\triangle(k,k',k'') &=& \nu(k,k',k'')\frac{N_\triangle(k,k',k'')}{6T},
\end{eqnarray}
where $W$ and $T$ are the total numbers of wedges and triangles in the network, and
\begin{equation}
\nu(k,k',k'') =
\begin{cases}
6 & \text{if $k=k'=k''$},\\
1 & \text{if $k \neq k' \neq k''$},\\
2 & \text{otherwise},
\end{cases}
\end{equation}
so that both $P_\wedge(k',k,k'')$ and $P_\triangle(k,k',k'')$ are normalized,
$\sum_{k,k',k''}P_\wedge(k',k,k'')=\sum_{k,k',k''}P_\triangle(k,k',k'')=1$.
The $3k$-distribution defines the $2k$-distribution (but not {\it vice versa}), by
\begin{eqnarray}
P(k,k') &=& \frac{1}{k+k'-2}\sum_{k''}\left\{ \frac{6T}{M}P_\triangle(k,k',k'') \right.\nonumber\\
&+& \left. \frac{W}{M}\left[P_\wedge(k',k,k'')+P_\wedge(k,k',k'')\right] \right\}.
\end{eqnarray}
The normalization of $2k$- and $3k$-distributions implies the following
identity between the numbers of triangles, wedges, edges, nodes, and the
second moment of the degree distribution $\bar{k^2}=\sum_k k^2 P(k)$:
\begin{equation}
2\frac{3T+W+M}{N}=\bar{k^2}.
\end{equation}
The degree-dependent clustering coefficient $\bar{c}(k)$ is the following
projection of the $3k$-distribution
\begin{equation}
\bar{c}(k) = \frac{6T}{N}\frac{\sum_{k',k''}P_\triangle(k,k',k'')}{k(k-1)P(k)}.
\end{equation}

\subsubsection{Subgraph frequencies}

The concentration of subgraphs of size $3$ is exactly fixed only by
the $3k$-distribution, or by the $3$-distribution,
Supplementary Note 4. 
 There are two non-isomorphic connected graphs of size $3$
(triangles and wedges), and their concentrations are defined as
\begin{equation}
\label{concentration}
C_{\wedge}=\frac{\wedge}{N_3},  \qquad  C_{\triangle}=\frac{\triangle}{N_3},
\end{equation}
where $\wedge$ is the number of wedges in the graph, $\triangle$ is the number of
triangles in the graph, and $N_3= \wedge + \triangle$ is the total number of connected subgraphs of size $3$ in the graph.

The concentration of subgraphs of size $4$ is exactly fixed only by the $4k$-distribution, or by the $4$-distribution.
There are six non-isomorphic connected graphs of size $4$,\\
\centerline{\includegraphics[width=4in]{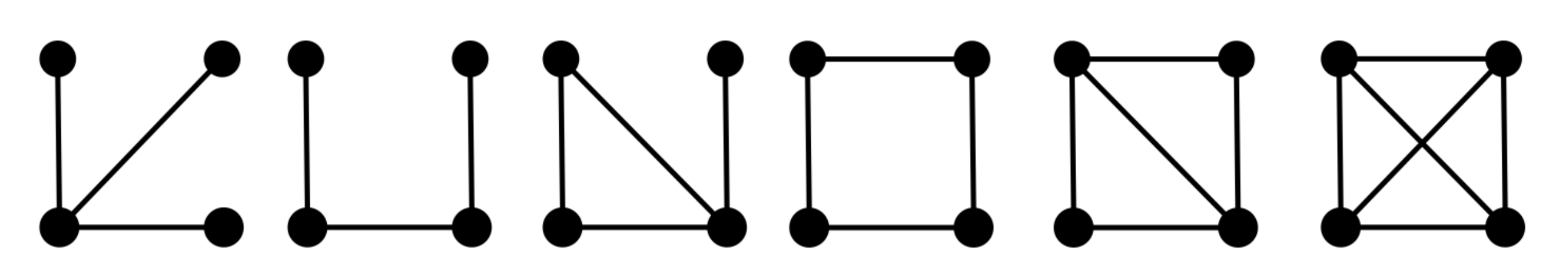}}.
and their concentrations are defined as the number of subgraphs of a particular type divided by the total number of connected subgraphs of size $4$.

In our comparisons of real networks and their $dk$-randomizations in
Supplementary Note 3
 we choose to compare the subgraph concentrations directly, versus computing z-scores, as common in the motif literature. The reasons for this decision is that z-scores are tailored for a fixed null model, while we are considered a series of null models parameterized by $d$ in $dk$-series. There is nothing in the z-score and $dk$-series definitions that could easily provide any estimates of how fast the subgraph frequency means and standard deviations in the z-score definition converge as functions of $d$. Therefore the comparisons of z-scores for different values of $d$ would be meaningless.

\subsubsection{Common neighbors}
The number $m_{ij}$ of common neighbors between two connected nodes $i$ and $j$ is the number of nodes
to which both $i$ and $j$ are connected, or equivalently the multiplicity of edge $(i,j)$:
\begin{equation}
m_{ij} = \sum_{l}A_{il}A_{lj}A_{ij},
\end{equation}
where $\{A_{ij}\}$ is the adjacency matrix of the graph.
The distribution $P(m)$ of the number of common neighbors $m$ is then
\begin{equation}
P(m) = {\sum_{i<j}\delta_{m_{ij},m} \over N(N-1)/2},
\end{equation}
where $\delta$ is the Kronecker delta.
The common neighbor distribution is thus the probability that two connected nodes in the graph have $m$ common neighbors. This property is exactly fixed only by the $3k$-distribution.

\subsubsection{$k$-coreness and $k$-denseness}

The $k$-core decomposition~\cite{AlvarezKcore2008} of a graph
is a set of nested subgraphs induced by nodes of the
same $k$-coreness. A node has $k$-coreness equal to $k$ if it belongs to a maximal
connected subgraph of the original graph, in which all nodes have degree at least
$k$, i.e., in which each node is connected to at least $k$ other nodes in the subgraph.

Similarly, the $k$-dense decomposition~\cite{SaitoKdensity2006} of a graph is
a set of nested subgraphs induced by edges of the same
$k$-denseness. An edge has $k$-denseness equal to $k$ if it belongs to a maximal
connected subgraph of the original graph, in which all edges have
multiplicity~\cite{SeBo06a,SeBo06b,Zlatic2012Multiplicities}
at least $k-1$, i.e., in which each pair of connected
nodes has at least $k-1$ common neighbors in the subgraph.

Both the $k$-core and $k$-dense decompositions rely on the analysis of
local properties of nodes and edges. However, due to the recursive
nature of these decompositions, the $dk$-distributions  with $d=0,1,2,2.1,2.5$
do not exactly fix either the $k$-core or $k$-dense distributions.

\subsubsection{Betweenness}

Betweenness $b(i)$ of node $i$ is a measure of how ``important'' $i$ is in terms of
the number of shortest paths passing through it. Formally, if $\sigma_{st}(i)$ is the
number of shortest paths between nodes $s \neq i$ and $t \neq i$ that pass through
$i$, and $\sigma_{st}$ is the total number of shortest paths between the two nodes
 $s \neq t$, then betweenness of $i$ is
\begin{equation}
b(i) = \sum_{s,t}\frac{\sigma_{s,t}(i)}{\sigma_{s,t}}.
\end{equation}
Averaging over all nodes of degree $k$, degree-dependent betweenness $\bar{b}(k)$ is
\begin{equation}
\bar{b}(k) = \sum_{i:\,\deg(i)=k}\frac{b(i)}{N(k)}.
\end{equation}

\subsubsection{Shortest path distance}

The distance distribution is the distribution of hop-lengths of shortest path
between nodes in a network. Formally, if $N(h)$ is the number of node pairs located
at hop distance $h$ from each other, then the distance distribution $P(h)$ is
\begin{equation}
P(h) = \frac{N(h)}{N(N-1)/2},
\end{equation}
where $N(N-1)/2$ is the total number of nodes pairs in the network.
The average distance is:
\begin{equation}\label{eq:bar_h}
\bar{h} = \sum_hP(h).
\end{equation}

Finally, the network diameter, i.e., the maximum hop distance between nodes in the network, is
\begin{equation}
d = \max(h).
\end{equation}

\subsubsection{Spectral properties}

The adjacency matrix of graph $A$ gives the full information on the structure of the graph. The largest eigenvalue of $A$ and the spectral gap, which is defined as the difference between the largest and second largest eigenvalue $A$, play important roles in the dynamic processes on networks. For instance, the largest eigenvalue of the adjacency matrix is related to the speed of the spreading processes on the network \cite{wang2003b,dagostino2012}, while the gap determines the speed of convergence of the random walk to its steady state \cite{van2010}.\\
The Laplacian matrix describes the diffusion of a random walker on the network and is defined as $L=D-A$, where $D$ is the diagonal matrix of degrees $D_{ij}=\delta_{ij}k_{i}$, $\delta_{ij}$ is Kronecker delta and $k_{i}$ is the degree of node $i$. The smallest eigenvalue of the Laplacian matrix is associated to stationary distribution of random walker and it is always equal to zero, while the smallest non-zero eigenvalue, Fiedler value, defines the time scale of the slowest mode of the diffusion \cite{van2010}.

\renewcommand{\thesubsection}{Supplementary Note \arabic{subsection}:}
\renewcommand{\thesubsubsection}{\arabic{subsection}.\arabic{subsubsection}}
\subsection{Considered networks}
\label{sec:networks}
\renewcommand{\thesubsection}{\arabic{subsection}}

We apply the $dk$-series analysis to the following six social, biological,
language, communication, and transportation networks, Table~\ref{tab:net-abbr}:
\begin{itemize}
\item AIR. The US air transportation network~\cite{CoPaSaVe07}.
The nodes are airports, and there is a link between two airports if there is
a direct flight between them.
\item BRAIN. The largest connected component of an fMRI map of the human
brain~\cite{EgCh05}. The nodes are voxels (small areas of a resting brain of
approximately $36$mm$^3$ volume each), and two voxels are connected if
the correlation coefficient of the fMRI activity of the voxels exceed $0.7$.
\item WORDS. The largest connected component of the network of adjacent words
in Charles Darwin's ``The Origin of Species''~\cite{MiItKaLeSoAyShAl04}. The nodes
are words, and two words are connected if they are adjacent in the text.
\item INTERNET. The topology of the Internet at the level of Autonomous
Systems~(ASes)~\cite{MaKrFo06}. The nodes are ASs (organizations owing
parts of the Internet infrastructure), and there is a link between two ASs if
 they have a business relationship to exchange Internet traffic.
\item PGP (\underline{considered in the main text}). The largest strongly
connected component of the technosocial web of trust relationships among
people extracted from the Pretty Good Privacy (PGP) data~\cite{BoPa04}. The
nodes are PGP certificates of users, and there is a link between two certificates
if their users mutually trust each other's certificate/user associations.
\item PPI. The largest connected component of the human protein interaction
network~\cite{Vidal2014HI}. The nodes are proteins, and there is a link between
two proteins if they interact.
\end{itemize}

Table~\ref{tab:targeting-parameters} reports the parameters used for each
network in the $dk$-randomization and $p$-targeting $dk$-preserving rewiring processes.

\renewcommand{\thesubsection}{Supplementary Note \arabic{subsection}:}
\renewcommand{\thesubsubsection}{\arabic{subsection}.\arabic{subsubsection}}
\subsection{Results}
\label{sec:results}
\renewcommand{\thesubsection}{\arabic{subsection}}

\paragraph{Degree distribution.}
We observe in Fig.~\ref{fig:k} that while $0k$-randomizations are way off, the $dk$-random graphs with $d\geq1$ reproduce the degree distributions in the real networks exactly, which is by definition: the $1k$-distribution is the degree distribution, and $dk$-random graphs with $d\geq1$ have exactly the same degree distributions as the real networks.

\paragraph{Average nearest neighbor degree (ANND).}
We observe in Fig.~\ref{fig:knn} that while $0k$-randomizations are way off,
the $1k$-random graphs tend to be closer to the real networks in terms of ANND, whereas the
$dk$-random graphs with $d\geq2$ have exactly the same average neighbor degrees as the
real networks, which is again by definition: the $dk$-random graphs with $d\geq2$ have exactly the same JDD
$P(k,k')$ as the real networks. In the WORDS, INTERNET, and PPI
cases, the ANNDs $\bar{k}_{nn}(k)$ even in the $1k$-random graphs do not noticeably differ from the ANNDs in the real networks.

\paragraph{Clustering.}
We observe in Fig.~\ref{fig:cc}
that degree-dependent average clustering in the $2.5k$-random graphs matches the one in the real networks,
which is again by definition. For $d<2.5$, degree-dependent
clustering differs sensibly in many cases. However, degree-dependent clustering in the AIR network does not exhibit
noticeable differences with its $2.1k$-randomizations, while in the WORDS case, even the $1k$-random graphs reproduce
degree-depended clustering nearly exactly.

\paragraph{Subgraph frequencies.}
We observe in Fig.~\ref{fig:motifssup} that the $2k$-random graphs
reproduce the subgraphs frequencies in most cases, but
the BRAIN and PGP require $d=2.5$ to reproduce these frequencies.

\paragraph{Common neighbors.}
We observe in Fig.~\ref{fig:cn} that the $1k$-random graphs reproduce the common neighbor distributions
in all the cases except the BRAIN, which requires $d=2$, and
PGP, which requires $d=2.5$.

\paragraph{$k$-coreness and $k$-denseness.}
We observe in Fig.~\ref{fig:kcore}  that the $2k$-random graphs reproduce the $k$-coreness distributions
in all the networks except the PGP and BRAIN that require $d=2.5$.
We observe in Fig.~\ref{fig:kdense} that the $2.5k$-random graphs reproduce the $k$-denseness distributions
in all the networks. The $k$-denseness distributions in the AIR and WORDS networks are reproduced even by their $2k$-random graphs.

\paragraph{Betweenness.}
We observe in Fig.~\ref{fig:bet} that betweenness in the BRAIN network cannot be approximated even
by its $2.5k$-random graphs. The INTERNET lies at the other extreme: even the $1k$-random graphs
reproduce its betweenness. The PGP network requires all the
constraints imposed by the $2.5k$-distribution, while betweenness in all the other networks
is similar to betweenness in their $2k$-random graphs.

\paragraph{Shortest path distance.}
We observe in Fig.~\ref{fig:distance} that the distance distributions in the INTERNET and WORDS networks are
correctly reproduced by their $1k$-random graphs. Even $d=2.5$ is not enough for the BRAIN, while the same value
of $d=2.5$ suffices for all the networks.

\paragraph{Spectral properties.}

We observe in Table~\ref{tab:largeigen} that the largest eigenvalue of the adjacency matrix is closely, although not exactly, reproduced by $d=2.5k$-random graphs for all six networks. Furthermore, we observe that the largest eigenvalues for $2k$-random graphs of AIR and WORDS networks are very close to the eigenvalues of the original networks.\\
The values of the spectral gaps for $2.5k$-random graphs shown in Table~\ref{tab:spgap} are relatively close to the values observed for the original networks, with relative difference for AIR, BRAIN and WORDS networks around $5\%$. The large values of the spectral gaps for $2k$ and $2.1k$-random graphs indicate that they are more robust, in the sense of being better connected and interlinked, compared to the original networks.

\paragraph{Kolmogorov-Smirnov distance.}
In Fig.~\ref{fig:ks} we quantify the convergence of $dk$-series
in terms of Kolmogorov-Smirnov (KS) distances between the distributions of per-node
values of a given property in the real networks and the same distributions in their
$dk$-random graphs. We report the KS distances for the following properties:
\begin{description}
\item[k] degree, cf.~Fig.~\ref{fig:k};
\item[knn] ANND, cf.~Fig.~\ref{fig:knn};
\item[c] clustering, cf.~Fig.~\ref{fig:cc};
\item[comm.neigh] common neighbors, cf.~Fig.~\ref{fig:cn};
\item[kcore] $k$-coreness, cf.~Fig.~\ref{fig:kcore};
\item[kdense] $k$-density, cf.~Fig.~\ref{fig:kdense};
\item[bet] betweenness, cf.~Fig.~\ref{fig:bet};
\item[path-len] shortest path distance, cf.~Fig.~\ref{fig:distance}.
\end{description}
The Kolmogorov-Smirnov distance between two cumulative distribution functions (CDFs) $F_1(x)$ and $F_2(x)$ is
\begin{equation}
D = \sup_{x} |F_{1}(x)-F_{2}(x)|.
\end{equation}
In our case, $F_1(x)$ is the per-node CDF of a given property in a real network,
and $F_2(x)$ is the per-node CDF for the same property computed across all different $dk$-random graph
realizations for the network with a given $d$.
We note that the KS distances provides more detailed statistics than the $dk$-distributions, because
the latter do not differentiate between nodes of the same degree, while the former do. For example,
even if the $2k$-distributions and consequently ANNDs $\bar{k}_{nn}(k)$ in two different networks are exactly the same,
the distributions of average degrees $\bar{k}_{i,nn}$ of neighbors of each individual node $i$, $i=1,\ldots,N$,
are in general different, so that the KS distance between the two per-node ANND CDFs is in general greater than zero.

\setcounter{subsection}{0}
\renewcommand{\thesubsection}{\arabic{subsection}}
\renewcommand{\thesubsubsection}{\arabic{subsection}.\arabic{subsubsection}}

\section*{Supplementary Discussion}
\label{sec:series}

We compare $dk$-series with the series based on subgraph frequencies, and show that
the latter cannot form a systematic basis for topology analysis.

The difference between $dk$-series and subgraph-based-series, which we can call
$d$-series, is that the former is the series of distributions of $d$-sized
subgraphs labeled with node degrees in a given network, while the $d$-series
is the distributions of such subgraphs in which this degree information is
ignored. This difference explains the mnemonic names for these two series:
`$d$' in `$dk$' refers to the subgraph size, while `$k$' signifies that they
are labeled by node degrees---`$k$' is a standard notation for node degrees.

This difference between the $dk$-series and $d$-series is crucial.
The $dk$-series are inclusive, in the sense that the $(d+1)k$-distribution
contains the full information about the $dk$-distribution, plus some additional
information, which is not true for $d$-series.

To see this, let us consider the first few elements of both series in
Table~\ref{tab:dk.vs.d}. In Supplementary Note 1 
 we show explicitly
 how the $(d+1)k$-distributions define the $dk$-distribution for $d=0,1,2$.
The key observation is that the $d$-series does not have this property.
The $0$'th element of $d$-series is undefined.
For $d=1$ we have the number
of subgraphs of size $1$, which is just $N$, the number of nodes in the network.
For $d=2$, the corresponding statistics is $M$, the number of links, subgraphs
of size $2$. Clearly, $M$ and $N$ are independent statistics, and the former does
not define the latter.
For $d=3$, the statistics are $W$ and $T$, the total number of wedges and triangles,
subgraphs of size $3$, in the network. These do not define the previous element $M$
either. Indeed, consider the following two networks of size $N$---the chain and the
star:\\
\centerline{\includegraphics[width=3in]{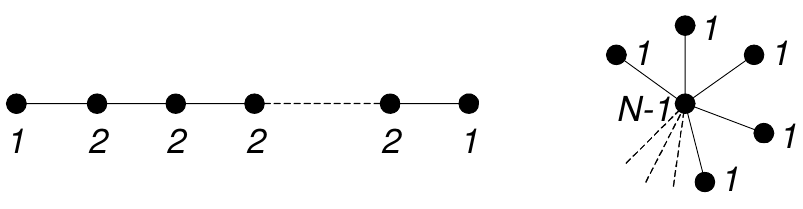}}
There are no triangles in either network, $T=0$.
In the chain network, the number of wedges is $W=N-2$, and in the star
$W=(N-1)(N-2)/2$. We see that even though $W$ ($d=3$) scales completely differently
with $N$ in the two networks, the number of edges $M=N-1$ ($d=2$) is the same.

In summary, $d$-series is not inclusive. For each $d$, the corresponding element
of the series reflects a differen kind of statistical information about the network
topology, unrelated or only loosely related to the information conveyed by the preceding
elements. At the same time, similar to $dk$-series, the $d$-series is also converging
since at $d=N$ it specifies the whole network topology. However, this convergence is
much slower that in the $dk$-series case. In the two networks considered above,
for example, neither $W=N-2,\; T=0$ nor $W=(N-1)(N-2)/2,\; T=0$, fix the network
topology as there are many non-isomorphic graphs with the same $(W,T)$ counts,
whereas the $3k$-distributions $N_\wedge(1,2,2)=2,\; N_\wedge(2,2,2)=N-4$ and
 $N_\wedge(1,N-1,1)=(N-1)(N-2)/2$ define the chain and star topologies exactly.

The node degrees thus provide necessary information about subgraph locations in the
original network, which significantly speeds up convergence as a function of $d$,
and more importantly makes the $dk$-series basis inclusive and systematic.

\section*{Supplementary Methods}
\label{sec:algs}

The methods that we use to sample $dk$-random graphs for a given graph representing a real network are based on
two different rewiring processes: $dk$-randomizing rewiring ($d=0,1,2$) and $p$-targeting
$dk$-preserving rewiring ($p=2.1k,2.5k$).

The first method (\textbf{$dk$-randomization}) consists of swapping random pairs of edges in the original network preserving its
$dk$-distribution, Algorithm~\ref{alg:randomization}. The following three input parameters are required: $G_T$ the original graph, $R$ the number of rewirings to apply, and $d$ index that indicates the $dk$-distribution to preserve.
The random edge selection function on line 4 
 and the rewiring function on line 5 
depend on $d$ as follows:
\begin{itemize}
\item if $d=0$, random edge $(i,j)$ and non-edge $(a,b)$ (disconnected nodes $a$ and~$b$) are selected, and the rewiring
  consists of removing edge $(i,j)$ and adding edge $(a,b)$.
\item if $d=1$, two random edges $(i,j)$ and $(a,b)$ are selected and discarded if either edge $(i,b)$ or edge $(j,a)$ exists; if neither edge $(i,b)$ nor edge $(j,a)$ exists,
  the rewiring consists of removing edges $(i,j)$ and $(a,b)$, and adding edges $(i,b)$ and $(j,a)$.
\item if $d=2$, two random edges $(i,j)$ and $(a,b)$ such that degrees $k_i = k_a$ are selected and discarded if either edge $(i,b)$ or edge $(j,a)$ exists; if neither edge $(i,b)$ nor edge $(j,a)$ exists,
  the associated rewiring consists of removing edges $(i,j)$ and $(a,b)$ and adding edges $(i,b)$ and $(j,a)$.
\end{itemize}

\begin{minipage}[t]{\textwidth}
\null
\begin{algorithm}[H]
 \caption{$dk$-randomization process.}
 \label{alg:randomization}
\end{algorithm}
\centerline{\includegraphics[width=\textwidth]{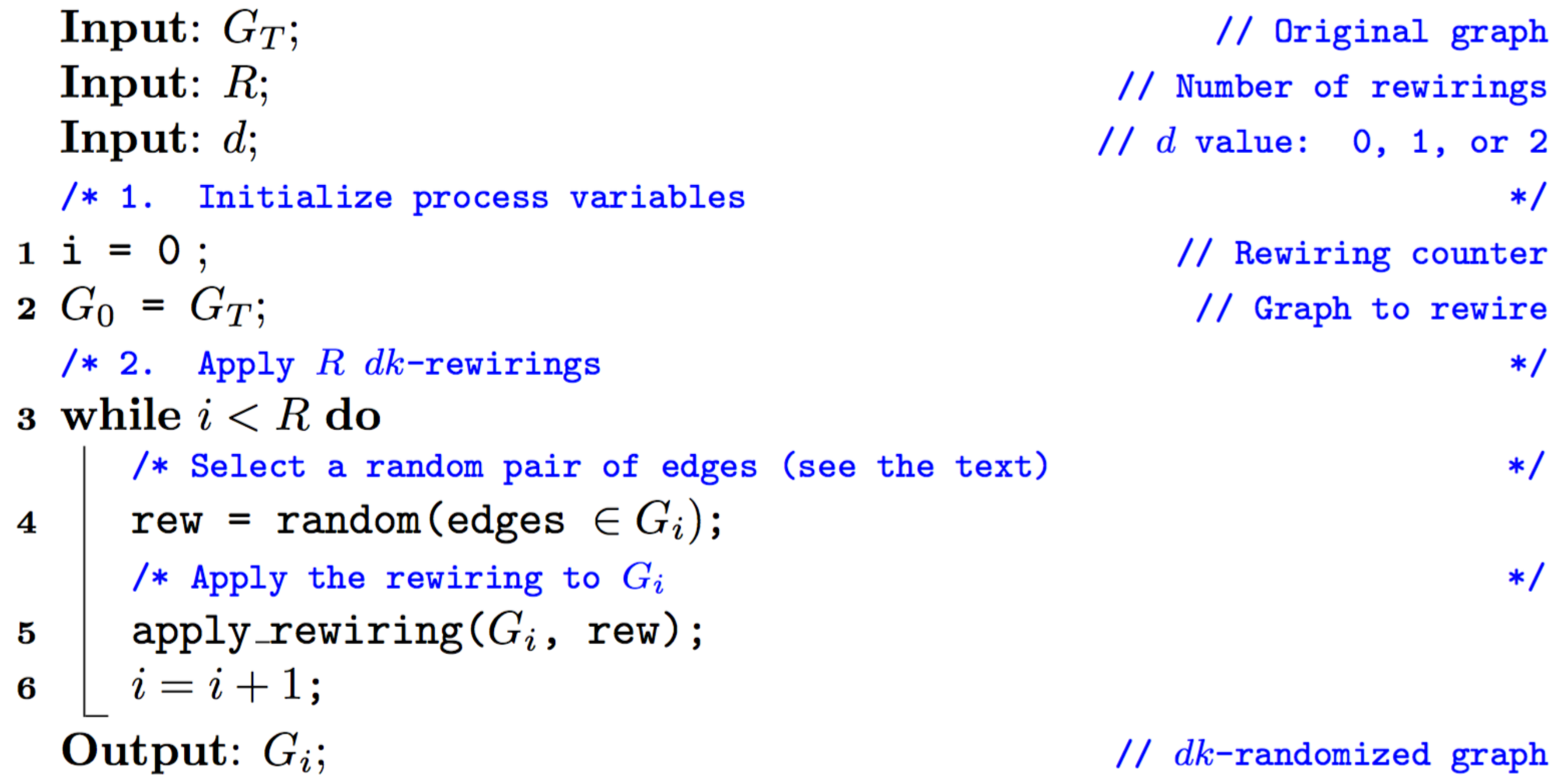}}
\end{minipage}

\vspace{.2in}

The second method of (\textbf{$p$-targeting $dk$-preserving rewiring}) is based on
simulated annealing, and consists of two phases:
randomization and targeting rewiring, Algorithm~\ref{alg:targeting}. The following input parameters are required:
$G_T$ the original graph, ${p}_{G_T}$ the property to target, $R$ the number of
$dk$-rewirings to apply at each value of temperature, $\beta_{0}$ the initial inverse temperature,  $\beta_{factor}$ the rate of temperature decrease, and $\alpha$ the acceptance threshold. In the first phase the original graph is $2k$-randomized by Algorithm~\ref{alg:randomization}.
In the second phase, the obtained $2k$-random graph is $2k$-rewired, but each rewiring is accepted with
probability $\min[\exp(-\beta H),1]$ which depends on current values of energy $H$ and temperature $1/\beta$. Energy
is defined as the distance between the values of property $p$ in the original and current rewired graphs.
Temperature is high initially, but each round of $R$ rewirings (line~9
), it decreases by factor $\beta_{factor}$,
thus decreasing the probability of accepting a rewiring that increases energy.
This second phase terminates when either energy is zero, meaning that the value of $p$-property in the rewired graph ${p}_{G_i}$ is equal to its value in the original graph ${p}_{G_T}$, or when the percentage of accepted rewirings during the last round falls below a user-specified threshold~$\alpha$.
Function \texttt{compute\_property($G$)} appearing on
lines~3 
and~12 
returns average clustering~$\bar{c}$ or average degree-dependent clustering $\bar{c}(k)$ of $G$
depending on whether $d=2.1$ or $d=2.5$, respectively.
Energy function \texttt{distance(${p}_{G_i},{p}_{G_T}$)} appearing on
lines~4
 and~13 
 depends on~$d$ as follows:
\begin{itemize}
\item if $d=2.1$, $\texttt{distance(${p}_{G_i},{p}_{G_T}$)}=\vert\overline{c}_{G_i}-\overline{c}_{G_T}\vert$,
\item if $d=2.5$, $\texttt{distance(${p}_{G_i},{p}_{G_T}$)}=\sum_k\vert\overline{c}_{G_i}(k)-\overline{c}_{G_T}(k)\vert$.
\end{itemize}

\paragraph{Code availability}
We release the software package that implements the $dk$-randomization
algorithms described above. The code is freely available at
\url{http://polcolomer.github.io/RandNetGen/}\cite{polcode}. 

\newpage

\begin{minipage}[t]{\textwidth}
\null
\begin{algorithm}[H]
 \caption{$p$-targeting $dk$-preserving rewiring process.}
 \label{alg:targeting}
\end{algorithm}
\centerline{\includegraphics[width=\textwidth]{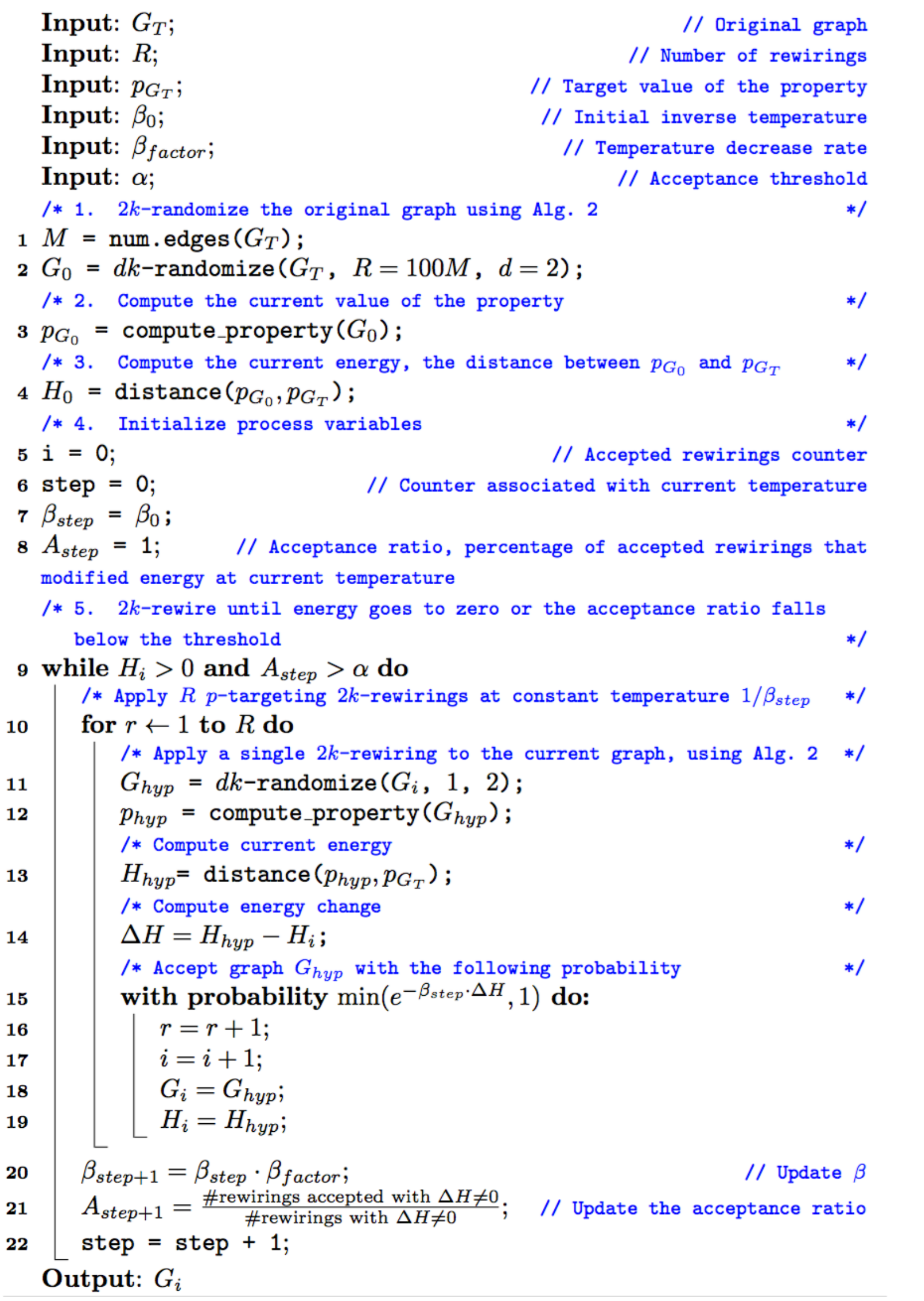}}
\end{minipage}

\clearpage
\newpage

\bibliographystyle{unsrturl}
\bibliography{./bib}

\begin{thebibliography}{10}

\bibitem{Newman10-book}
M~E~J Newman.
\newblock {\em {Networks: An Introduction}}.
\newblock Oxford University Press, Oxford, 2010.

\bibitem{BarratVespignani-book08}
Alain Barrat, Marc Barth\'{e}lemy, and Alessandro Vespignani.
\newblock {\em {Dynamical Processes on Complex Networks}}.
\newblock Cambridge University Press, Cambridge, 2008.

\bibitem{newman03c-review}
M~E~J Newman.
\newblock {The Structure and Function of Complex Networks}.
\newblock {\em SIAM Rev}, 45(2):167--256, 2003.
\newblock \href {http://dx.doi.org/10.1137/S003614450342480}
  {\path{doi:10.1137/S003614450342480}}.

\bibitem{MotterOptimization2007}
Adilson~E Motter and Zolt\'{a}n Toroczkai.
\newblock {Introduction: optimization in networks}.
\newblock {\em Chaos}, 17(2):026101, July 2007.
\newblock \href {http://dx.doi.org/10.1063/1.2751266}
  {\path{doi:10.1063/1.2751266}}.

\bibitem{BurdaMotifs2011}
Z~Burda, A~Krzywicki, O~C Martin, and M~Zagorski.
\newblock {Motifs emerge from function in model gene regulatory networks.}
\newblock {\em Proc Natl Acad Sci}, 108(42):17263--8, October 2011.
\newblock \href {http://dx.doi.org/10.1073/pnas.1109435108}
  {\path{doi:10.1073/pnas.1109435108}}.

\bibitem{VazquezTopologicalRelationship2004}
A~V\'{a}zquez, R~Dobrin, D~Sergi, J-P Eckmann, Z~N Oltvai, and A-L
  Barab\'{a}si.
\newblock {The topological relationship between the large-scale attributes and
  local interaction patterns of complex networks.}
\newblock {\em Proc Natl Acad Sci}, 101(52):17940--5, December 2004.
\newblock \href {http://dx.doi.org/10.1073/pnas.0406024101}
  {\path{doi:10.1073/pnas.0406024101}}.

\bibitem{GuimeraClasses2007}
Roger Guimer\`{a}, Marta Sales-Pardo, and Luis~A. Amaral.
\newblock {Classes of complex networks defined by role-to-role connectivity
  profiles.}
\newblock {\em Nat Phys}, 3(1):63--69, January 2007.
\newblock \href {http://dx.doi.org/10.1038/nphys489}
  {\path{doi:10.1038/nphys489}}.

\bibitem{TakemotoStructure2007}
K~Takemoto, C~Oosawa, and T~Akutsu.
\newblock {Structure of n-clique networks embedded in a complex network}.
\newblock {\em Physica A}, 380:665--672, July 2007.
\newblock \href {http://dx.doi.org/10.1016/j.physa.2007.02.042}
  {\path{doi:10.1016/j.physa.2007.02.042}}.

\bibitem{FosterClustering2011}
David~V. Foster, Jacob~G. Foster, Peter Grassberger, and Maya Paczuski.
\newblock {Clustering drives assortativity and community structure in ensembles
  of networks}.
\newblock {\em Phys Rev E}, 84(6):066117, December 2011.
\newblock \href {http://dx.doi.org/10.1103/PhysRevE.84.066117}
  {\path{doi:10.1103/PhysRevE.84.066117}}.

\bibitem{ColomerClustering2013}
Pol Colomer-de Sim\'{o}n, M.~\'{A}ngeles Serrano, Mariano~G Beir\'{o},
  Jos\'{e}~Ignacio Alvarez-Hamelin, and Mari\'{a}n Bogu\~{n}\'{a}.
\newblock {Deciphering the global organization of clustering in real complex
  networks.}
\newblock {\em Sci Rep}, 3:2517, January 2013.
\newblock \href {http://dx.doi.org/10.1038/srep02517}
  {\path{doi:10.1038/srep02517}}.

\bibitem{MiSh02-motifs}
R~Milo, S~Shen-Orr, S~Itzkovitz, N~Kashtan, D~Chklovskii, and U~Alon.
\newblock {Network motifs: simple building blocks of complex networks}.
\newblock {\em Science}, 298(5594):824--7, October 2002.
\newblock \href {http://dx.doi.org/10.1126/science.298.5594.824}
  {\path{doi:10.1126/science.298.5594.824}}.

\bibitem{Markov2013}
N.T Markov, M.~Ercsey-Ravasz, D.C. {Van Essen}, K.~Knoblauch, Z.~Toroczkai, and
  H.~Kennedy.
\newblock {Cortical high-density counterstream architectures}.
\newblock {\em Science}, 342(6158):1238406, November 2013.
\newblock \href {http://dx.doi.org/10.1126/science.1238406}
  {\path{doi:10.1126/science.1238406}}.

\bibitem{AmaralLies2006}
Luis~A. Amaral and Roger Guimera.
\newblock {Complex networks: Lies, damned lies and statistics}.
\newblock {\em Nat Phys}, 2(2):75--76, February 2006.
\newblock \href {http://dx.doi.org/10.1038/nphys228}
  {\path{doi:10.1038/nphys228}}.

\bibitem{CoFlSeVe06}
Vittoria Colizza, Alessandro Flammini, M.~\'{A}ngeles Serrano, and Alessandro
  Vespignani.
\newblock {Detecting rich-club ordering in complex networks}.
\newblock {\em Nat Phys}, 2(2):110--115, January 2006.
\newblock \href {http://dx.doi.org/10.1038/nphys209}
  {\path{doi:10.1038/nphys209}}.

\bibitem{Zscore2015}
J.~Trevino~III, A.~Nyberg, C.I. Del~Genio, and K.E. Bassler.
\newblock {Fast and accurate determination of modularity and its effect size}.
\newblock {\em J Stat Mech}, 15:P02003, February 2015.
\newblock \href {http://dx.doi.org/10.1088/1742-5468/2015/02/P02003}
  {\path{doi:10.1088/1742-5468/2015/02/P02003}}.

\bibitem{MaKrFaVa06-phys}
Priya Mahadevan, Dmitri Krioukov, Kevin Fall, and Amin Vahdat.
\newblock {Systematic Topology Analysis and Generation Using Degree
  Correlations}.
\newblock {\em Comput Commun Rev}, 36(4):135--146, 2006.
\newblock \href {http://dx.doi.org/10.1145/1151659.1159930}
  {\path{doi:10.1145/1151659.1159930}}.

\bibitem{PrzuljHiddenLanguage2014}
\"{O}mer~Nebil Yavero{\u g}lu, No\"{e}l Malod-Dognin, Darren Davis, Zoran
  Levnajic, Vuk Janjic, Rasa Karapandza, Aleksandar Stojmirovic, and Nata\v{s}a
  Pr\v{z}ulj.
\newblock {Revealing the hidden language of complex networks}.
\newblock {\em Sci Rep}, 4:4547, January 2014.
\newblock \href {http://dx.doi.org/10.1038/srep04547}
  {\path{doi:10.1038/srep04547}}.

\bibitem{NewmanDseries2010}
B~Karrer and M~E~J Newman.
\newblock {Random graphs containing arbitrary distributions of subgraphs}.
\newblock {\em Phys Rev E}, 82(6):066118, December 2010.
\newblock \href {http://dx.doi.org/10.1103/PhysRevE.82.066118}
  {\path{doi:10.1103/PhysRevE.82.066118}}.

\bibitem{Coolen2011}
A~C~C Coolen, F~Fraternali, A~Annibale, L~Fernandes, and J~Kleinjung.
\newblock {\em {Modelling Biological Networks via Tailored Random Graphs}},
  pages 309--329.
\newblock John Wiley \& Sons, Ltd, 2011.
\newblock \href {http://dx.doi.org/10.1002/9781119970606.ch15}
  {\path{doi:10.1002/9781119970606.ch15}}.

\bibitem{CoMa09}
A~C~C Coolen, A~Martino, and A~Annibale.
\newblock {Constrained Markovian Dynamics of Random Graphs}.
\newblock {\em J Stat Phys}, 136(6):1035--1067, September 2009.
\newblock \href {http://dx.doi.org/10.1007/s10955-009-9821-2}
  {\path{doi:10.1007/s10955-009-9821-2}}.

\bibitem{AnCo09}
A~Annibale, A~C~C Coolen, L~Fernandes, F~Fraternali, and J~Kleinjung.
\newblock {Tailored graph ensembles as proxies or null models for real networks
  I: tools for quantifying structure}.
\newblock {\em J Phys A-Math Gen}, 42(48):485001, December 2009.
\newblock \href {http://dx.doi.org/10.1088/1751-8113/42/48/485001}
  {\path{doi:10.1088/1751-8113/42/48/485001}}.

\bibitem{Roberts2011a}
E~S Roberts, T~Schlitt, and A~C~C Coolen.
\newblock {Tailored graph ensembles as proxies or null models for real networks
  II: results on directed graphs}.
\newblock {\em J Phys A Math Theor}, 44(27):275002, July 2011.
\newblock \href {http://dx.doi.org/10.1088/1751-8113/44/27/275002}
  {\path{doi:10.1088/1751-8113/44/27/275002}}.

\bibitem{Roberts2012}
E~S Roberts and A~C~C Coolen.
\newblock {Unbiased degree-preserving randomization of directed binary
  networks}.
\newblock {\em Phys Rev E}, 85(4):046103, April 2012.
\newblock \href {http://dx.doi.org/10.1103/PhysRevE.85.046103}
  {\path{doi:10.1103/PhysRevE.85.046103}}.

\bibitem{LovaszBook2012}
L\'{a}szl\'{o} Lov\'{a}sz.
\newblock {\em {Large Networks and Graph Limits}}.
\newblock American Mathematical Society, Providence, RI, 2012.

\bibitem{ErRe59}
P~Erd\"{o}s and A~R\'{e}nyi.
\newblock {On Random Graphs}.
\newblock {\em Publ Math}, 6:290--297, 1959.

\bibitem{BeCa78}
E~Bender and E~Canfield.
\newblock {The Asymptotic Number of Labeled Graphs with Given Degree
  Distribution}.
\newblock {\em J Comb Theory A}, 24:296--307, 1978.

\bibitem{NewStrWat01}
M~E~J Newman, S~H Strogatz, and D~J Watts.
\newblock {Random Graphs with Arbitrary Degree Distributions and Their
  Applications}.
\newblock {\em Phys Rev E}, 64:26118, 2001.
\newblock \href {http://dx.doi.org/10.1103/PhysRevE.64.026118}
  {\path{doi:10.1103/PhysRevE.64.026118}}.

\bibitem{ChatterjeeDegreeSequences2011}
Sourav Chatterjee, Persi Diaconis, and Allan Sly.
\newblock {Random graphs with a given degree sequence}.
\newblock {\em Ann Appl Probab}, 21(4):1400--1435, August 2011.
\newblock \href {http://dx.doi.org/10.1214/10-AAP728}
  {\path{doi:10.1214/10-AAP728}}.

\bibitem{StantonJDD2012}
Isabelle Stanton and Ali Pinar.
\newblock {Constructing and sampling graphs with a prescribed joint degree
  distribution}.
\newblock {\em J Exp Algorithmics}, 17(1):3.1, July 2012.
\newblock \href {http://dx.doi.org/10.1145/2133803.2330086}
  {\path{doi:10.1145/2133803.2330086}}.

\bibitem{BianconiEntropy2008}
Ginestra Bianconi.
\newblock {The entropy of randomized network ensembles}.
\newblock {\em Eur Lett}, 81(2):28005, January 2008.
\newblock \href {http://dx.doi.org/10.1209/0295-5075/81/28005}
  {\path{doi:10.1209/0295-5075/81/28005}}.

\bibitem{BarvinokNumberOfGraphs2013}
A~Barvinok and J~A Hartigan.
\newblock {The number of graphs and a random graph with a given degree
  sequence}.
\newblock {\em Random Struct Algorithms}, 42(3):301--348, May 2013.
\newblock \href {http://dx.doi.org/10.1002/rsa.20409}
  {\path{doi:10.1002/rsa.20409}}.

\bibitem{HollandExponentialFamily1981}
Paul~W Holland and Samuel Leinhardt.
\newblock {An Exponential Family of Probability Distributions for Directed
  Graphs}.
\newblock {\em J Am Stat Assoc}, 76(373):33--50, 1981.

\bibitem{PaNe04}
J~Park and M~E~J Newman.
\newblock {Statistical Mechanics of Networks}.
\newblock {\em Phys Rev E}, 70:66117, 2004.
\newblock \href {http://dx.doi.org/10.1103/PhysRevE.70.066117}
  {\path{doi:10.1103/PhysRevE.70.066117}}.

\bibitem{ChatterjeeERGM2013}
Sourav Chatterjee and Persi Diaconis.
\newblock {Estimating and understanding exponential random graph models}.
\newblock {\em Ann Stat}, 41(5):2428--2461, October 2013.
\newblock \href {http://dx.doi.org/10.1214/13-AOS1155}
  {\path{doi:10.1214/13-AOS1155}}.

\bibitem{Horvat2014Degeneracy}
Sz. Horv\'{a}t, \'{E}. Czabarka, and Z.~Toroczkai.
\newblock {Reducing Degeneracy in Maximum Entropy Models of Networks}.
\newblock {\em Phys Rev Lett}, 114(15-17):158701, April 2015.
\newblock \href {http://dx.doi.org/10.1103/PhysRevLett.114.158701}
  {\path{doi:10.1103/PhysRevLett.114.158701}}.

\bibitem{GarlaschelliLikelihood2011}
Tiziano Squartini and Diego Garlaschelli.
\newblock {Analytical maximum-likelihood method to detect patterns in real
  networks}.
\newblock {\em New J Phys}, 13(8):083001, August 2011.
\newblock \href {http://dx.doi.org/10.1088/1367-2630/13/8/083001}
  {\path{doi:10.1088/1367-2630/13/8/083001}}.

\bibitem{Squartini2015Sampling}
Tiziano Squartini, Rossana Mastrandrea, and Diego Garlaschelli.
\newblock {Unbiased sampling of network ensembles}.
\newblock {\em New J Phys}, 17(2):023052, 2015.
\newblock \href {http://dx.doi.org/10.1088/1367-2630/17/2/023052}
  {\path{doi:10.1088/1367-2630/17/2/023052}}.

\bibitem{MaSne03}
Sergei Maslov, Kim Sneppen, and Uri Alon.
\newblock {\em {Handbook of Graphs and Networks}}, chapter~8.
\newblock Wiley-VCH, Berlin, 2003.

\bibitem{MaSneZa04}
Sergei Maslov, Kim Sneppen, and Alexei Zaliznyak.
\newblock {Detection of topological patterns in complex networks: Correlation
  profile of the Internet}.
\newblock {\em Physica A}, 333:529--540, February 2004.
\newblock \href {http://dx.doi.org/10.1016/j.physa.2003.06.002}
  {\path{doi:10.1016/j.physa.2003.06.002}}.

\bibitem{Gjoka2.5K2013}
Minas Gjoka, Maciej Kurant, and Athina Markopoulou.
\newblock {2.5K-graphs: From sampling to generation}.
\newblock In {\em 2013 Proc IEEE INFOCOM}, pages 1968--1976. IEEE, April 2013.
\newblock \href {http://dx.doi.org/10.1109/INFCOM.2013.6566997}
  {\path{doi:10.1109/INFCOM.2013.6566997}}.

\bibitem{Kim2009}
H.~Kim, Z.~Toroczkai, P.L. Erd\H{o}s, I.~Mikl\'{o}s, and L.A. Sz\'{e}kely.
\newblock {Degree-based graph construction}.
\newblock {\em J Phys A Math Theor}, 42(39):392001, October 2009.
\newblock \href {http://dx.doi.org/10.1088/1751-8113/42/39/392001}
  {\path{doi:10.1088/1751-8113/42/39/392001}}.

\bibitem{ZoltanSampling2010}
C.I {Del Genio}, H.~Kim, Z.~Toroczkai, and K.E. Bassler.
\newblock {Efficient and exact sampling of simple graphs with given arbitrary
  degree sequence}.
\newblock {\em PLoS One}, 5(4):e10012, January 2010.
\newblock \href {http://dx.doi.org/10.1371/journal.pone.0010012}
  {\path{doi:10.1371/journal.pone.0010012}}.

\bibitem{Kim2012}
H.~Kim, C.I. Del~Genio, K.E. Bassler, and Z.~Toroczkai.
\newblock Constructing and sampling directed graphs with given degree
  sequences.
\newblock {\em New J Phys}, 14:023012, 2012.
\newblock \href {http://dx.doi.org/10.1088/1367-2630/14/2/023012}
  {\path{doi:10.1088/1367-2630/14/2/023012}}.

\bibitem{Bassler2015}
K.E. Bassler, C.I. Del~Genio, P.L. Erd\H{o}s, I.~Mikl\'{o}s, and Z.~Toroczkai.
\newblock {Exact sampling of graphs with prescribed degree correlations}.
\newblock {\em New J Phys}, 17:083052, August 2015.
\newblock \href {http://dx.doi.org/10.1088/1367-2630/17/8/083052}
  {\path{doi:10.1088/1367-2630/17/8/083052}}.

\bibitem{Zlatic2009RichClub}
V.~Zlatic, G.~Bianconi, A.~D\'{\i}az-Guilera, D.~Garlaschelli, F.~Rao, and
  G.~Caldarelli.
\newblock {On the rich-club effect in dense and weighted networks}.
\newblock {\em Eur Phys J B}, 67(3):271--275, 2009.
\newblock \href {http://dx.doi.org/10.1140/epjb/e2009-00007-9}
  {\path{doi:10.1140/epjb/e2009-00007-9}}.

\bibitem{ErdosJDD2015}
\'{E}va Czabarka, Aaron Dutle, P\'{e}ter~L. Erd{\H o}s, and Istv\'{a}n
  Mikl\'{o}s.
\newblock {On realizations of a joint degree matrix}.
\newblock {\em Discret Appl Math}, 181:283--288, January 2015.
\newblock \href {http://dx.doi.org/10.1016/j.dam.2014.10.012}
  {\path{doi:10.1016/j.dam.2014.10.012}}.

\bibitem{JaMa09}
Almerima Jamakovic, Priya Mahadevan, Amin Vahdat, Mari\'{a}n Bogu\~{n}\'{a},
  and Dmitri Krioukov.
\newblock {How small are building blocks of complex networks}. {Preprint at
  \url{http://arxiv.org/abs/0908.1143}}, 2009.

\bibitem{MiloUniform2003}
R.~Milo, N.~Kashtan, S.~Itzkovitz, M~E~J Newman, and U.~Alon.
\newblock {On the uniform generation of random graphs with prescribed degree
  sequences}. {Preprint at \url{http://arxiv.org/abs/cond-mat/0312028}}, 2003.

\bibitem{BasslerSparse2011}
Charo {Del Genio}, Thilo Gross, and Kevin~E Bassler.
\newblock {All Scale-Free Networks Are Sparse}.
\newblock {\em Phys Rev Lett}, 107(17):1--4, October 2011.
\newblock \href {http://dx.doi.org/10.1103/PhysRevLett.107.178701}
  {\path{doi:10.1103/PhysRevLett.107.178701}}.

\bibitem{Gjoka2015JDD}
Minas Gjoka, Balint Tillman, and Athina Markopoulou.
\newblock {Construction of Simple Graphs with a Target Joint Degree Matrix and
  Beyond}.
\newblock In {\em 2015 Proc IEEE INFOCOM}, pages 1553--1561. IEEE, 2015.
\newblock \href {http://dx.doi.org/10.1109/INFOCOM.2015.7218534}
  {\path{doi:10.1109/INFOCOM.2015.7218534}}.

\bibitem{polcode}
Pol~Colomer de~Simon.
\newblock {RandNetGen: a Random Network Generator}.
\newblock URL: \url{http://polcolomer.github.io/RandNetGen/}.

\bibitem{DoMeSa01}
S~N Dorogovtsev, J.~Mendes, and A.~Samukhin.
\newblock {Size-dependent degree distribution of a scale-free growing network}.
\newblock {\em Phys Rev E}, 63(6):062101, May 2001.
\newblock \href {http://dx.doi.org/10.1103/PhysRevE.63.062101}
  {\path{doi:10.1103/PhysRevE.63.062101}}.

\bibitem{KlEg02}
Konstantin Klemm and V\'{\i}ctor Egu\'{\i}luz.
\newblock {Highly clustered scale-free networks}.
\newblock {\em Phys Rev E}, 65(3):036123, February 2002.
\newblock \href {http://dx.doi.org/10.1103/PhysRevE.65.036123}
  {\path{doi:10.1103/PhysRevE.65.036123}}.

\bibitem{Vazquez2003Growing}
Alexei V\'{a}zquez.
\newblock {Growing network with local rules: Preferential attachment,
  clustering hierarchy, and degree correlations}.
\newblock {\em Phys Rev E}, 67(5):056104, May 2003.
\newblock \href {http://dx.doi.org/10.1103/PhysRevE.67.056104}
  {\path{doi:10.1103/PhysRevE.67.056104}}.

\bibitem{Serrano2005Clustering}
M.~{\'{A}ngeles Serrano} and Mari\'{a}n Bogu\~{n}\'{a}.
\newblock {Tuning clustering in random networks with arbitrary degree
  distributions}.
\newblock {\em Phys Rev E}, 72(3):036133, September 2005.
\newblock \href {http://dx.doi.org/10.1103/PhysRevE.72.036133}
  {\path{doi:10.1103/PhysRevE.72.036133}}.

\bibitem{PaBoKr11}
Fragkiskos Papadopoulos, Maksim Kitsak, M.~\'{A}ngeles Serrano, Mari\'{a}n
  Bogu\~{n}\'{a}, and Dmitri Krioukov.
\newblock {Popularity versus similarity in growing networks}.
\newblock {\em Nature}, 489:537--540, September 2012.
\newblock \href {http://dx.doi.org/10.1038/nature11459}
  {\path{doi:10.1038/nature11459}}.

\bibitem{bianconi2014b}
Ginestra Bianconi, Richard~K. Darst, Jacopo Iacovacci, and Santo Fortunato.
\newblock {Triadic closure as a basic generating mechanism of communities in
  complex networks}.
\newblock {\em Phys Rev E}, 90:042806, 2014.
\newblock \href {http://dx.doi.org/10.1103/PhysRevE.90.042806}
  {\path{doi:10.1103/PhysRevE.90.042806}}.

\bibitem{KuBa06}
P~Kuo, W~Banzhaf, and A~Leier.
\newblock {Network Topology and the Evolution of Dynamics in an Artificial
  Genetic Regulatory Network Model Created by Whole Genome Duplication and
  Divergence}.
\newblock {\em Biosystems}, 85:177--200, 2006.
\newblock \href {http://dx.doi.org/10.1016/j.biosystems.2006.01.004}
  {\path{doi:10.1016/j.biosystems.2006.01.004}}.

\bibitem{MeLaMaBy01-phys}
A~Medina, A~Lakhina, I~Matta, and J~Byers.
\newblock {BRITE: An approach to universal topology generation}.
\newblock In {\em MASCOTS 2001, Proc Ninth Int Symp Model Anal Simul Comput
  Telecommun Syst}, pages 346--353, 2001.
\newblock \href {http://dx.doi.org/10.1109/MASCOT.2001.948886}
  {\path{doi:10.1109/MASCOT.2001.948886}}.

\bibitem{DiKr09}
Xenofontas Dimitropoulos, Dmitri Krioukov, George Riley, and Amin Vahdat.
\newblock {Graph Annotations in Modeling Complex Network Topologies}.
\newblock {\em ACM T Model Comput S}, 19(4):17, 2009.
\newblock \href {http://dx.doi.org/10.1145/1596519.1596522}
  {\path{doi:10.1145/1596519.1596522}}.

\bibitem{Foster2010Hysteresis}
David Foster, Jacob Foster, Maya Paczuski, and Peter Grassberger.
\newblock {Communities, clustering phase transitions, and hysteresis: Pitfalls
  in constructing network ensembles}.
\newblock {\em Phys Rev E}, 81(4):046115, April 2010.
\newblock \href {http://dx.doi.org/10.1103/PhysRevE.81.046115}
  {\path{doi:10.1103/PhysRevE.81.046115}}.

\bibitem{Coolen2014Loops}
E~S Roberts and A~C~C Coolen.
\newblock {Random Graph Ensembles with Many Short Loops}.
\newblock In {\em ESAIM Proc Surv}, volume~47, pages 97--115, 2014.
\newblock \href {http://dx.doi.org/10.1051/proc/201447006}
  {\path{doi:10.1051/proc/201447006}}.

\bibitem{Bollobas2011Sparse}
B\'{e}la Bollob\'{a}s and Oliver Riordan.
\newblock {Sparse graphs: Metrics and random models}.
\newblock {\em Random Struct Algorithms}, 39:1--38, 2011.
\newblock \href {http://dx.doi.org/10.1002/rsa.20334}
  {\path{doi:10.1002/rsa.20334}}.

\bibitem{Borgs2014LpI}
Christian Borgs, Jennifer~T. Chayes, Henry Cohn, and Yufei Zhao.
\newblock {An $L^p$ theory of sparse graph convergence I: Limits, sparse random
  graph models, and power law distributions}. {Preprint at
  \url{http://arxiv.org/abs/1401.2906}}, 2014.

\bibitem{AlvarezLaNet-vi2008}
M~G Beir\'{o}, J~I Alvarez-Hamelin, and J~R Busch.
\newblock {A low complexity visualization tool that helps to perform complex
  systems analysis}.
\newblock {\em New J Phys}, 10(12):125003, December 2008.
\newblock \href {http://dx.doi.org/10.1088/1367-2630/10/12/125003}
  {\path{doi:10.1088/1367-2630/10/12/125003}}.

\bibitem{CoPaSaVe07}
Vittoria Colizza, Romualdo Pastor-Satorras, and Alessandro Vespignani.
\newblock {Reaction-diffusion Processes and Metapopulation Models in
  Heterogeneous Networks}.
\newblock {\em Nat Phys}, 3:276--282, 2007.
\newblock \href {http://dx.doi.org/10.1038/nphys560}
  {\path{doi:10.1038/nphys560}}.

\bibitem{EgCh05}
Victor Egu\'{\i}luz, Dante Chialvo, Guillermo Cecchi, Marwan Baliki, and
  A.~Vania Apkarian.
\newblock {Scale-Free Brain Functional Networks}.
\newblock {\em Phys Rev Lett}, 94(1):018102, January 2005.
\newblock \href {http://dx.doi.org/10.1103/PhysRevLett.94.018102}
  {\path{doi:10.1103/PhysRevLett.94.018102}}.

\bibitem{MiItKaLeSoAyShAl04}
R~Milo, S~Itzkovic, N~Kashtan, R~Levitt, S~Shen-Orr, I~Ayzenshtat, M~Sheffer,
  and U~Alon.
\newblock {Superfamilies of Evolved and Designed Networks}.
\newblock {\em Science}, 303:1538--1542, 2004.
\newblock \href {http://dx.doi.org/10.1126/science.1089167}
  {\path{doi:10.1126/science.1089167}}.

\bibitem{MaKrFo06}
Priya Mahadevan, Dmitri Krioukov, Marina Fomenkov, Bradley Huffaker, Xenofontas
  Dimitropoulos, Kc~Claffy, and Amin Vahdat.
\newblock {The Internet AS-Level Topology: Three Data Sources and One
  Definitive Metric}.
\newblock {\em Comput Commun Rev}, 36(1):17--26, 2006.
\newblock \href {http://dx.doi.org/10.1145/1111322.1111328}
  {\path{doi:10.1145/1111322.1111328}}.

\bibitem{BoPa04}
Mari\'{a}n Bogu\~{n}\'{a}, Romualdo Pastor-Satorras, Albert D\'{\i}az-Guilera,
  and Alex Arenas.
\newblock {Models of social networks based on social distance attachment}.
\newblock {\em Phys Rev E}, 70(5):056122, November 2004.
\newblock \href {http://dx.doi.org/10.1103/PhysRevE.70.056122}
  {\path{doi:10.1103/PhysRevE.70.056122}}.

\bibitem{Vidal2014HI}
Thomas Rolland, Murat Tas, Nidhi Sahni, Song Yi, Irma Lemmens, Celia
  Fontanillo, Roberto Mosca, Atanas Kamburov, Susan~D Ghiassian, Xinping Yang,
  Lila Ghamsari, Dawit Balcha, Bridget~E Begg, Pascal Braun, Marc Brehme,
  Martin~P Broly, Anne-ruxandra Carvunis, Dan Convery-zupan, Roser Corominas,
  Changyu Fan, Eric Franzosa, Jasmin Coulombe-huntington, Elizabeth Dann,
  Matija Dreze, Fana Gebreab, Bryan~J Gutierrez, Madeleine~F Hardy, Mike Jin,
  Shuli Kang, Ruth Kiros, Guan~Ning Lin, Ryan~R Murray, Alexandre Palagi,
  Matthew~M Poulin, Katja Luck, Andrew Macwilliams, Xavier Rambout, John Rasla,
  Patrick Reichert, Viviana Romero, Elien Ruyssinck, Julie~M Sahalie, Annemarie
  Scholz, Akash~a Shah, Amitabh Sharma, Yun Shen, Kerstin Spirohn, Stanley Tam,
  Alexander~O Tejeda, Shelly~a Trigg, Jean-claude Twizere, Kerwin Vega, and
  Jennifer Walsh.
\newblock {A Proteome-Scale Map of the Human Interactome Network}.
\newblock {\em Cell}, 159:1212--1226, 2014.
\newblock \href {http://dx.doi.org/10.1016/j.cell.2014.10.050}
  {\path{doi:10.1016/j.cell.2014.10.050}}.

\bibitem{SeBo06a}
M.~\'{A}ngeles Serrano and Mari\'{a}n Bogu\~{n}\'{a}.
\newblock {Clustering in complex networks. I. General formalism}.
\newblock {\em Phys Rev E}, 74(5):056114, November 2006.
\newblock \href {http://dx.doi.org/10.1103/PhysRevE.74.056114}
  {\path{doi:10.1103/PhysRevE.74.056114}}.

\bibitem{SeBo06b}
M.~\'{A}ngeles Serrano and Mari\'{a}n Bogu\~{n}\'{a}.
\newblock {Clustering in complex networks. II. Percolation properties}.
\newblock {\em Phys Rev E}, 74(5):056115, August 2006.
\newblock \href {http://dx.doi.org/10.1103/PhysRevE.74.056115}
  {\path{doi:10.1103/PhysRevE.74.056115}}.

\bibitem{AlvarezKcore2008}
Jos\'{e} Alvarez-Hamelin, Luca Dall'Asta, Alain Barrat, and Alessandro
  Vespignani.
\newblock {K-core decomposition of Internet graphs: hierarchies,
  self-similarity and measurement biases}.
\newblock {\em Networks Heterog Media}, 3(2):371--393, March 2008.
\newblock \href {http://dx.doi.org/10.3934/nhm.2008.3.371}
  {\path{doi:10.3934/nhm.2008.3.371}}.

\bibitem{SaitoKdensity2006}
Kazumi Saito and Takeshi Yamada.
\newblock {Extracting Communities from Complex Networks by the k-dense Method}.
\newblock In {\em Sixth IEEE Int Conf Data Min - Work}, pages 300--304. IEEE,
  2006.
\newblock \href {http://dx.doi.org/10.1109/ICDMW.2006.76}
  {\path{doi:10.1109/ICDMW.2006.76}}.

\bibitem{Zlatic2012Multiplicities}
Vinko Zlati\'{c}, Diego Garlaschelli, and Guido Caldarelli.
\newblock {Networks with arbitrary edge multiplicities}.
\newblock {\em EPL}, 97(2):28005, January 2012.
\newblock \href {http://dx.doi.org/10.1209/0295-5075/97/28005}
  {\path{doi:10.1209/0295-5075/97/28005}}.

\bibitem{wang2003b}
Yang Wang, Deepayan Chakrabarti, Chenxi Wang, and Christos Faloutsos.
\newblock {Epidemic spreading in real networks: An eigenvalue viewpoint}.
\newblock In {\em Proceedings of the 22nd International Symposium on Reliable
  Distributed Systems}, pages 25--34. IEEE, 2003.
\newblock \href {http://dx.doi.org/10.1109/RELDIS.2003.1238052}
  {\path{doi:10.1109/RELDIS.2003.1238052}}.

\bibitem{dagostino2012}
Gregorio D'Agostino, Antonio Scala, Vinko Zlati{\'c}, and Guido Caldarelli.
\newblock {Robustness and assortativity for diffusion-like processes in
  scale-free networks}.
\newblock {\em EPL}, 97(6):68006, 2012.
\newblock \href {http://dx.doi.org/10.1209/0295-5075/97/68006}
  {\path{doi:10.1209/0295-5075/97/68006}}.

\bibitem{van2010}
Piet Van~Mieghem.
\newblock {\em {Graph spectra for complex networks}}.
\newblock Cambridge University Press, Cambridge, 2010.

\end{thebibliography}

\end{document}